\documentclass[onecolumn, numberedappendix]{aastex6}
\usepackage{color}

\newcommand{\eg}{{\it e.g.}}
\newcommand{\etal}{et~al.}

\newcommand{\ks}{$K_{\rm s}$}
\newcommand{\vmkz}{$(V-K_{\rm s})_0$}
\newcommand{\vmk}{$(V-K_{\rm s})$}

\newcommand{\teff}{$T_{\rm eff}$}

\begin{document}

\title{Rotation of Late-Type Stars in Praesepe with K2 }

\slugcomment{Version from \today}

\author{L.~M.~Rebull\altaffilmark{1,2},
J.~R.~Stauffer\altaffilmark{2},
L.~A.~Hillenbrand\altaffilmark{3}, 
A.~M.~Cody\altaffilmark{4},
J.~Bouvier\altaffilmark{5}, 
D.~R.~Soderblom\altaffilmark{6}, 
M.~Pinsonneault\altaffilmark{7}, 
L.~Hebb\altaffilmark{8}}

\altaffiltext{1}{Infrared Science Archive (IRSA), Infrared Processing
and Analysis Center (IPAC), 1200 E.\ California Blvd., California
Institute of Technology, Pasadena, CA 91125, USA; rebull@ipac.caltech.edu}
\altaffiltext{2}{Spitzer Science Center (SSC), Infrared Processing and
Analysis Center (IPAC), 1200 E.\ California Blvd., California
Institute of Technology, Pasadena, CA 9112, USA5}
\altaffiltext{3}{Astronomy Department, California Institute of
Technology, Pasadena, CA 91125, USA}
\altaffiltext{4}{NASA Ames Research Center, Kepler Science Office,
Mountain View, CA 94035, USA}
\altaffiltext{5}{Universit\'e de Grenoble, Institut de Plan\'etologie
et d'Astrophysique de Grenoble (IPAG), F-38000 Grenoble, France;
CNRS, IPAG, F-38000 Grenoble, France}
\altaffiltext{6}{Space Telescope Science Institute, 3700 San Martin Drive,
Baltimore, MD 21218, USA; Center for Astrophysical Sciences, Johns Hopkins University,
3400 North Charles St., Baltimore, MD 21218, USA and Center for
Astrophysical Sciences, Johns Hopkins University, Baltimore, MD 21218}
\altaffiltext{7}{Department of Astronomy, The Ohio State University,
Columbus, OH 43210, USA; Center for Cosmology and Astroparticle
Physics, The Ohio State University, Columbus, OH 43210, USA}
\altaffiltext{8}{Department of Physics, Hobart and William Smith
Colleges, Geneva, NY 14456, USA}

\begin{abstract}  

We have Fourier analyzed 941 K2 light curves of likely members of
Praesepe, measuring periods for 86\% and increasing the number of
rotation periods ($P$) by nearly a factor of four. The distribution of
$P$ vs.\ \vmk, a mass proxy, has three different regimes: \vmk$<$1.3,
where the rotation rate rapidly slows as mass decreases; 1.3$<$ \vmk
$<$4.5, where the rotation rate slows more gradually as mass
decreases; and \vmk$>$4.5, where the rotation rate rapidly increases
as mass decreases. In this last regime, there is a bimodal
distribution of periods, with few between $\sim$2 and $\sim$10 days.
We interpret this to mean that once M stars start to slow down, they
do so rapidly. The K2 period-color distribution in Praesepe ($\sim$790
Myr) is much different than in the Pleiades ($\sim$125 Myr) for late
F, G, K, and early-M stars; the overall distribution moves to longer
periods, and is better described by 2 line segments. For mid-M stars,
the relationship has similarly broad scatter, and is steeper in
Praesepe. The diversity of lightcurves and of periodogram types is
similar in the two clusters; about a quarter of the periodic stars in
both clusters have multiple significant periods. Multi-periodic stars
dominate among the higher masses, starting at a bluer color in
Praesepe (\vmk$\sim$1.5) than in the Pleiades (\vmk$\sim$2.6). In
Praesepe, there are relatively more light curves that have two widely
separated periods, $\Delta P >$6 days. Some of these could be examples
of M star binaries where one star has spun down but the other
has not.

\end{abstract}

\section{Introduction}
\label{sec:intro}

Praesepe (M44 = NGC~2632 = the Beehive cluster) and the Hyades are
often considered as twin open clusters because they appear to have
essentially the same age ($\sim$790 Myr) and metallicity
([Fe/H]$\sim$0.15).   Some authors attribute these similarities to the
two clusters having been born within the same molecular cloud or cloud
complex (Schwarzschild \& Hertzsprung 1913, Klein-Wassink 1927);  the
space motions appear to support this picture (see, \eg, Eggen 1992). 
Because the Hyades is much closer than Praesepe (44 pc vs.\ $\sim$180
pc; Perryman \etal\ 1998, van Leeuwen 2007),  its stars are much
brighter; the Hyades has therefore received much more attention in the
published literature. However, proximity also has its disadvantages. 
The Hyades is spread out over a  very large area on the sky, making it
difficult to take advantage of multi-object spectrographs or
wide-field cameras to efficiently survey many cluster members at
once.   Praesepe's four times greater distance can become an advantage
in that sense, if a suitable wide-area facility becomes available. We
have taken advantage of one such circumstance, the unexpected
availability of the 100 square degree FOV Kepler space telescope to
obtain synoptic photometry of ecliptic plane fields for $\sim$75 days.

There have been four prior large surveys to determine rotation periods
for low mass  stars in Praesepe.  Scholz \etal\ (2011) reported
$\sim$50 periods; Delorme \etal\ (2011) used data from the WASP
telescope network to monitor $\sim$70 members; Kov\'acs \etal\ (2014)
monitored $\sim$400 members with the HAT telescope network, and a pair
of papers by Agueros \etal\ (2011) and Douglas \etal\ (2014) used data
from the PTF telescope to monitor $\sim$500 members.  Those groups
reported periods for 49, 52, 180, and 40 Praesepe members,
respectively (and a total of 220 unique stars with periods), primarily
covering stars with spectral types G, K and early M.

The K2 data for Praesepe were obtained in 2015 in Campaign 5.   K2
provides precision, sensitivity and long duration, continuous coverage
that yields superb light curves (LCs) of much greater quality than can
be obtained from the ground, and extending to much lower signal
amplitudes and masses. At least three groups have already obtained
rotation periods for Praesepe members based on these data.  Libralato
\etal\ (2016) derived rotation periods as a step in searching the
light curves for exoplanet transits, but they did not discuss the
periods nor the shapes of the phased light curves. Mann \etal\ (2016)
similarly derive rotation periods while searching for exoplanet
transits; they illustrate the distribution of rotation period as a
function of $M_k$ and color, but do not otherwise discuss the periods
or light curve morphology.   Finally, our group plotted an early
version of the Praesepe period distribution in Stauffer \etal\ (2016)
in the context of comparing the Pleiades rotation periods to those in
other clusters.  However, no paper has yet published a complete
analysis of the K2 Praesepe rotational data nor a discussion of the
light curve morphologies shown by those stars.   This paper is devoted
to those two topics.

Much of our current analysis is very similar to that we conducted in
the Pleiades (Rebull \etal\ 2016a,b, Stauffer \etal\ 2016), and we
often make comparisons to the Pleiades analysis and results.  In
Section~\ref{sec:obs}, we present the observations and data reduction,
as well as the range of data from the literature that we assembled,
and define the final sample of members that we analyze in the rest of
the paper. The overall distribution of K2-derived rotation rates is
discussed in Section~\ref{sec:rotationdistrib}. 
Section~\ref{sec:LCPcats} places these LCs in the same categories as
we defined in the Pleiades (Rebull \etal\ 2016b).
Section~\ref{sec:singleandmulti} compares the locations of the single-
and multi-period stars in a variety of ways, and
Section~\ref{sec:deltaP} looks at the spacings between periods for
those stars with clear multiple periods. Finally, we summarize our
results in Section~\ref{sec:concl}.

\section{Observations and Methods}
\label{sec:obs}

\subsection{K2 Data}

\begin{figure}[ht]
\epsscale{0.7}
\plotone{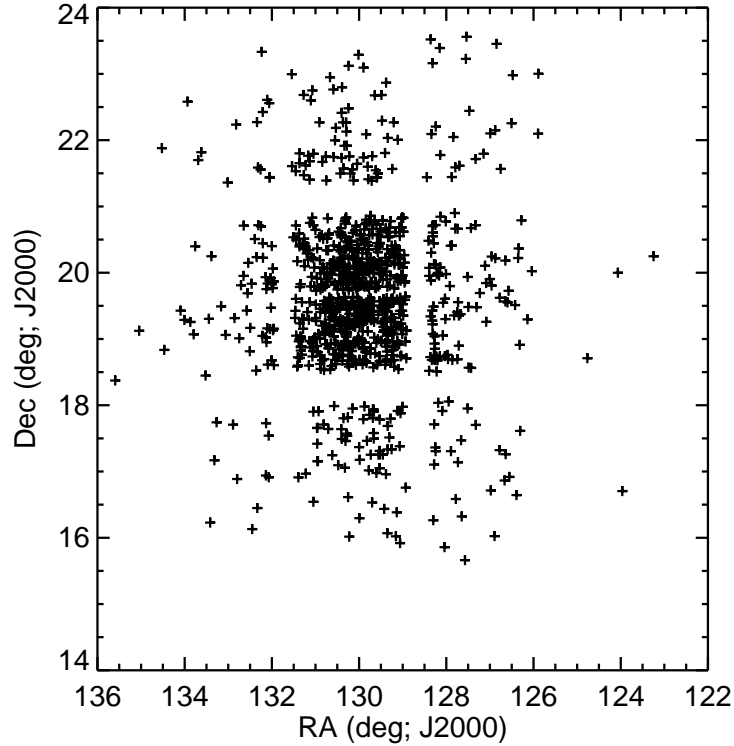}
\caption{All 984 members or candidate members  (see
Sec~\ref{sec:membership})  of Praesepe with K2 LCs projected onto the
sky. Note the gaps between K2 detectors. Stars that are $>$5$\arcdeg$
from the cluster center are highlighted in App.~\ref{app:halo}.}
\label{fig:where}
\end{figure}

Praesepe members and candidate members were observed in K2 campaign 5,
which lasted for 75 d between April and October 2015.
Fig.~\ref{fig:where} shows the distribution of objects with K2 LCs on
the sky; note the gaps between detectors.   All of the stars shown
were observed in the long-cadence ($\sim$30 min exposure) mode. Some
of these stars were additionally observed in fast cadence ($\sim$1 min
exposure), but those data are beyond the scope of the present paper. 
There are 984 unique K2 long-cadence light curves corresponding to
members or candidate members of Praesepe (see Sec~\ref{sec:membership}
below). The tidal radius of Praesepe is 12.1 pc (Holland \etal\ 2000),
which at a distance of 184 pc (see Sec.~\ref{sec:suppdata} below) is
$\sim$3.8$\arcdeg$  across, which is easily covered by K2. (Stars that
are more than 5$\arcdeg$ away from the cluster center are highlighted
in App.~\ref{app:halo}.)

Kepler pixel sizes are relatively large, $3.98\arcsec \times 3.98
\arcsec$, and the 95\% encircled energy diameter ranges from 3.1 to
7.5 pixels with a median value of 4.2 pixels.  During the K2 portion
of the mission, because only two reaction wheels can be used, the
whole spacecraft slowly drifts and then repositions regularly every
0.245 d. This drift is $\sim$0.1$\arcsec$ per hour (Cody \etal\ 2017).

Since these data were reduced at the same time as we reduced our
Pleiades data (Rebull \etal\ 2016a), we have the same sets of LCs
available to us.  (1) The pre-search data conditioning (PDC) version
generated by the Kepler project and obtained from MAST, the Mikulski
Archive for Space Telescopes. (2) A version with moving apertures
obtained following Cody \etal\ in prep. (3) The version using a
semiparametric Gaussian process model used by Aigrain \etal\ (2015,
2016). (4) The `self-flat-fielding' approach used by Vanderburg \&
Johnson (2014) as obtained from MAST. To this, late in the process, we
added (5) the LCs from the EVEREST pipeline (Luger \etal\ 2016),
which uses pixel level decorrelation.  We removed any data points
corresponding to thruster firings and any others with bad data flags
set in the corresponding data product.

Again, following the same approach as in the Pleiades, we inspected
LCs from each reduction approach, and we selected the visually `best'
LC from among the original four LC versions. Since the EVEREST LCs
became available late in our analysis process, for the most part, we
used the EVEREST LCs to break ties or clarify what the LC was doing.
About 7\% had such severe artifacts that no `best' LC could be
identified; as for the Pleiades, this is often a result of
instrumental (non-astrophysical) artifacts (because the star is too
bright or too faint or adversely affected by nearby stars, etc.). As
for the Pleiades, though, the periods we report here are not generally
ambiguous, and are detected in all the LC versions; they just look
cosmetically better in one version or another. Our approach of
comparing several different LC versions also minimizes the liklihood
that any of these LC reductions removes stellar signal.

Out of the 984 LCs, only one pair of stars for which a K2 LC was
requested were within 4$\arcsec$ of each other (within a Kepler
pixel). These objects are discussed further in App.~\ref{app:pair}.

\subsection{Finding Periods}
\label{sec:periods}

Our approach for finding periods was identical to
that which we used in the Pleiades in Rebull \etal\ (2016a). In
summary, we used the Lomb-Scargle (LS; Scargle 1982) approach as
implemented by the NASA Exoplanet Archive Periodogram
Service\footnote{http://exoplanetarchive.ipac.caltech.edu/cgi-bin/Periodogram/nph-simpleupload}
(Akeson \etal\ 2013). We looked for periods between 0.05 and 35 d,
with the upper limit being set by roughly half the campaign length.

\begin{figure}[ht]
\epsscale{1.0}
\plotone{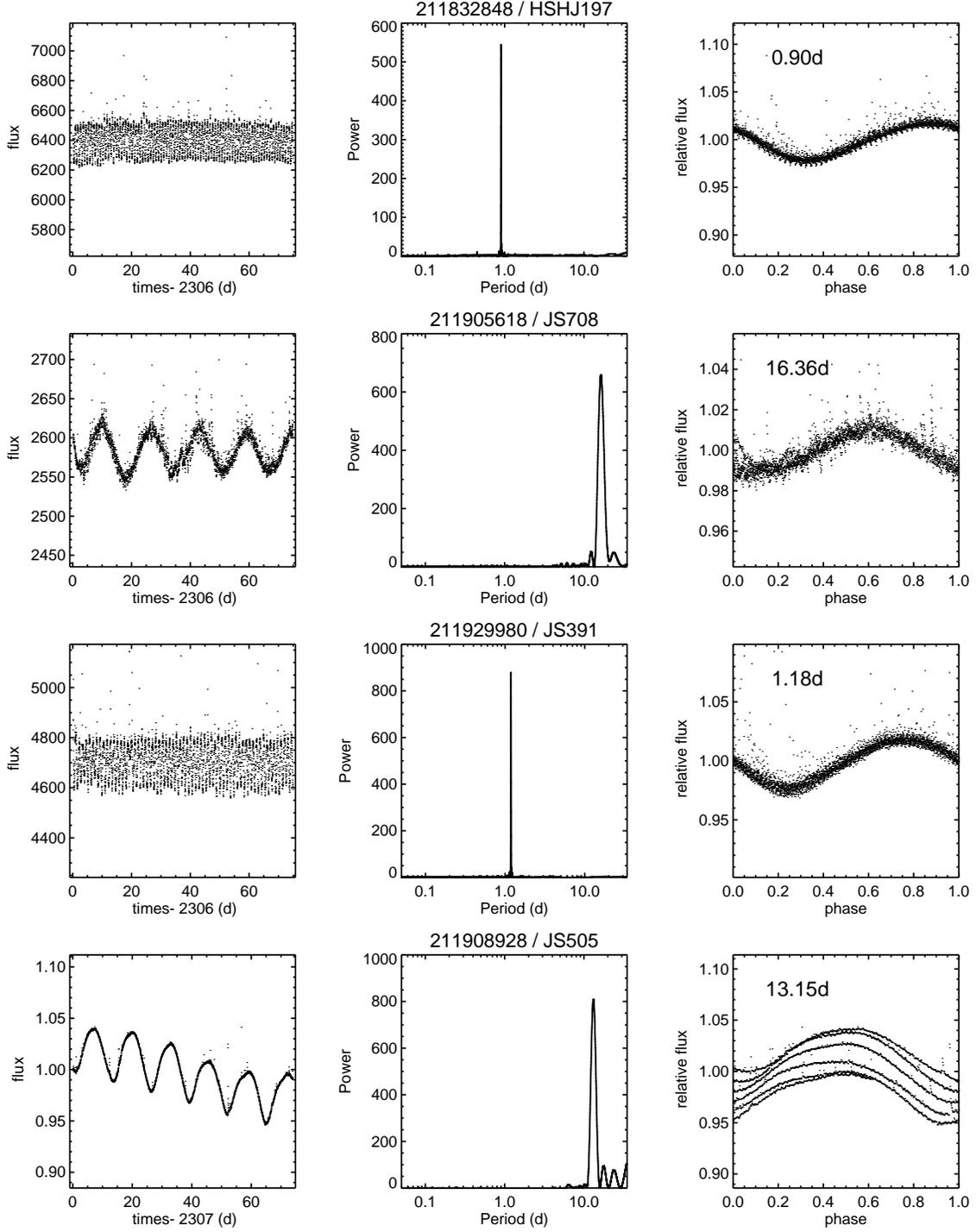}
\caption{Five examples of finding single sinusoidal periods in the K2
Praesepe data. Left column: full LC; middle column: LS periodogram;
right column: phased LC, with best period (in days) as indicated.
Rows, in order: EPIC 211832848/HSHJ197,
211905618/JS708,
211929980/JS391, and
211908928/JS505. These are
representatives from a range of brightnesses and periods. Note that in
each case, the power spectrum indicates unambiguously periodic signals
-- the peak is so high that little structure other than the peak  can
be seen in the power spectrum. These LCs are best interpreted as large
spots or spot groups rotating into and out of view. }
\label{fig:singlefreq}
\end{figure}

Fig.~\ref{fig:singlefreq} shows LCs, periodograms, and phased LCs for
some Praesepe stars with single, unambiguous periods. As we
found in the Pleiades, the overwhelming majority of them have
zero, or effectively zero, false alarm probability (FAP); a
significant fraction have more than one peak with 0 FAP; see
Section~\ref{sec:LCPcats} below.

\subsection{Interpretation of Periods}
\label{sec:interpofperiods}

For stars of the mass range considered here, the phased LCs are mostly
sinsuoidal and therefore best attributed to star spot-modulated
rotation.  We find periods for 828/984 K2 light curves, or 84\% of all
K2 light curves of candidate or confirmed Praesepe members (see
Sec.~\ref{sec:membership}). There is no significant trend with color
(as a proxy for mass) for the fraction of periodic stars. 

As for the Pleiades, we removed from this distribution any periods
that are likely to be eclipsing binaries (see Appendix
\ref{app:binaries}) or those whose waveforms did not seem to be
rotation periods (see Appendix~\ref{app:timescales}). 

There are a few multi-periodic stars which we suspect to be pulsating
variables (see Appendix~\ref{app:pulsator}); many of them are also
reported in the literature as $\delta$ Scuti-type variables. For some
pulsators, the first period is still likely to be similar to the
rotation period (see discussion in Rebull \etal\ 2016b). For the
remaining multi-periodic stars ($\sim$25\% of the sample; see
\S\ref{sec:LCPcats} below), for the most part, we took the period
corresponding to the strongest peak in the periodogram as the rotation
period to be used for subsequent analysis. In a few cases (e.g.,
double dip stars; see Sec.~\ref{sec:singleandmulti}), a secondary peak is the
right $P_{\rm rot}$ to use.

\clearpage
\subsection{Comparison to Literature Periods}
\label{sec:lit}

\begin{figure}[h]
\epsscale{0.4}
\plotone{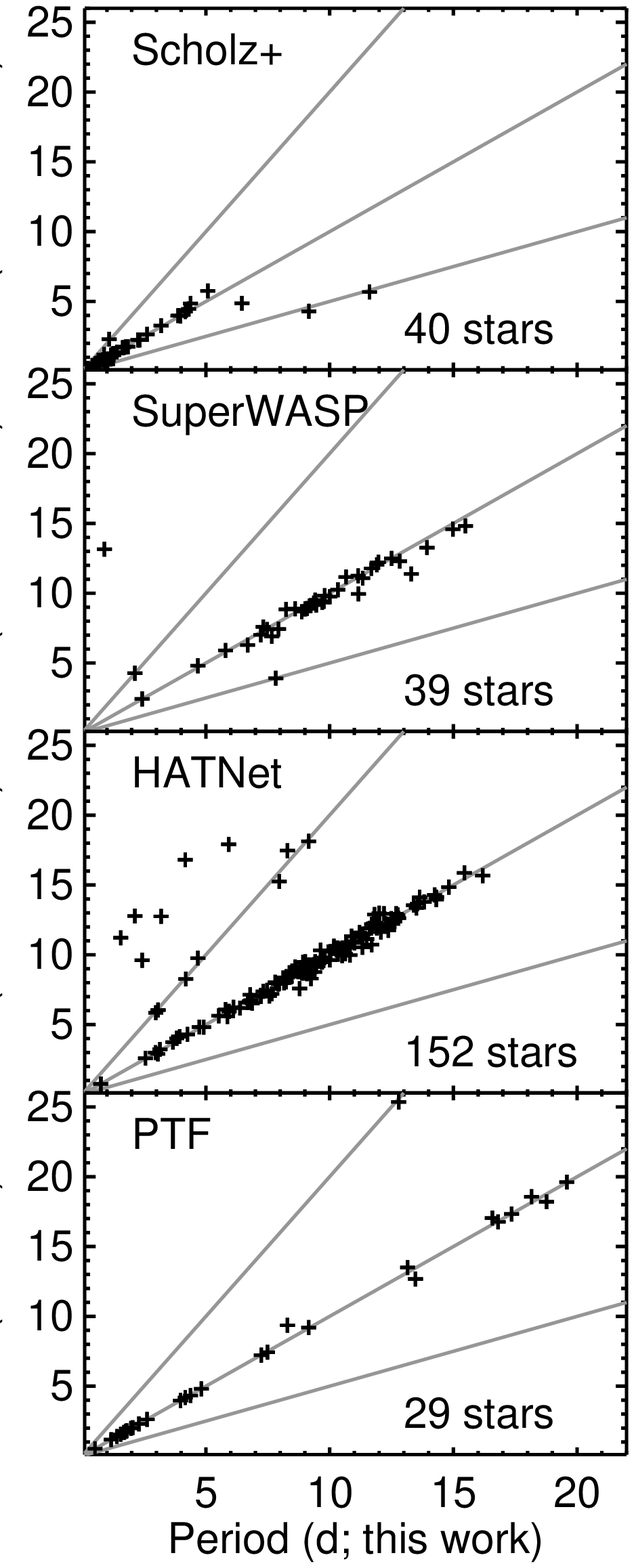}
\caption{Comparison of periods obtained here to periods obtained in
the literature. First panel: Scholz \etal\ 2001, 40 stars in common;
second panel: SuperWASP (Delorme \etal\ 2011, 39 stars in common);
third panel: HATNet  (Kov\'acs \etal\ 2014, 151 stars in common);
fourth panel: PTF (Ag\"ueros \etal\ 2011, 29 stars in common). The
grey lines are at 1-to-1, $P$/2, and 2$P$. Most of the periods match
well (see text). }
\label{fig:compareliterature}
\end{figure}

We have chosen four of the most recent surveys looking for rotation
rates in Praesepe for a detailed comparison of our results; they are
summarized in Figure~\ref{fig:compareliterature}. There are 220
Praesepe stars in the literature with at least one estimate of $P_{\rm
rot}$; 60 of those do not have K2 LCs.

Scholz \etal\ (2011) used the Isaac Newton Telescope to monitor very
low mass members, reporting 49 periods. We have periods for 40 of the
stars in common between the two studies; the remaining Scholz \etal\
stars were not monitored (either because they are too faint for K2, or
because they are in the gaps between chips).  The periods derived for
these 40 are shown in Figure~\ref{fig:compareliterature}. There is
generally very good agreement; the median fractional difference
($|P_{\rm Scholz+} - P_{\rm here}|/ P_{\rm here}$)  is 2.4\%. There
are 8 stars for which their period and ours do not match, and most of
them are harmonics, because they are factors of two off (too small or
large). In four of these cases, the period we derive is very close to
1 day, which is hard to recover with a ground-based telescope. We
believe our periods to be correct in all cases for the epochs in which
we observed this cluster. Two stars are of special note. EPIC
211951438/HSHJ396 has a period of 0.593 d in Scholz \etal, but we
cannot recover any period for this star, so we do not retain it
as periodic. EPIC
211984058/2MASSJ08384128+1959471 was not a star for which we initially
found a period, because all five light curve versions are different,
and it was not clear which LC version is the `best' (see discussion
above in \S\ref{sec:obs}).  However, upon examination of the K2
thumbnail and POSS images of the region, several of the LC versions
appear to have been dragged off to a nearby bright star. The Aigrain
\etal\ version is the only one that both stays on the target star and
has a periodic signal. The period we derive from this is identical to
that from Scholz \etal\ (2011), so we retained this star as periodic
with that period.

Delorme \etal\ (2011) used SuperWASP to monitor 71 cluster members,
looking for periods from 1.1 to 20 d, determining  that 52 were
periodic. Of these periodic stars, we have 39 in common (the remaining
13 fall in the gaps between K2 chips, or are otherwise off the K2
FOV). The periods derived for these 39 are shown in
Figure~\ref{fig:compareliterature}. There is generally very good
agreement; the median fractional difference ($|P_{\rm SuperWASP} -
P_{\rm here}|/ P_{\rm here}$) is 2.7\%. There are three stars for
which our periods do not agree.  One is EPIC
212013132/JS379=2MASSJ08404426+2028187, where it is likely that one of
the surveys detected an harmonic. SuperWASP obtains a period of 4.27
d; we obtain three periods, including one that is close to 4.4 d, but
the primary period we obtained for this star is 2.129 d. The second
star is EPIC 211950227/2MASSJ08402554+1928328, where the K2 light
curve is very, very messy. SuperWASP obtains 13.15 d. The period we
adopted for this one is 0.8984 d, and comes from only part of the LC,
with most of the LC having been corrupted by instrumental effects. The
third star is EPIC 211995288/KW30=2MASSJ08372222+2010373, where we
find a period of 7.8 d, a factor of 2 larger than the 3.9 d found in
SuperWASP, so another harmonic. Having tested these periods by
phasing our LCs at these alternate periods, we believe our periods to
be correct in all cases for the epochs in which we observed these
stars.  There are no stars in common where SuperWASP has a period and
we do not have a period.

Kov\'acs \etal\ (2014) used HATNet (Hungarian Automated Telescope
Networks) to monitor 381 members, finding 180 rotation periods, all
ranging between 2.5 and 15 d.  They identified 10 more stars as having
what we would call `timescales'; that is, repeated patterns that may
or may not correspond to periods.  There are 152 stars (150 out of
their rotation periods, and 2 more out of their `timescales') in
common between the surveys, shown in
Figure~\ref{fig:compareliterature}; most of the remaining stars fall
in the gaps between chips, while a few could have been but simply were
not observed. For the stars that we have in common, the median
fractional difference ($|P_{\rm HATNet} - P_{\rm here}|/ P_{\rm
here}$) is 2.2\%. There are 13 stars for which the periods do not
match, 7 of which are likely harmonics; we believe our period to be
correct. One of the two from their `timescales' category is  EPIC
211918335/HAT-269-0000582/KW244=TX Cnc, which Kov\'acs \etal\ identify
as an eclipsing binary and Whelan \etal\ (1973) identifies it as a
W~UMa-type; we drop this as $P_{\rm orb}$.
The other is 211947631/HAT-269-0000465/BD+19d2087 (which Kov\'acs
\etal\ simply call `miscellaneous'); we retain $P$=4.74 as a rotation
period. There are no stars for which we both have light curves but
HATNet has a period and we do not.

Ag\"ueros \etal\ (2011) and  Douglas \etal\ (2014) are a 2-part study
on the Palomar Transient Factory (PTF) observations of Praesepe. They
used PTF to monitor 534 members, 40 of which were periodic between
0.52 and 35.85 d. There are 30 stars in common between the studies
(see Figure~\ref{fig:compareliterature}), with the remaining 10
falling in the gaps between chips. The derived periods are generally
in very good agreement; the median fractional difference ($|P_{\rm
PTF} - P_{\rm here}|/ P_{\rm here}$) is just 0.36\%.  There is one
star (EPIC 211989299/AD2552=2MASSJ08392244+2004548) for which one of
us is likely to have found an harmonic, though the light curve is
messy (we obtain 12.791 and they obtain 25.36). We believe our
periods to be correct for the epochs in which we observed
these stars. For another one (EPIC 211886612/JS525), they found a
period of 22.6 d; we retrieved a LS peak at 17.2 d for this object,
but relegated it to `timescales' (in  Appendix~\ref{app:timescales};
this is why only 29 stars appear in Fig.~\ref{fig:compareliterature}).
For the light curve we have, it just does not look as nice as the rest
of the clear $P_{\rm rot}$ here.   

We note here that, using K2, we find periods for a far higher number
and higher fraction of Praesepe members than has been accomplished
before. Moving to space enables higher precision and continuous
photometry, resulting in a far higher fraction of detectable rotation
periods.

\subsection{Membership and Supporting Data}
\label{sec:suppdata}

We started with the list of Praesepe members that we had proposed for
this campaign, and added targets from other related programs. From
this list, we then amassed supporting data and assessed membership for
each star. For some objects, we obtained additional Keck/HIRES
spectroscopy; see App.~\ref{app:keck}.

Table~\ref{tab:bigperiods} includes the supporting photometric data we
discuss in this section, plus the periods we derive here (in
Section~\ref{sec:periods}), for members of Praesepe. For completeness,
non-members (NM) appear in Appendix~\ref{app:nm}.

\floattable
\begin{deluxetable}{cl}
\tabletypesize{\scriptsize}
\tablecaption{Contents of Table: Periods and Supporting Data for
Praesepe Members with K2 Light Curves\label{tab:bigperiods}}
\tablewidth{0pt}
\tablehead{\colhead{Label} & \colhead{Contents}}
\startdata
EPIC & Number in the Ecliptic Plane Input Catalog (EPIC) for K2\\
coord & Right ascension and declination (J2000) for target \\
othername & Alternate name for target \\
Vmag & V magnitude (in Vega mags), if observed\\
Kmag & \ks\ magnitude (in Vega mags), if observed\\
vmk &  \vmk, as directly observed (if $V$ and \ks\
exist), or as inferred (see text)\\
P1 & Primary period, in days (taken to be rotation period)\\
P2 & Secondary period, in days\\
P3 & Tertiary period, in days\\
P4 & Quaternary period, in days\\
LC & LC used as `best'\tablenotemark{a}\\
single/multi-P & indicator of whether single or multi-period star; if
object has a `timescale' (see App~\ref{app:timescales}), that is 
indicated\\
dd &  indicator of whether or not it is a double-dip LC \\
ddmoving & indicator of whether or not it is a moving double-dip LC\\
shch & indicator of whether or not it is a shape changer\\
beat & indicator of whether or not the full LC has beating visible\\
cpeak & indicator of whether or not the power spectrum has a complex,
structured peak and/or has a wide peak\\
resclose & indicator of whether or not there are resolved close
periods in the power spectrum\\
resdist & indicator of whether or not there are resolved distant
periods in the power spectrum\\
pulsator & indicator of whether or not the power spectrum and period
suggest that this is a  pulsator\\
\enddata
\tablenotetext{a}{LC1=PDC, from MAST; LC2=version following Cody
\etal\ in prep; LC3=version following Aigrain \etal\ (2015, 2016);
LC4=version reduced by Vanderburg \& Johnson (2014) and downloaded
from MAST; LC5=version reduced by EVEREST code (Luger \etal\ 2016) and
downloaded from MAST.}
\end{deluxetable}


\subsubsection{Metallicity, Reddening, Age}

Despite Praesepe's proximity and richness, its age and metallicity are
a subject of considerable debate.  Age estimates range from 600 to 900
Myr: Harris (1976), 630 Myr; Mermilliod (1981), 660 Myr; vandenBerg \&
Bridges (1984), 900 Myr; Janes \& Phelps (1994), 900 Myr; Bonatto
\etal\ (2004), 729 Myr; Salaris \etal\ (2004), 700 Myr; Brandt \&
Huang (2015), 790 Myr.  For this paper, we adopt the Brandt \& Huang
age, both because it is based on the most recent physics and because
it is a reasonable average of the other estimates. Metallicity
estimates for Praesepe also span a somewhat surprisingly large range: 
Friel \& Boesgaard (1992), [Fe/H] = +0.04; An \etal\ (2007), [Fe/H] =
+0.19; Pace, Pasquini \& Francois (2008), [Fe/H] = +0.27; Boesgaard,
Roper \& Lum (2013), [Fe/H] = +0.12; Blanco-Cuaresma \etal\ (2015),
[Fe/H] = 0.20; and Yang, Chen \& Zhao (2015), [Fe/H] = +0.16).   We
adopt [Fe/H] = +0.15 as a reasonable compromise value.  The difference
in metallicity should not matter significantly in the context of
comparisons to other Kepler and K2 light curves, but we note it
nonetheless, as it might affect stellar structure and angular momentum
evolution.

Despite the lack of agreement on age and metallicity, there is
actually little dispute for the distance to the cluster, with modern
values of $(m-M)_0$ = 6.32 (Salaris \etal\ 2004),  6.33 (An \etal\
2007), and 6.30 (van Leeuwen 2009).  We adopt $(m-M)_0$ = 6.32,
corresponding to a distance of 184 pc. 

The reddening to the cluster is quite small,
though possibly as large as $E(B-V)$ = 0.027 (Taylor 2006).   For
convenience, we adopt $E(B-V)$ = 0.0. To the accuracy that we have
determined and are using $V-K_s$, the error from reddening is
negligible.

In subsequent analysis here, we often compare Praesepe results to
those from the Pleiades (Rebull \etal\ 2016a,b). We take the age of
the Pleiades to be 125 Myr (Stauffer \etal\ 1998), the distance to be
136 pc (Melis \etal\ 2014), and the metallicity to be [Fe/H]=+0.04
(Soderblom \etal\ 2009).

\subsubsection{Target List}

We started with a list of members based on an in-house/private open
cluster database originally created by Prosser \& Stauffer in the
1990s.   For Praesepe, this was a merger of membership lists in
Klein-Wassink (1927), Jones \& Cudworth (1983), and Jones \& Stauffer
(1991), with some candidate members deleted due to discrepant
photometry or radial velocities.  This list was then merged with  half
a dozen recent proper motion membership studies (Adams \etal\ 2002;
Kraus \& Hillenbrand 2007; Baker \etal\ 2010; Boudreault \etal\ 2012;
Khalaj \& Baumgardt 2013; Wang \etal\ 2014),  retaining stars
considered as likely members in those papers.   We then merged this
Praesepe membership catalog with the list of all stars observed in K2
Campaign 5 within programs targeting Praesepe.  About 600 did
not have K2 LCs, sometimes due to the star falling in CCD gaps or just
completely outside the K2 FOV; in other cases, the star may have been
observable, but no light curve was obtained.

At this point, then, we have a set of 984 Praesepe members or
candidate members with K2 light curves. 

\subsubsection{Literature Photometry}
\label{sec:litphotom}

$BVRI$ photometry on the Kron system has been published by Upgren,
Weis \& DeLuca (1979), Weis (1981), Stauffer (1982), and Mermilliod
\etal\ (1990).   For stars not in those papers, we also queried the
APASS database (Henden \etal\ 2016), particularly for $V$ magnitudes.
We similarly queried the Gaia DR1 release (Gaia Collaboration 2016)
for their $G$ magnitudes (the released proper motions do not go faint
enough for this work). Finally, we queried the recently released
Pan-STARRS1 database (Chambers \etal\ 2016). 

We added to this  data from the Two-Micron All Sky Survey (2MASS;
Skrutskie \etal\ 2006), from the Spitzer Space Telescope (Werner
\etal\ 2004), including measurements from  the Spitzer Enhanced
Imaging Products,
SEIP\footnote{http://irsa.ipac.caltech.edu/data/SPITZER/Enhanced/SEIP/overview.html},
and from the Widefield Infrared Survey Explorer (WISE; Wright \etal\
2010).

\subsubsection{Membership}
\label{sec:membership}

To refine the list of members, we queried the data from the United
States Naval Observatory (USNO) Robotic Astrometric Telescope (URAT;
Zacharias \etal\ 2015) astrometric catalog to extract the ``$f$"
magnitude (a broad band optical/red magnitude) and the URAT proper
motions for these stars.   

We then plotted the candidate Praesepe members in a vector-point
diagram (VPD) and in two color-magnitude diagrams (CMDs), $f$
vs.\ $f-K_s$ and $G$ vs.\ $G-K_s$.   In the VPD, we took the star as a
member  if the star's proper motion was within 15 mas yr$^{-1}$ of 
$\mu_{\rm RA} = -35$ mas yr$^{-1}$  and $\mu_{\rm Dec} = -15$ mas
yr$^{-1}$ (which is a mean motion consistent with the URAT proper
motions for the cluster).   For the two CMDs, we fit a polynomial
curve to the single star locus and then took the star as a member if
the star was displaced $<$1.3 magnitude above the locus or 0.7
magnitudes below the locus.   For the stars included in our member
catalog, we required that two of three of these flags (one flag
for the VPD and one for each of the CMDs) be ``true".   

We removed stars flagged as non-members in the Mermilliod \etal\
(1990) radial velocity and photometry survey for Praesepe halo
members. For some objects identified as outliers in the analysis
below, we obtained additional Keck/HIRES spectroscopy; see
App.~\ref{app:keck}. For those objects whose radial velocities (RV)
were inconsistent with membership, we identifed them as non-members
(see App.~\ref{app:nm} for a list).

From the initial sample of 984 LCs, then, 943 are members of Praesepe
by our criteria (and 41 are non-members; see Appendix~\ref{app:nm}).

\subsubsection{Obtaining \vmk}
\label{sec:vmk}

Since we want to compare to the results we obtained in the Pleiades,
we wanted to use \vmk\ as a mass proxy in the same fashion as we did
in the Pleiades.

We originally approached this the same way we had for the Pleiades,
and used data from URAT.  However, Gaia data are now available. Since
$V-K_s$ is available (with $V$ and $K_s$ directly observed) for
$\sim$250 Praesepe stars with K2 LCs, we derived a formula to convert
$G-K_s$ colors to $V-K_s$ colors by comparing the $V-K_s$ to the
$G-K_s$.  However, there are no observed $V$ photometry of known
members for colors redder than about $V-K_s$=5.5.  To extend the
calibration of $G-K_s$, we have used faint Pleiades and Hyades stars
(where $V$ photometry was obtained for very low mass stellar and brown
dwarf candidates in Pleiades and Hyades respectively by Stauffer
\etal\ 1989, 1994 and Bryja \etal\ 1992, 1994). To these, we added
GJ512B, a
field star. These stars are listed in Table~\ref{tab:redanchors}. The
final relationship between $V-K_s$ and $G-K_s$ is:
\begin{equation}
(V-K_s)= 0.27354 + 0.7336\times (G-K_s) + 0.1646\times (G-K_s)^2 -
0.000922\times (G-K_s)^3
\end{equation}
We used this to interpolate \vmk\ for those Praesepe stars lacking
\vmk, but having a Gaia measurement. For about a dozen more stars,
there is no Gaia measurement, but there is a URAT measurement;
we used the relationship we derived for the Pleiades to obtain a \vmk\
estimate from the URAT data for these remaining stars. 

All of the stars with a measured period have a measurement or estimate
of \vmk; five of the non-periodic stars are missing a \vmk. For
\vmk$<$2.5, median uncertainties (based on the scatter in the Gaia
calibration above) are $\sim$0.017 mag; for 2.5$<$\vmk$>$5.5, median
uncertainties are $\sim$0.085 mag.  

\begin{deluxetable}{lllll}
\tabletypesize{\scriptsize}
\floattable
\tablecaption{Red Stars from the Pleiades and Hyades Used to Extend
the Relation Between $V-K_s$ and $G-K_s$\label{tab:redanchors}}
\tablewidth{0pt}
\tablehead{\colhead{Star} 
& \colhead{$V$ (mag)}& \colhead{$G$ (mag)}& \colhead{$K_s$ (mag)}
& \colhead{Source of $V$}}
\startdata
HHJ3     &  22.14  & 18.78  &  14.13  &   Stauffer \etal\ (1998)\\
HHJ5     &  20.71  & 18.45  &  13.93  &   Stauffer \etal\ (1998)\\
HHJ8     &  20.60  & 18.38  &  14.23  &   Stauffer \etal\ (1998)\\
HHJ10    &  20.74  & 18.39  &  13.77  &   Stauffer \etal\ (1998)\\
PPL1     &  22.08  & 19.49  &  14.31  &   Stauffer \etal\ (1989)\\
PPL2     &  22.12  & 19.32  &  14.37  &   Stauffer \etal\ (1989)\\
PPL14    &  21.56  & 19.12  &  14.48  &   Stauffer \etal\ (1994)\\
Bry804   &  19.27  & 16.83  &  12.37  &   Bryja \etal\ (1994)\\
Bry816   &  18.83  & 16.61  &  12.20  &   Bryja \etal\ (1994)\\
GJ512B   &  13.70  & 12.10  &  8.30   &   field star\\
\enddata
\end{deluxetable}

\subsection{Bright and Faint Limits}

In the Pleiades, we discarded stars with \ks$\lesssim$6 and
\ks$\gtrsim$14.5 as being too bright and faint, respectively, for the
K2 light curves to be reliable. Here in Praesepe, the appropriate
limits are less obvious; we have dropped the brightest
(\ks$\lesssim$6) and retained the rest. 

There are two stars with \ks$<$6, one of which we determined to be
periodic, and both of which are listed in Appendix~\ref{app:bright}. 
Both of them are discarded from our sample as too bright. 

There are 21 stars with 6$<$\ks$<$8; \ks=8 is roughly an F5 spectral
type. At least 11 of them are likely pulsators (with 6 more likely pulsators 
that have fainter \ks); see 
Appendix~\ref{app:pulsator}. We have left these in the sample to allow
for comparison to our Pleiades work (which also included likely
pulsators), but have identified those pulsators where necessary in the
remaining discussion.

In the Pleiades, there were very few sources with K2 light curves in
the optical CMD below \ks=14.5;  as we will see below in
Fig.~\ref{fig:optcmd}, such a cutoff is not as obvious here. There are
many targets with 14.5$<$\ks$<$15.5 that have clear periods, and we
have left those in the sample.

\subsection{Final Definition of Sample}

Final star counts are as follows. From an initial sample of 984
candidate members, we find 941 members that are also not too bright. 
Limiting it further, there are 809 stars that we find to be periodic
in these K2 data. 

\begin{figure}[ht]
\epsscale{0.8}
\plotone{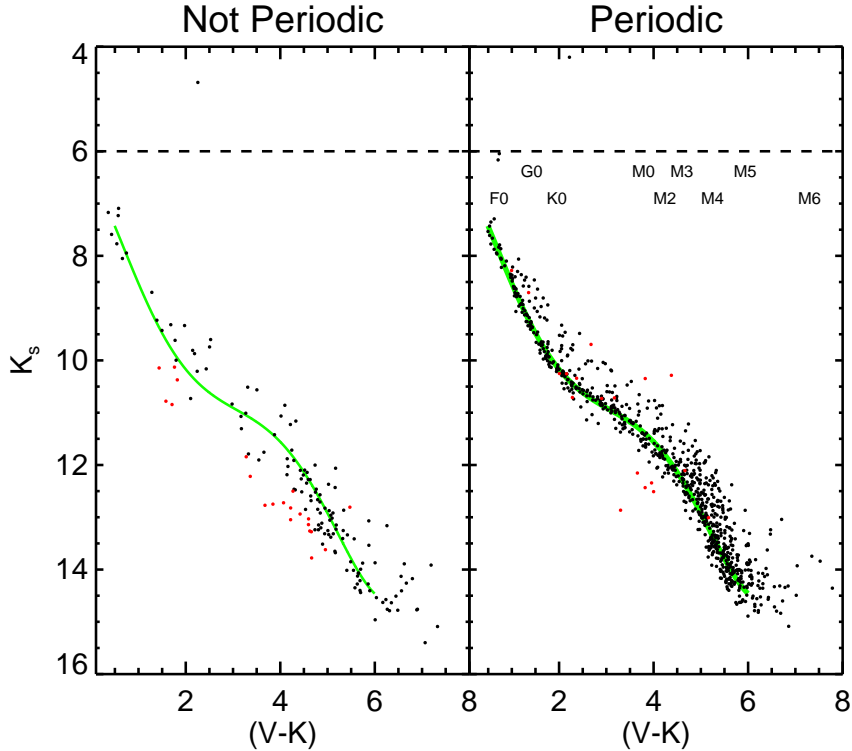}
\caption{Color-magnitude diagram (\ks\ vs.\vmk) for targets with K2
LCs and for which we had or could infer \vmk.  Left panel is stars not
detected as periodic, and right panel is stars for which we could
measure periods. In both panels, red circles denote NM (see
Sec.~\ref{sec:membership}), and the green line is an empirical
fit to the single star locus
for Praesepe.  Spectral types for a given \vmk\ are as
shown in the right panel. The periodic stars for the
most part follow the expected main sequence relation for Praesepe. The
stars we do not detect as periodic appear to have a less-well-defined
main sequence relation, which would be consistent with those stars
being less likely members.}
\label{fig:optcmd}
\end{figure}

Figure~\ref{fig:optcmd} shows color-magnitude diagrams for the stars
detected as periodic and not detected as periodic.  The periodic stars
for the most part follow the expected main sequence relation for
Praesepe. The stars we do not detect as periodic appear to have a
less-well-defined main sequence relation, which would be consistent
with those stars more likely to be non-members, despite satisfying the
membership criteria described in Sec.~\ref{sec:membership} above.  

In the Pleiades, we determined 92\% of the sample to be periodic
(Rebull \etal\ 2016a); here, we obtain 809/941=86\% of the member
sample to be periodic\footnote{The 84\% in
Sec.~\ref{sec:interpofperiods} refers to the fraction of the entire
initial sample that is periodic; the 86\% refers to the fraction of
members that are periodic.}. If we have more non-members inadvertently
included in the sample for Praesepe than we did for the Pleiades
(despite very similar selection methods), a lower fraction of periodic
stars might be expected. However, Praesepe is considerably older than
the Pleiades, so the stars are expected to have fewer spots (hence
lower amplitude signals) and rotate more slowly. Both of these factors
would contribute to a lower fraction of detectably periodic members in
Praesepe.

\section{Period and Period-Color Distributions}
\label{sec:rotationdistrib}

Although many of the Praesepe stars, like many in the Pleiades, are
multi-periodic (see Sec.~\ref{sec:LCPcats} below), we have selected
only one $P$ and color to be representative of the star in order to
investigate the distribution of rotation rates.  Now, we explore the
overall distribution of $P$, and the distribution of $P$ as a function
of \vmk\ as a proxy for mass.

\subsection{Distribution of $P$}

\begin{figure}[ht]
\epsscale{0.7}
\plotone{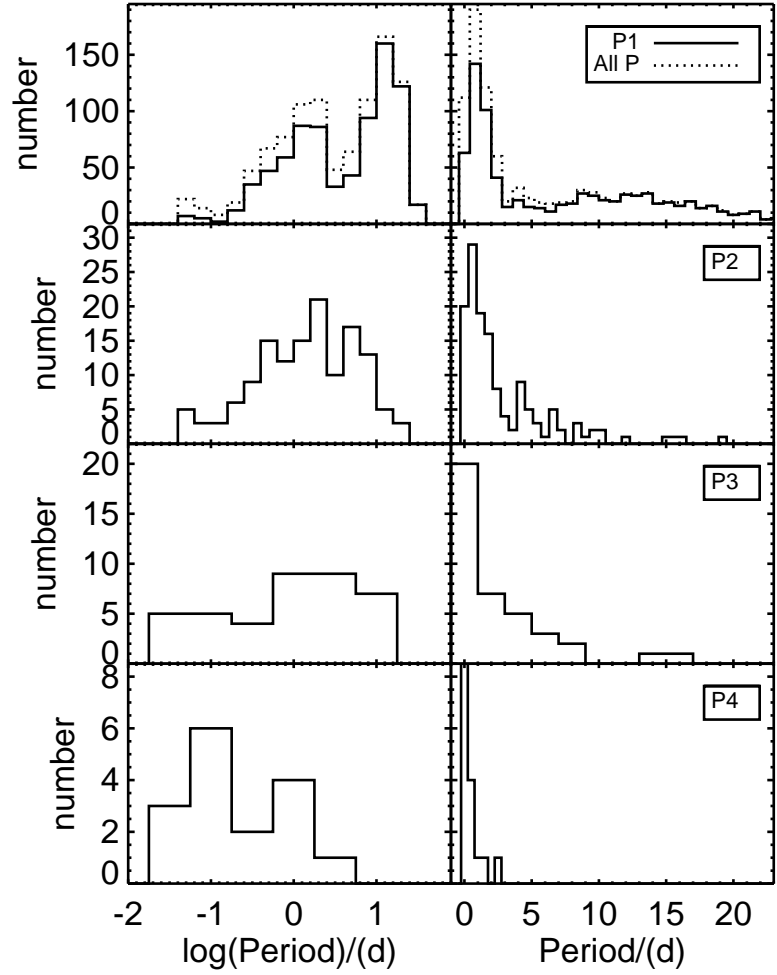}
\caption{Histograms, on the left of the log of periods, and on the
right of the linear periods, found by our analysis, in days.  Solid
line is the primary period (that which we take to be the rotation
period of the star), and dotted line is (for reference) a histogram of
all the periods found here, including the secondary, tertiary, and
quaternary periods found here (see Section~\ref{sec:LCPcats}).   There
are two peaks, one near $\sim$1 d and one near $\sim$10-15 d. The
faster peak corresponds largely to M stars, and the slower peak is
primarily more massive stars. A significant fraction of the tertiary
and quaternary periods are from likely pulsators.}
\label{fig:phisto}
\end{figure}

The distribution of periods we found is shown in
Figure~\ref{fig:phisto}; note that this excludes the stars with
periods that were determined to be non-members (see
Sec.~\ref{sec:membership} and App.~\ref{app:nm}). The stars are
rotating more slowly, on average, than the analogous figure from the
Pleiades (see Fig.~3 in Rebull \etal\ 2016a or Fig.~9 in Rebull \etal\
2016b). Whereas the Pleiades is strongly peaked at $<$1 day, only
$\sim$20\% of the Praesepe stars with rotation periods rotate faster
than a day, with $\sim$42\% rotating between 1 and 10 d, and
$\sim$37\% rotating between 10 and 35d. 

There seems to be a bimodal distribution of periods in Praesepe, with
one peak near $\sim$1 d and another peak near $\sim$10 d.  Assuming a
Skumanich law (Skumanich 1972), $v_{\rm rot} \propto t^{-0.5}$,
assuming that the Pleiades is 125 Myr and Praesepe is 800 Myr, we
would expect the peak at $\sim$0.3 d from the Pleiades to become the
peak at $\sim$1 d in Praesepe.  This is the case for the M stars, which
compose most of the Pleiades peak at $\sim$0.3 d and most of the
Praesepe peak at $\sim$1 d.

There is a substantial number of Praesepe stars, however, with periods
near 10-15 d, about 35\% of the distribution. The stars contributing
to this peak are for the most part more massive than those stars
composing the other, faster rotating peak. Using the Skumanich law,
these stars should correspond to Pleiades stars with periods of 4-6 d.
Indeed, about 20\% of the Pleiades stars with K2 rotation rates have
periods greater than about 4 d; presumably, these go on to populate
the slower peak in older clusters. However, there are fractionally
more stars in the slower peak in Praesepe, which suggests that this
peak includes stars with a larger range of masses than we have assumed
in the rough Skumanich calculation.

Note that the 17 stars we have identified as likely pulsators are
still in this sample shown in Fig.~\ref{fig:phisto}; their periods do
not make a significant difference to the histograms of $P_1$ (first
period, e.g., the $P_{\rm rot}$) or $P_2$ (secondary period), just
because so many stars are represented in these plots. But, they do
make a significant difference to the histograms of $P_3$ and $P_4$
(tertiary and quaternary periods), where they are largely responsible
for the periods in these histograms.

\clearpage

\begin{figure}[ht]
\epsscale{1.0}
\plottwo{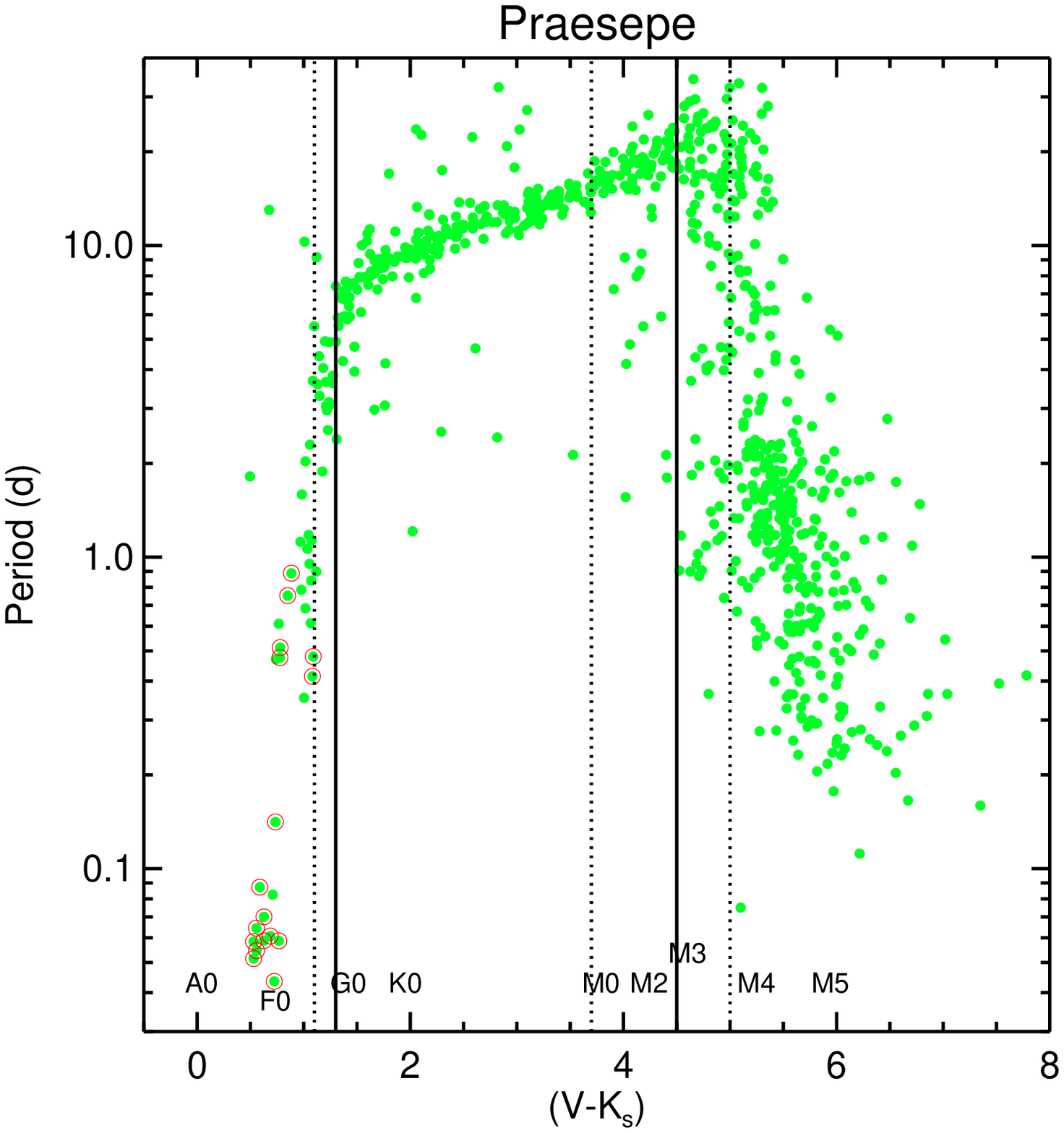}{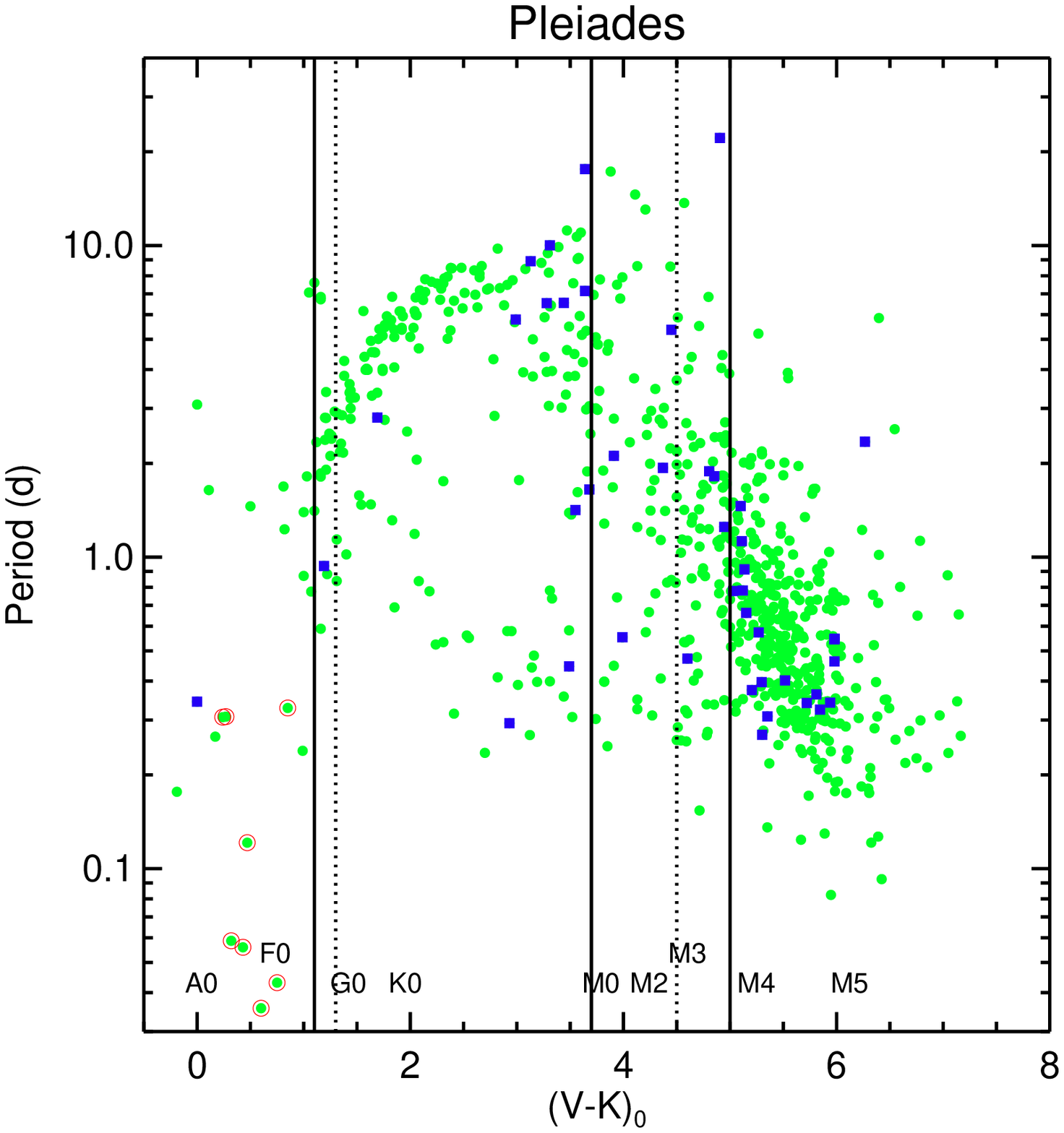}
\caption{Plot of $P$ vs.~\vmk\ for Praesepe (left) and the Pleiades
(right). The Pleiades plot is from Rebull \etal\ (2016a), Fig.~9. In
the Pleiades, we had best members (green dots) and the lower
confidence members (blue squares). Likely pulsators (see Rebull \etal\
2016b for the Pleiades) have an additional red circle in both panels. 
There is clearly considerable change between the age of the Pleiades
and that of Praesepe.  In each plot, the solid vertical lines denote
different regions of the diagram defined in the discussion; the dotted
vertical lines are the lines from the other cluster (the solid lines
in the Praesepe plot are the dotted lines in the Pleiades plot and
vice versa). }
\label{fig:pvmk}
\end{figure}

\begin{figure}[ht]
\epsscale{1.0}
\plottwo{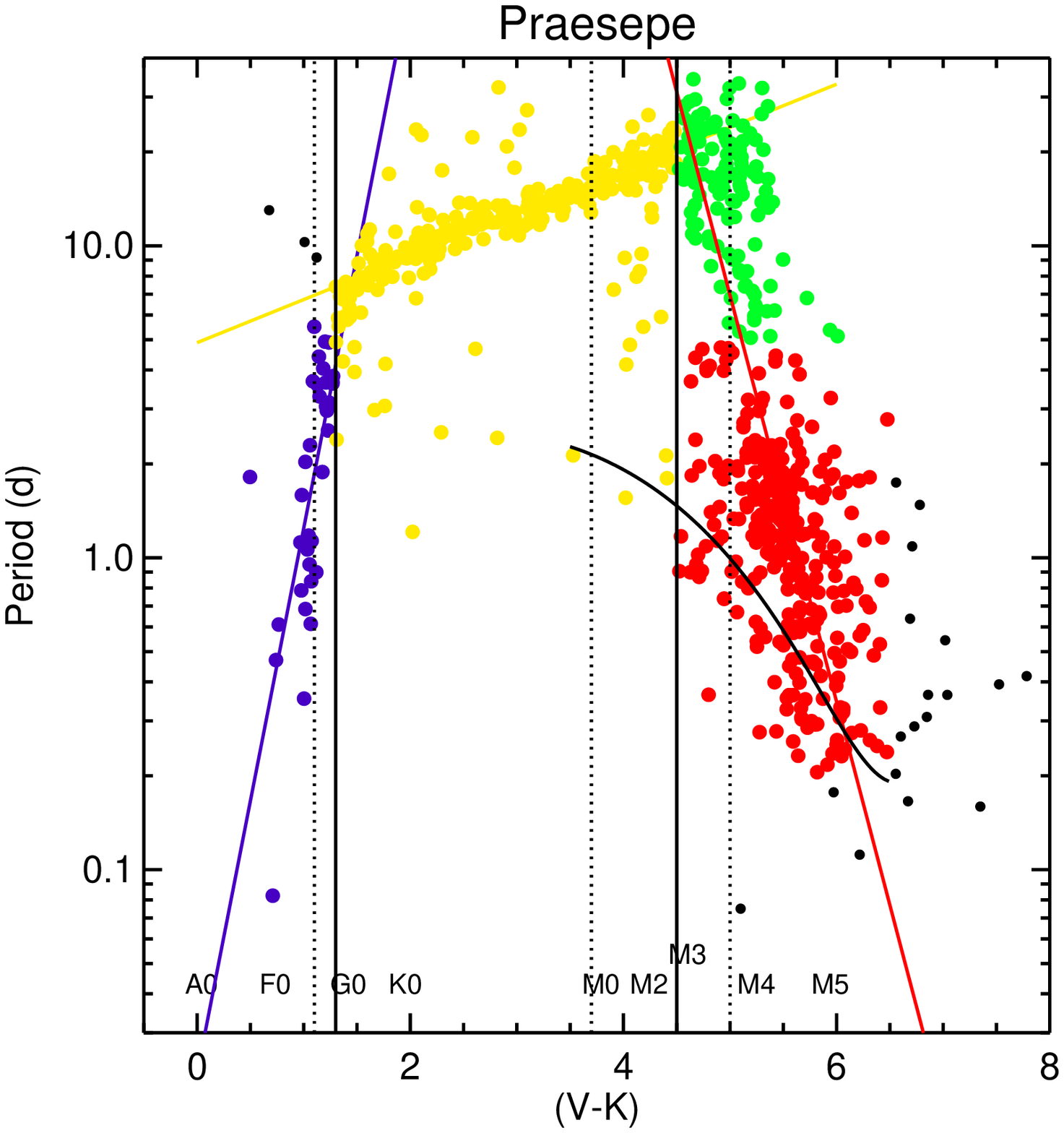}{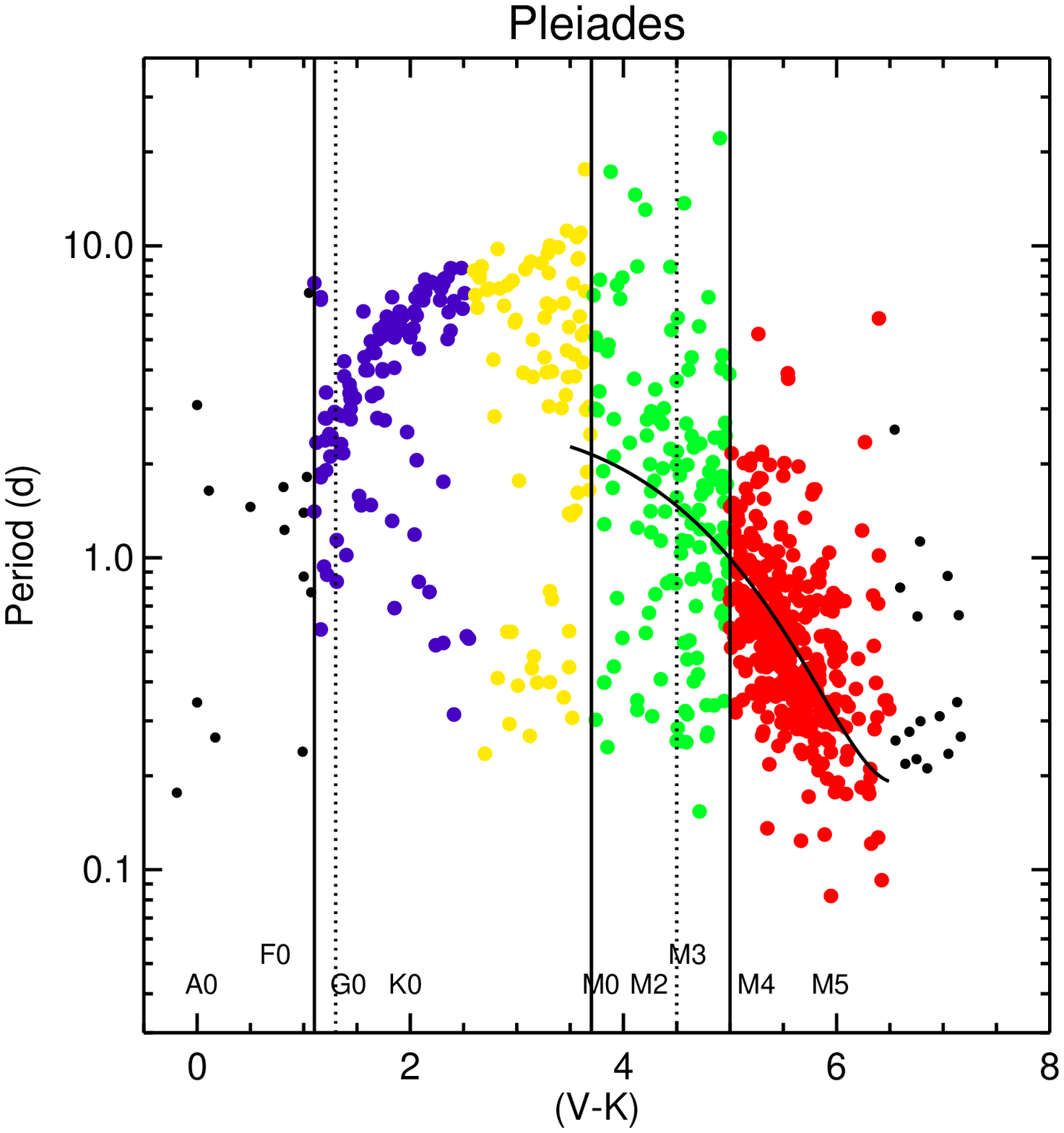}
\caption{Plot of $P$ vs.~\vmk\ for Praesepe (left) and the Pleiades
(right), with additional color coding indicating regions discussed
here and in Rebull \etal\ (2016a) and Stauffer \etal\ (2016). Obvious
pulsators
have been removed in both plots.  Black points in both plots have been
omitted as outliers  for the purpose of fitting the lines.
For Praesepe, colored points are as follows. Blue is \vmk$<$1.3; 
yellow is 1.3$<$\vmk$<$4.5;
green and red are both \vmk$>$4.5, but those with $P>$5
are green and those with $P<$ are red. Linear fits are also shown in
the blue, yellow, and red+green regimes.
For the Pleiades, colored points follow Stauffer \etal\ (2016),
Fig.~24: blue is 1.1$<$\vmk0$<$2.6, and yellow
is 2.5$<$\vmk0$<$3.7. The bulk of the
distribution through both of these regions trace out the `slowly
rotating sequence'; the transition to yellow is placed at the location
of the `kink' in the slowly rotating sequence. Green is
3.7$<$\vmk0$<$5, the `disorganized region'; red
is \vmk0$>$5, the `fast sequence.'
The vertical lines are the same as the prior figure.
The curved black line is the fit to the Pleiades M stars from Stauffer
\etal\ (2016), and is the same in both panels. The M star relation is 
steeper in Praesepe.}
\label{fig:pvmk2e}
\end{figure}

\clearpage

\subsection{Morphology of $P$ vs. \vmk}

Qualitatively, the morphology of the period vs.\ color diagram
for  Praesepe, and the evolution of that distribution from young ages
to Praesepe age, has been well-documented in the literature.  The fact
that by Praesepe's age G and K dwarfs have a very narrow distribution
in period at a given mass was first shown for Hyades stars in the
1980s (Duncan \etal\ 1984; Radick \etal\ 1987).  That F and early G
stars arrive on the main sequence with higher rotational velocities
and subsequently spin down on the main sequence was first shown by
Robert Kraft via spectroscopic rotational velocities for stars in a
number of open clusters (Kraft 1967, and references therein). Later
spectroscopic studies of those same clusters with more modern
spectrographs and detectors showed that low mass stars of all masses
arrive on the main sequence with a wide range in rotational velocities
and that angular momentum loss on the main sequence causes those stars
to converge over time to a much narrower range in rotation at a given
mass, with the convergence time being longer for lower masses
(Stauffer \&  Hartmann 1986; Stauffer \etal\ 1987; Stauffer \etal\
1989).  The  subsequent development of wide-format CCDs made it
possible to obtain rotation periods for large samples of stars in open
clusters, allowing the distribution of rotation rate as a function of
mass to be determined for many of Kraft's open clusters (and other
clusters), largely confirming the spectroscopic results but with
better precision and larger samples of stars; see Gallet \& Bouvier
(2015) and Coker \etal\ (2016) for a review of the
rotation period data and theoretical models of angular momentum loss
which attempts to explain the data. The light curves for open clusters
and star-forming regions from K2 builds on this heritage, but also
adds the benefits in photometric stability and signal-to-noise
provided by space-based observations.  For the open clusters observed
with K2, these data for the first time allow the rotation periods of
both stars in most binary systems to be determined and for all cluster
members observed the light curves allow an assessment of the shape of
the phased light curve, which provides insight into the size and
location of star spots on their surfaces.

Figure~\ref{fig:pvmk} shows the relationship between $P$ and \vmk\ for
Praesepe using K2 data. For comparison, Figure~\ref{fig:pvmk}
also shows the $P$ vs.~\vmk\ plot for the Pleiades K2 data.
There has clearly been significant changes in these distributions
between the ages of the Pleiades and Praesepe. As we noted above,
Praesepe stars are on the whole rotating more slowly. However, the M
stars in both clusters are primarily rapid rotators.

Figure~\ref{fig:pvmk2e} has the $P$ vs.~\vmk\ plots again for both
Praesepe and the Pleiades, but with color coding to aid in this
discussion. In Praesepe, there are 3 relatively well-defined and
apparently linear sequences. At the blue end (omitting the pulsators),
for $(V-K)<1.3$ ($\lesssim$F8), $\log P = 1.748 \times (V-K) -1.653$.
$P$ is changing rapidly over a very small range of \vmk. There is an
abrupt transition between the reddest end of this blue sequence and
the bluest end of the next sequence; this is the Kraft break (Kraft
1967), where magnetic braking becomes less efficient for bluer stars.
In the middle (yellow portion) of Fig.~\ref{fig:pvmk2e}, $1.3 \leq (V-K) <
4.5$ ($\sim$F8 to $\sim$M3), $\log P = 0.138 \times (V-K) + 0.692$. In
this regime, the points are tightly clumped around this relationship.
For $4.5 \leq (V-K) < 6.5$, $\sim$M3 to $\sim$M6, (which encompasses
both the green and red points in Fig.~\ref{fig:pvmk2e}), $\log P =
-1.303 \times (V-K) + 7.360$.  Here again, $P$ is changing rapidly
with \vmk. This end of the distribution contributes substantially to
the bimodal nature of the $P$ distribution (Fig.~\ref{fig:phisto});
there are many stars with $4.5 \leq (V-K) < 6.5$ and relatively few
stars with $2<P<10$. The transition between the blue and yellow
regions in Fig.~\ref{fig:pvmk2e} is abrupt. The transition between the
yellow and green regions is less obvious; the green points, at least
the ones with $P>10$, could justifiably be included in the linear fit
of the yellow points, and the bulk of these points are already
consistent with that fit.  The green points, however, are also
consistent with the relationship delineated by the red points (or even
a slightly steeper relation). The end of the slow sequence (yellow
through the green points consistent with the relationship) in Praesepe
is \vmk$\sim$5.2, or M4.

Figure~\ref{fig:pvmk2e} also includes the Pleiades (Rebull \etal\
2016a) for context. The color-coding in Fig.~\ref{fig:pvmk2e} follows
Stauffer \etal\ (2016), Fig.~24, and is {\em not} meant to trace
exactly the same populations as seen in Praesepe, but simply
illustrates different sections of the diagram that we call out in the
text.  In the Pleiades, we discussed  the `slowly rotating sequence'
for $1.1\lesssim$\vmk$\lesssim 3.7$ ($\sim$F5 to
$\sim$K9\footnote{Technically, some investigators have no K8 or K9
class defined; the dividing line here is based on linearly
interpolating the \vmk\ colors and the corresponding spectral type. 
This division might formally be at K7, but we use `K9' here to
indicate succinctly `just before G0'.}, 2$\lesssim P \lesssim$11 d),
which is the blue and yellow points together in
Fig.~\ref{fig:pvmk2e}.  The transition of blue to yellow is placed at
the location of the `kink' in the slowly rotating sequence ($\sim$K3;
see Stauffer \etal\ 2016). The green points delineate a region in
which there seems to be a `disorganized relationship' between $P$ and
\vmk\ between $3.7\lesssim$\vmk$\lesssim 5.0$  ($\sim$K9 to $\sim$M3,
0.2$\lesssim P \lesssim$ 15 d). Finally, the red points represent the
`fast sequence,'  with \vmk$\gtrsim 5.0$ ($\gtrsim$M3, 0.1 $\lesssim P
\lesssim$ 2 d).   

Now between the Pleiades ($\sim$120 Myr) and Praesepe ($\sim$800 Myr),
parts of this diagram see tremendous change (earlier types), and parts
show more subtle changes (later types). All of the G and K (and even
early M) stars (yellow in the Praesepe plot) have spun down into a
well-defined, linear relationship from the curved relationship (blue
and yellow in the Pleiades plot). The `disorganized region' (green in
the Pleiades plot, and even to some extent the yellow) is no longer
quite so disorganized by Praesepe, where many of the earlier M stars
have spun down into a relationship consistent with the G and K stars.
The Praesepe early M stars are on average slow rotators, and the later
M stars are on average more rapid rotators, but there is clear overlap
between the rotation rates of early and late M stars; one cannot
divide the stars by mass and have them also divided by period.  In the
Pleiades, there is less obviously a bimodal distribution in M star
periods, and (in contrast to Praesepe) the division between fast and
slow rotators is also roughly a division in mass. The overall  M-star
relationship (the red points in both panels in Fig.~\ref{fig:pvmk2e})
sees less obvious changes compared to the large changes for the more
massive stars, but there are differences for the M stars too. The
slope of the M star relationship between period and color is overall
much steeper in Praesepe than in the Pleiades; this is most easily
seen from the black curved line in Fig.~\ref{fig:pvmk2e}, which is the
relationship for Pleiades M stars derived in Stauffer \etal\ (2016).

In the Pleiades, some of the M stars are still contracting (still
spinning up); this is not the case in Praesepe. The M3 and M4 stars in
Praesepe have spun down considerably, and they have a shorter
contraction time than the M5 and later stars. The distribution of M5
stars has not changed much between the Pleiades and Praesepe. In the
older cluster, angular momentum loss via wind braking has had more
time to counteract the contraction in the M3-4 stars, and the balance
between contraction/spin up and angular momentum loss must be
different in the M5 (and later) stars. 

The existence of a well-defined slow sequence of late F and even early
G stars (blue points in Fig.~\ref{fig:pvmk2e}) in Praesepe presumably
points to their having at least some amount of angular momentum loss.
If they had no angular momentum loss at all, there would be a scatter
in rotation reflecting a range in initial angular momentum and a range
in disk lifetimes.  However, presumably their angular momentum loss
rate is quite small since they have so little outer convective
envelope. As a speculation, it is possible that the F dwarfs in
Praesepe still have rapidly rotating radiative cores, with their
observed rotation periods representing a balance between the angular
momentum feeding up from below with the angular momentum lost from
their winds.   The G/K/M stars have much larger angular momentum loss
rates, and have had time to spin down their cores.  So, the two
sequences reflect that dichotomy: core/envelope still decoupled for
the F dwarfs, core/envelope coupled for the G, K, and early Ms.

The bimodal period distribution of M stars in Praesepe (but not the
Pleiades) is interesting because field M stars are found to have a
bimodal $P$ distribution in the Kepler field (\eg, McQuillan \etal\
2013, Davenport 2016) and in nearby M field stars (\eg, Newton \etal\
2016, Kado-Fong \etal\ 2016). These field stars are older, on average,
than Praesepe, and the locations of the two peaks are slower ($\sim$19
and $\sim$33 d) compared to Praesepe with $\sim$1 and $\sim$17 d for
the  $\sim$500 Praesepe stars with \vmk$\geq$3.79, the color
corresponding to M0. Assuming Skumanich evolution, these two peaks
cannot both evolve together in lock-step. However, it is interesting
that both distributions are bimodal.  It may be that the mechanism
which causes some M stars to rapidly spin down (\eg, Newton \etal\
2016, Brown 2014)  has started to operate in some of the Praesepe
stars.  We note, of course, that our data cannot constrain the
behavior of M stars later than about M5 or M6.

\subsection{Outliers in $P$ vs. \vmk}
\label{sec:outliers}

While the relationships delineated by the majority of the Praesepe
stars in Figures~\ref{fig:pvmk} and~\ref{fig:pvmk2e} are striking, it
is worth looking at some of the outliers in this distribution. Notes
on specific stars appear in Appendix~\ref{app:pvmkoutliers}, as well
as optical color-magnitude and $P$ vs.~\vmk\ diagrams with these
objects highlighted.

The G and K stars in young open clusters like the Pleiades have
bimodal rotational velocity distributions, with a majority of stars on
a slowly rotating branch and a minority in a rapidly rotating branch. 
The latter stars are generally believed to descend from
pre-main-sequence stars that lost their circumstellar disks (and hence
their ability to rapidly shed angular momentum) very early.  Our $P$
vs.\ color plot also shows several rapidly rotating G/K stars.   Are
these the descendents of the rapidly rotating G/K stars in the
Pleiades?   For most or all of these stars, we believe not.  Instead,
they are best explained as tidally locked short-period binaries. Most
of them are known short-period binaries; many of the remaining ones
have little to no spectroscopic information, but tidally locked
binaries are a logical explanation for stars that are rotating too
quickly in comparison to the other stars of similar colors.

The stars with periods much longer than average for their \vmk\
color are also curious.   For a few of them, tidal synchronization (or
pseudo-synchronization) in a short-period binary may be the
explanation, as has been advocated for similar outliers in M35 (Meibom
\etal\ 2006) and other clusters.  Because of the steep  dependence of
the synchronization time on the binary-star separation, for the longer
period outliers ($P>\sim$20 days), this becomes a less viable
possibility. If these stars are all Praesepe members, and if we are
correct that all of them are rotation periods, then something about
these stars made them spin down more than most other stars in the
cluster.  For these longest-period stars, interpretation of their LCs
as rotation periods is more fraught than other shorter-period stars,
as there is more variation from cycle to cycle, and it becomes more of
a judgement call as to whether or not the star has a $P_{\rm rot}$, or
just a repeated pattern that may or may not be tied to rotation.
Slowly rotating stars may undergo more significant spot evolution, so
this may be a real astrophysical effect. Having fewer complete cycles
within the K2 campaign, coupled with significant changes every cycle,
makes it hard to assess. Many of the long-$P$ G and K stars from the
bulk of the distribution have more obviously sinusoidal LCs, and many
of the longest $P$ M stars (ones that are still with the bulk of the
distribution)  share LC characteristics with these long-$P$ outlier G
and K stars; there is a continuum of LC properties such that it is not
always easy to draw a line between $P_{\rm rot}$ and just a repeated
pattern.  

Specifically because of this ambiguity, we obtained follow-up
Keck/HIRES spectroscopy of many of these very slowly rotating stars
(see App.~\ref{app:keck}). Of the ones we observed, about half of them
have radial velocities inconsistent with cluster membership. We
identified these as non-members (Sec.~\ref{sec:membership} and
App.~\ref{app:nm}).  Four of the longest $P$ objects (for which we
have yet to obtain spectra) appear just below the single-star main
sequence, suggesting that they may also be non-members; given the
uncertainties in \vmk, we have  provisionally left them in  the list
of members (see App.~\ref{app:pvmkoutliers}).

\section{Light Curve and Periodogram Categories}
\label{sec:LCPcats}

\subsection{Identification of Categories}
\label{sec:definecategories}

\begin{figure}[ht]
\epsscale{1.0}
\plotone{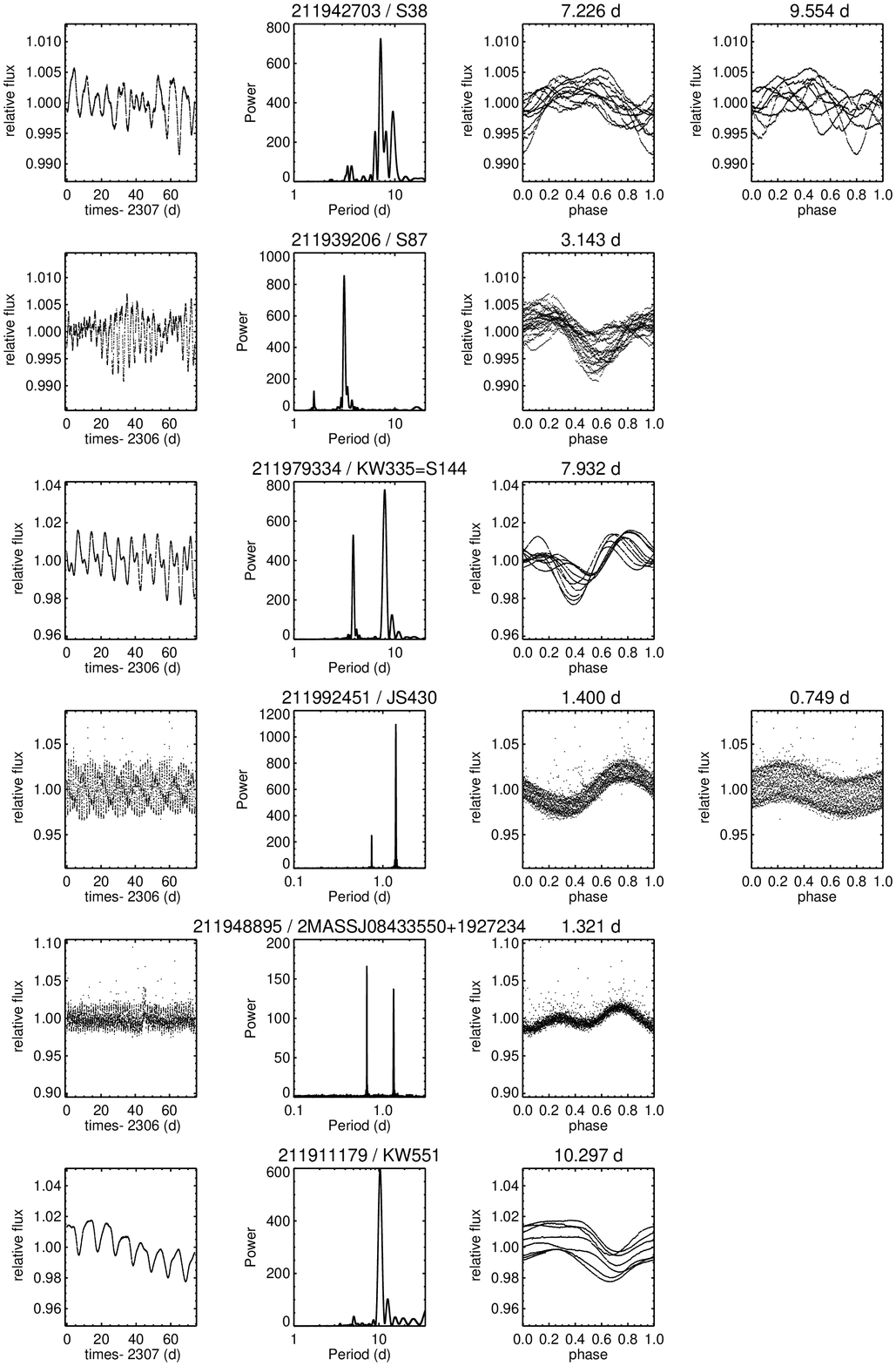}
\caption{ Six examples of LC/periodogram categories in the K2
Praesepe data. Left column: full LC; second column: LS periodogram;
third column: phased LC, with best period (in days) as indicated;
fourth column: phased LC to second period (if available, in days) as
indicated.
Rows, in order: 211942703/S38 (shape changer, beater, complex peak),
211939206/S87  (beater, complex peak),
211979334/KW335=S144 (moving double dip [=double dip, shape changer]),
211992451/JS430 (beater, resolved distant peaks),
211948895/2MASSJ08433550+1927234 (double dip; appears to have two
significant peaks in the periodogram, but really has only one real $P$),
211911179/KW551 (shape changer). These are
representatives from a range of categories and periods. 
\label{fig:multifreq}}
\end{figure}

In Rebull \etal\ (2016b), we presented a set of empirical structures
in the K2 Pleiades LCs and periodograms, finding via visual
inspection 11 different categories. These categories are discussed in
detail in Rebull \etal\ (2016b), and we do not repeat that discussion
here. All but one of the categories of light curves can be found in
Praesepe as well; see Table~\ref{tab:countclasses} for a list. Many of
the categories have more than one significant period.  
Figure~\ref{fig:multifreq} shows some examples of these classes in
Praesepe, two of which have more than one measured $P$; these examples
span a range of brightnesses, periods, and categories. (For examples
of each of the categories in the Pleiades, see Rebull \etal\ 2016a.) 

\subsection{Comparison to Pleiades}
\label{sec:comparetypestopleiades}

\begin{figure}[ht]
\epsscale{0.5}
\plotone{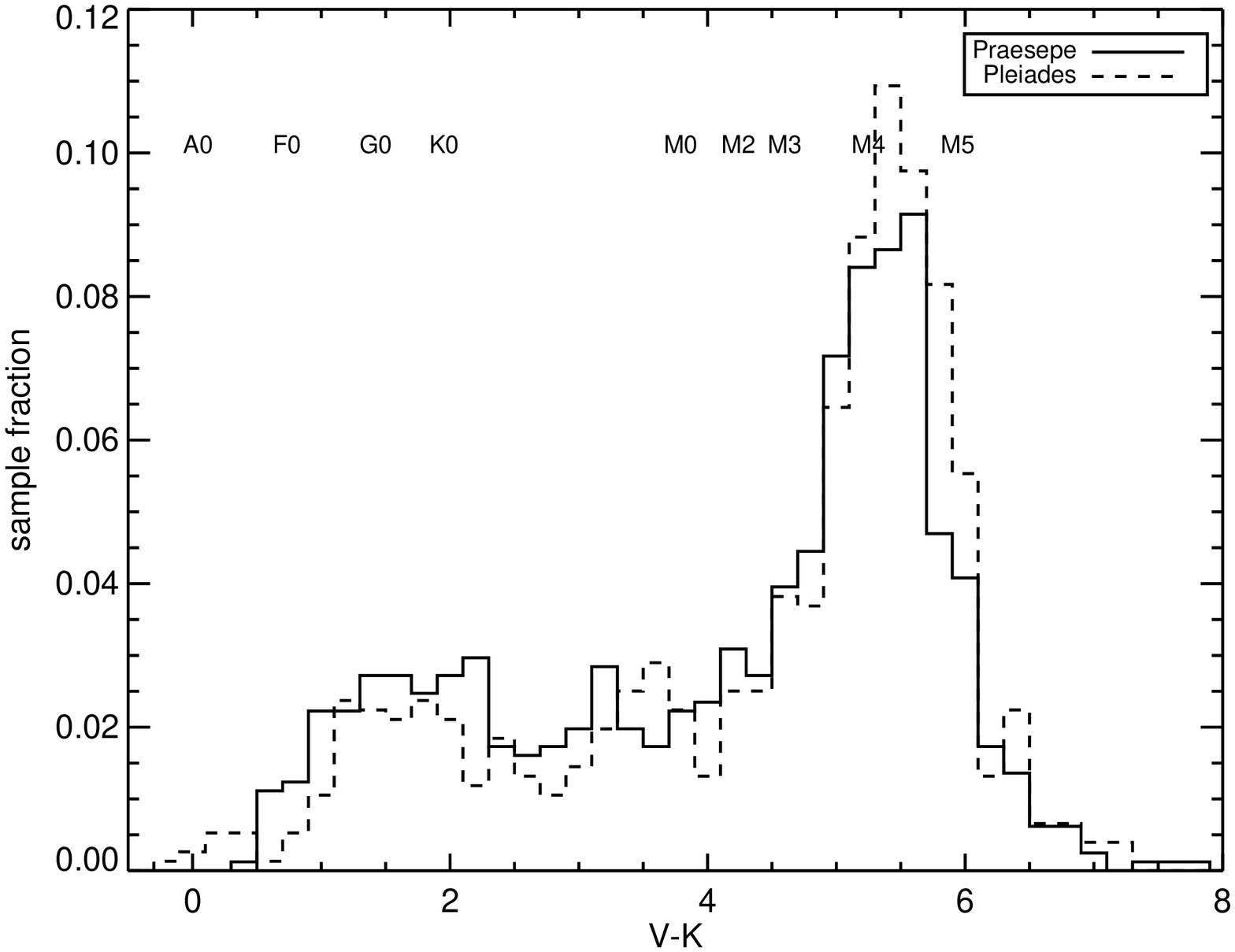}
\caption{Distribution of \vmk\ values for the member sample with
rotation periods in both Praesepe (solid) and the Pleiades (dotted).
The y-axis is in units of sample fraction. We are sampling the same
range of \vmk\ values overall in the two clusters; there are notable
differences at the bluest and reddest ends. 
\label{fig:vkdistrib4}}
\end{figure}

Before we can compare the total counts of objects in the various
periodogram categories, we need to be sure that we are sampling the
same range of masses in the two clusters. Figure~\ref{fig:vkdistrib4}
shows the sample fraction for the periodic member samples as a
function of \vmk\ as a proxy for mass. The two samples are comparable
over most of the range of \vmk; there are sample completeness effects
at the reddest and bluest bins, which are the most poorly populated. 
The most notable differences are at
the bluest end (\vmk$\lesssim$1), which will affect primarily stars
whose measured K2 periods are most likely to be pulsation rather than
rotation. 

Now with the knowledge that the samples are of comparable mass ranges,
we can compare sample fractions of LC/periodogram categories. 
Table~\ref{tab:countclasses} summarizes the total counts of objects in
each LC/periodogram category for Praesepe and the Pleiades (in both
cases, only cluster members are included). There are some significant
differences between the clusters. Overall, the fraction of stars that
are singly periodic and multiply periodic are roughly comparable in
the two clusters ($\sim$75-80\% and $\sim$20-25\%, respectively).
However, there is a slightly higher fraction of singly periodic
sources (and slightly lower fraction of multi-periodic sources) in
Praesepe.  

In the context of this work, we identified an inconsistency in
the way that we were identifying close and distant resolved peaks in
our earlier Pleiades work. As stated there, we calculated $\Delta
P/P_1$ (see also Sec.~\ref{sec:deltaP} below), and used that value to
identify close ($\Delta P/P_1<$0.45) and distant ($\Delta P/P_1>$0.45)
peaks. However, we went on to say (without explanation) that some
objects could be identified as both close and distant peaks. We now
more consistently identify an object as both close and distant peaks
if the difference in any two periods divided by the $P_{\rm rot}$ is
$<$0.45 for two of at least three periods, and $>$0.45 for (a
different) two of at least three periods (unless the star is already
identified as a pulsator, in which case it is not identified as either
close or distant peaks). As a result, three objects from the Pleiades
that were already identified as resolved close peaks should also have
been tagged as resolved distant peaks (EPIC 210877423, 211112974, and
211128979). This has been corrected in the statistics in
Table~\ref{tab:countclasses}.

\floattable
\begin{deluxetable}{lcccccc}
\tabletypesize{\scriptsize}
\floattable
\tablecaption{Star/Light Curve/Periodogram
Categories\label{tab:countclasses}\label{tab:multipstats}}
\tablewidth{0pt}
\tablehead{\colhead{Category} & \multicolumn{3}{c}{{\bf Praesepe}} 
& \multicolumn{3}{c}{{\bf Pleiades}} \\
& \colhead{Number} & \colhead{Frac.~of }
& \colhead{Frac.~of }& 
\colhead{Number} & \colhead{Frac.~of }
& \colhead{Frac.~of }\\
&  & \colhead{sample}
& \colhead{periodic sample}& 
 & \colhead{sample}
& \colhead{periodic sample}}
\startdata
Periodic & 809 & 0.86 & 1.00  & 759 & 0.92 & 1.00 \\
Single period & 645 & 0.69 & 0.80 & 559 & 0.68 & 0.74 \\
Multi-period & 164 & 0.17 & 0.20 & 200 & 0.24 & 0.26 \\
Double-dip & 163 & 0.17 & 0.20 & 107 & 0.13 & 0.14 \\
Moving double-dip & 121 & 0.13 & 0.15 & 31 & 0.04 & 0.04 \\
Shape changer & 297 & 0.32 & 0.37 & 114 & 0.14 & 0.15 \\
Orbiting clouds? & 0 & 0 & 0 &  5 & $<$0.01 & $<$0.01\\
Beater &  77 & 0.08 & 0.10 & 135 & 0.16 & 0.18 \\
Complex peak &  68 & 0.07 & 0.08 & 89 & 0.11 & 0.12 \\
Resolved, close peaks &  68 & 0.07 & 0.08 & 126 & 0.15 & 0.17 \\
Resolved, distant peaks &  71 & 0.08 & 0.09 & 39 & 0.05 & 0.05 \\
Pulsator &  17 & 0.02 & 0.02 & 8 & 0.01 & 0.01 \\
\enddata
\end{deluxetable}

Among the LC/periodogram categories, there is a higher fraction of
moving double-dip and shape changers in Praesepe than there are in the
Pleiades; there are more than twice the fraction of shape changers as
in the Pleiades ($\sim$40\% vs.\ $\sim$15\%).  In Rebull \etal\
(2016b) and Stauffer \etal\ (2016), we suggested that the shape
changers and moving double dips are due to latitudinal differential
rotation and/or spot/spot group evolution. These LC types were found
primarily in the slower rotators in the Pleiades; with more slower
rotators in Praesepe, it may not be surprising that we have more shape
changers, because this may reflect a real difference in the incidence
rate of differential rotation.  

Because there is a slightly lower fraction of multi-period sources in
Praesepe, there is a lower fraction of nearly all the multi-period
subcategories in Praesepe. However, the most discrepant fractions are
found in the resolved close peaks category, where the fraction is half
what it was in the Pleiades. We postulated in Rebull \etal\ (2016b)
and Stauffer \etal\ (2016) that the resolved peaks categories could be
a result of latitudinal differential rotation and/or spot/spot group
evolution among the G and K stars. 

The shape changers, moving double dips, and close resolved peaks (in
the G and K stars) may all be subject to an observational bias in the
following sense.  We noted in Rebull \etal\ (2016b) that, particularly
for the shape changers/moving double dips, it was possible that if
there had been much more data, $>>$70 d, encompassing more cycles,
then it might have been possible to differentiate the two (or more?)
periods contributing to the changing shape. In Praesepe, since the
periods are on average longer than in the Pleiades, consequently there
are (on average) fewer complete cycles encompassed in the K2 campaign.
Thus, perhaps the higher occurrence rate of shape changers/moving
double dips and the lower occurrence rate of resolved close peaks (for
G and K stars) in Praesepe may be attributable to the number of
complete cycles available for stars in Praesepe (as compared to the
Pleiades).  

For the M stars in the Pleiades, we postulated that the close resolved
peaks were binaries, since most of them appeared above the single-star
main sequence. The distant resolved peaks we thought were most likely
to be binaries. The same basic result is true in Praesepe as well --
the M stars with resolved peaks are above the single-star main
sequence (see Sec.~\ref{sec:pvmkmultip} below).  From theory, we expect
stronger differential rotation in earlier stars (\eg, Kitchatinov \&
Olemskoy 2012), and effectively solid body rotation in the M stars. 
Praesepe has a comparable if not slightly greater fraction of resolved
distant peaks, though many more instances where the difference between
the peaks is very large indeed ($>$6 d; see \S\ref{sec:deltaP}
below).

\subsection{Unusual LC Shapes}
\label{sec:batwingy}

The phased light curves of most spotted stars are simple, showing a
morphology that is more or less sinusoidal or one that has one or two
broad ``humps" or dips.   That is as expected, because spots located
at most positions on a stellar surface and when viewed from most
vantage points will be visible for half or more of the rotation
period. That means that most spots will have a contribution to the
phased light curve that spans 180$\arcdeg$ or more in phase.  Flux
dips or other structures that cover less than 90$\arcdeg$ in phase are
very hard to produce with spots. The only way they can be produced is
by placing the spot very near the upper or lower limb of the star. 
Because of limb darkening and geometric fore-shortening, light from
such a location contributes little to the total integrated brightness
of the star; therefore, flux dips from such spots cannot yield light
curve features that are very deep.  

In the Pleiades (Rebull \etal\
2016b), we nevertheless identified six stars with short-duration flux
dips (full width at zero intensity, FWZI, $<$0.2 in phase);  one star
had three such dips, another had two, and the remaining four had just
one flux dip.  The dips were to first order constant in shape and
depth over the duration of the K2 campaign.  All six stars were mid-M
dwarfs with very short ($P <$0.7 day) periods.   We identified another
19 rapidly rotating, mid-to-late M dwarfs in the 8 Myr old Upper Sco
association whose K2 phased light curves seemed to show more structure
than could be explained by spots (Stauffer \etal\ 2017).   We have
attributed the light curve structure to warm clouds of coronal gas
orbiting the stars near their Keplerian co-rotation radius, following
models by Jardine \& van Ballegooijen (2005) and Townsend \& Owocki
(2005).

\begin{figure}[ht]
\epsscale{1.0}
\plotone{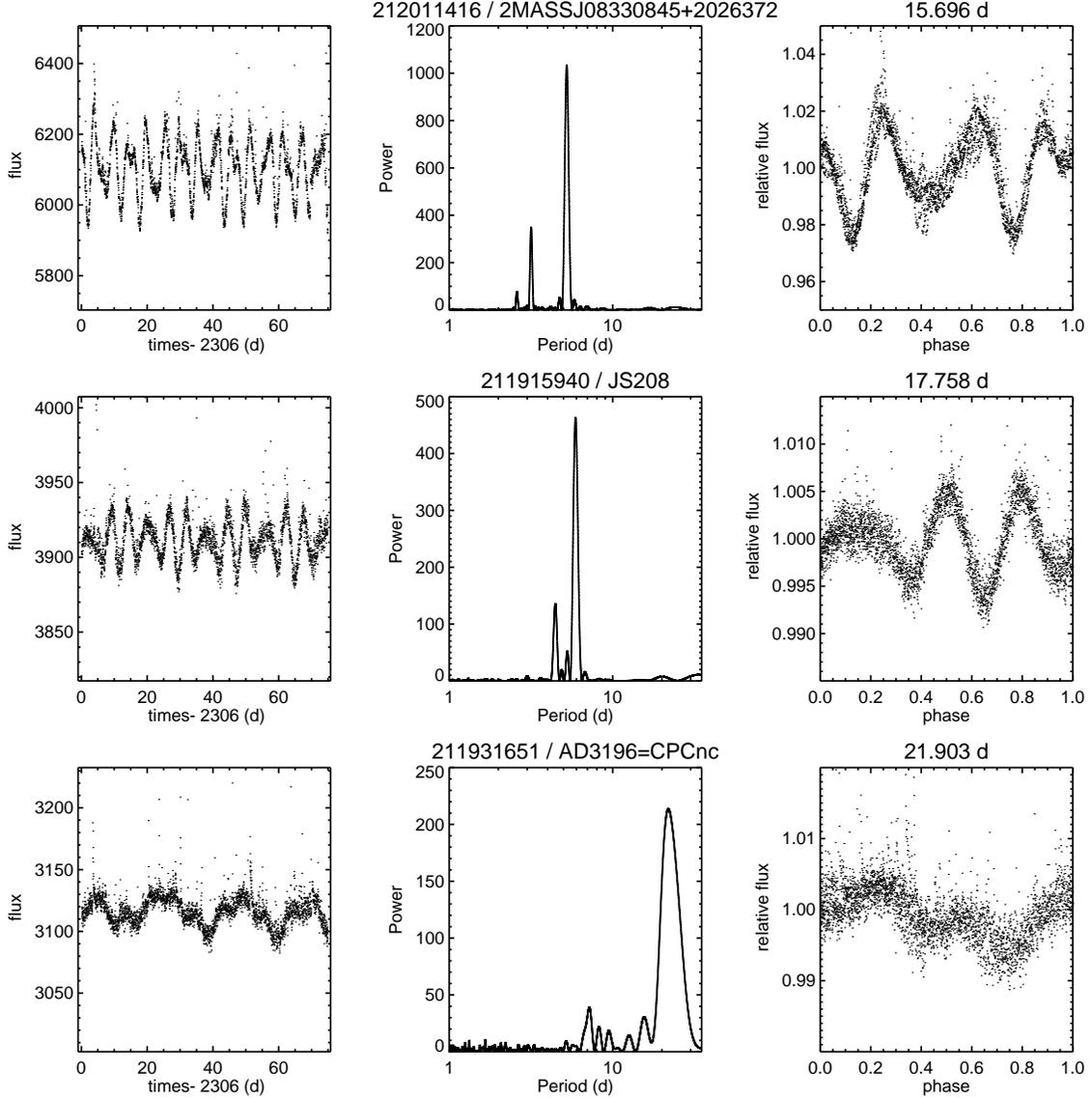}
\caption{ Three stars that have unusual shapes in their phased light
curves. Columns: LC, periodogram, phased LC. Rows, in order:  
212011416/2MASSJ08330845+2026372 (note $x$-axis is just one phase,
though it looks repeated), 211915940/JS208, 211931651/AD3196=CP Cnc.
These light curves all have structures in them uncharacteristic of
most of the rest of the LCs in this study. They are all too slowly
rotating to be of the unusual sorts found in the Pleiades or Upper
Sco. See text for additional discussion.
\label{fig:batwingy}}
\end{figure}

We have found no Praesepe stars with the same properties
(short-duration flux dips or other structure, rapid rotation,
mid-to-late M dwarf spectral type).   However, there are three slowly
rotating Praesepe M dwarfs (\vmk$\sim$5) that do have more structured
light curves than we expect; these are shown in
Figure~\ref{fig:batwingy}: 212011416/2MASSJ08330845+2026372,
211915940/JS208, and 211931651/AD3196=CP Cnc. One (EPIC
212011416/2MASSJ08330845+2026372) has a nearby saturated column in the
K2 data, but multiple LC versions obtain the same shape and (large!)
depth of the features.  Note that in the first two cases, the highest
peak in the periodogram results in a phased light curve that has more
scatter than when phased at three times the peak value; this is
similar to the double-dips (see discussion in Rebull \etal\
2016b).   These stars have periods $\gtrsim$10$\times$
longer than the stars in Pleiades or USco.

We do not believe the physical mechanism producing these light curve
morphologies is the same as for the Pleiades and Upper Sco. Because
the flux dips are broader than in the Pleiades and are generally low
amplitude, it is (barely) possible that spots could be responsible.
However, the full amplitude of one of them (EPIC 212011416) is about
4\%, which would be very difficult to produce using spots located near
the limb of the visible hemisphere.

In order to help constrain the nature of these stars, we obtained
single-epoch spectroscopy of these sources with Keck/HIRES  (see
App.~\ref{app:keck}). All are consistent with being radial velocity
members of Praesepe, and all are narrow-lined (and single-lined).  

The phased light curve morphologies for these three stars are much
more complex than most of the rest of the LCs in Praesepe.   Because
they represent a tiny minority of the Praesepe sample, we do not
believe that our failure to understand their properties should affect
any of the other conclusions in the paper.  We plan a future paper
(Hebb \etal) that will discuss modelling of these LCs.

\subsection{Comparison to Other Classes in the Literature}

Kov\'acs \etal\ (2014) also identified LC/power spectrum classes in
their study of Praesepe. They had (a) monoperiodic sinusoids (13\% of
their periodic sample); (b) single but unstable sinusoid (39\%); (c) 
two peaks interpreted as latitudinal differential rotation (7\%); (d)
power moved into first harmonic (41\%).  

(a) The monoperiodic sinusoids of their periodic sample can be matched
to those in our sample with single periods but that are not double-dip
nor shape-changer. There are 366 of those, or $\sim$45\% of the
periodic sources, in our member sample. (b) Single but unstable
sinusoids would be analogous to our shape changers, without the
double-dips; we find 164 of them, 20\% of the periodic sources.  (c)
They rarely find two discrete peaks in the power spectrum; this would
be analogous to our resolved close and distant peaks combined, and we
have 15\% of our periodic sample falling into this category. We agree
that in some cases, at least, this could be attributable to
latitudinal differential rotation, but we suspect (also see discussion
in Rebull \etal\ 2016b) that at least some (most of the multi-peaked M
stars) are binaries.  (d) Lastly, their category where power is moved
into the first harmonic is analogous to our double-dip category, which
is 20\% of our periodic sample.

Perhaps a more fair comparison is to work just with the 152 stars that
are in common between the two studies. For that sample, we obtain  (a)
monoperiodic sinusoids: 6\%; (b) single unstable sinusoids: 23\%; (c)
resolved peaks: 16\%; and (d) double dips: 47\%.  The closest match in
terms of sample fraction is this last category. 

Assuming that we have correctly captured the relationship between our
classes and those in Kov\'acs \etal\ (2014), we have roughly similar
sample fractions for the same characteristics.

\section{Comparison of the Single- and Multi-Periodic Sources}
\label{sec:singleandmulti}

In the prior section, we identified the $\sim$20\% of stars with
multiple periods.  In this section, we focus on where the
multi-periodic sources fall with respect to the single-period sources
in a variety of parameter spaces. We compare the single- and
multi-period sources within Praesepe, and also to our analysis of the
same phenomenon in the Pleiades (from Rebull \etal\ 2016b).

\subsection{Amplitudes}
\label{sec:amplitudes}

\begin{figure}[ht]
\epsscale{1.0}
\plotone{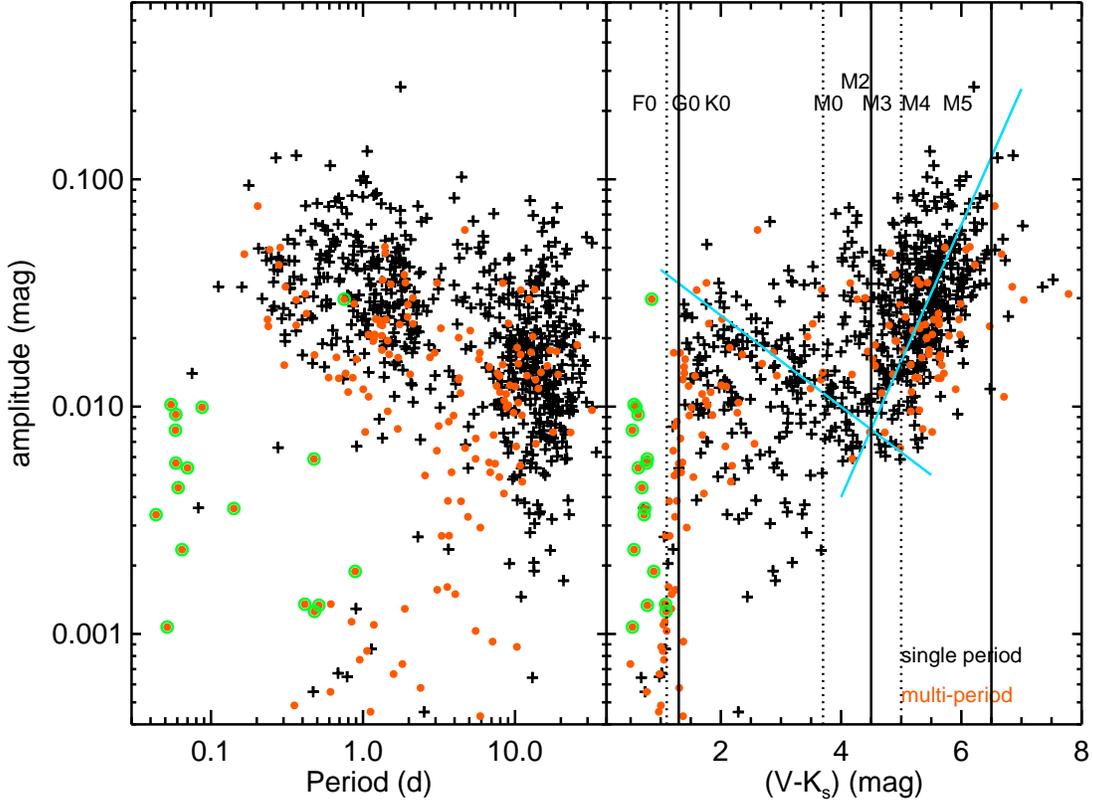}
\caption{The amplitude (from the 10th to the 90th percentile), in
magnitudes, of the periodic light curves, against $P$ (left) and \vmk\
(right).  The vertical lines are the divisions from
Fig.~\ref{fig:pvmk2e} -- solid lines are for Praesepe and dotted lines
are for the Pleiades. Black points are those stars with single
periods, and orange dots highlight those with multiple periods. An
additional green circle highlights the likely pulsators. Blue lines in
the second panel are lines just to guide the eye in the text
discussion. Longer periods have smaller amplitudes.
}
\label{fig:ampldistrib}
\end{figure}

We calculated the amplitude of the periodic signal in the same fashion
as we did in the Pleiades; we assembled the distribution of all points
in the light curve, took the log of the 90th percentile flux,
subtracted from that the log of the 10th percentile flux, and
multiplied by 2.5 to convert to magnitudes.
Figure~\ref{fig:ampldistrib} plots that amplitude against both $P$ and
\vmk\ for the periodic light curves. Note that this is not necessarily
the amplitude of a sinusoid overlaid or fit to the periodic signal,
but the amplitude of the overall light curve, which necessarily
includes long-term trends.

In the Pleiades, there was no obvious trend of amplitude with color or
period. However, in Praesepe, there is a trend in both of these
panels. On the left of Fig.~\ref{fig:ampldistrib}, longer period stars
have lower amplitudes. This is consistent with expectations in that,
certainly by this age, more slowly rotating stars should be less
active and therefore have smaller spots. Note, however, that nearly
all of the turn-down at the longest periods are higher-mass G and K
stars; likewise, the shorter $P$ stars are primarily mid to late-M
stars. The bimodal distribution of periods seen in
Fig.~\ref{fig:phisto} is apparent here too. There are more
multi-period low-amplitude sources than single period low-amplitude
sources (though several of these are likely pulsators, which explains
the low amplitude in those bluest and smallest $P$ cases). On the
whole, however, there is no clear distinction in the left panel of
Fig.~\ref{fig:ampldistrib} between the distributions of singly and
multi-periodic sources.

On the right of Fig.~\ref{fig:ampldistrib}, it can be seen that bluer
stars (corresponding to \vmk$<$1.3, blue points in
Fig.~\ref{fig:pvmk2e}) are generally multiperiodic and have smaller
amplitudes, not all of which are likely pulsators.   There are several
lower amplitude LCs with colors redder than \vmk$\sim$1.3.  This, and
the trend towards larger amplitudes at even redder colors
(\vmk$>$4.5), could be an observational bias in that stars need to be
`bright enough' for a period to be derivable; fainter stars need
larger amplitudes to be seen as periodic, and brighter stars with
smaller amplitudes are more easily detected. Note, however, that there
is substructure within the right panel of Fig.~\ref{fig:ampldistrib}
--  there is a `clumping' of the distribution for 1.3$\leq$\vmk$<$4.5
(corresponding to the yellow points in Fig.~\ref{fig:pvmk2e}) that
moves to lower amplitude as the color gets redder. Then, it turns
around and moves to larger amplitude as the color gets redder for
\vmk$>$4.5 (the green/red points in Fig.~\ref{fig:pvmk2e}). For {\em
both} of these color regimes, the amplitude gets smaller as the period
gets longer.  This makes sense in the standard rotation-activity
sense, if the spot filling factor (at least for the
non-axisymmetrically distributed component) decreases for longer
periods.

\subsection{Distribution with Color}
\label{sec:color}

\begin{figure}[ht]
\epsscale{0.8}
\plotone{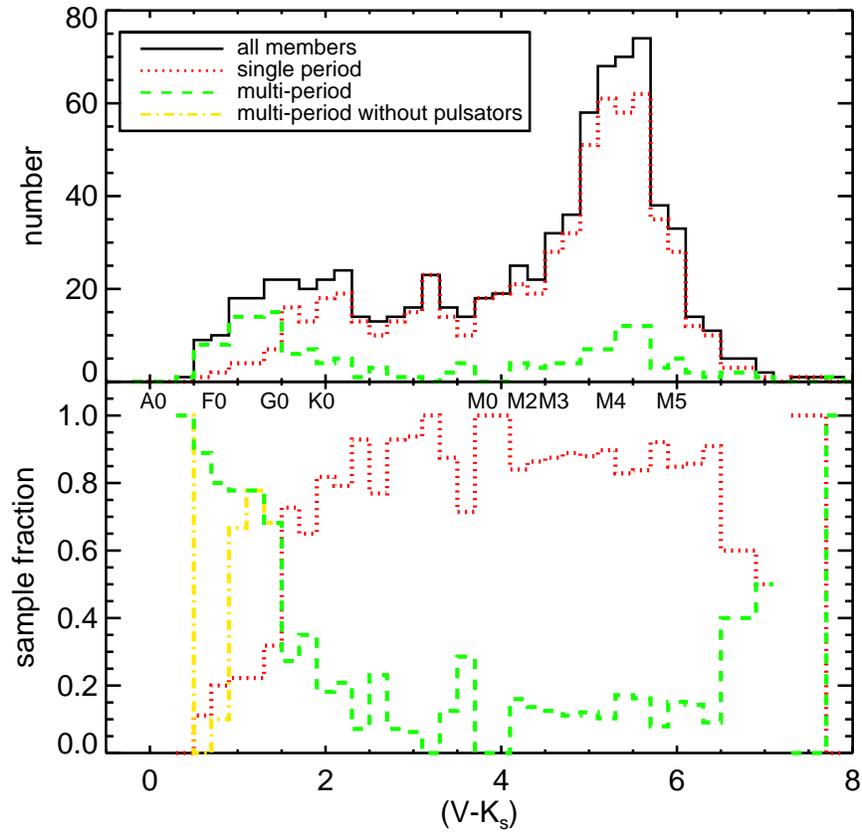}
\caption{Distribution of \vmk\ for the ensemble, with the single (red
dotted) and multiple (green dashed) populations called out. The top
panel is absolute numbers, and the bottom is the sample fraction.
(The yellow dash-dot line is the sample fraction that is
multi-periodic with the likely pulsators removed.) The transition
between where multiple periods dominate is much bluer here
(\vmk$\sim$1.5) than it was in the Pleiades (\vmk$\sim$2.6), and
corresponds roughly to the point where the bluest `branch' turns down
in the $P$ vs.\ \vmk\ diagram in Fig.~\ref{fig:pvmk} above. The
fraction of the rest of the sample that has multiple periods is
roughly constant with color through the rest of the sample. 
\label{fig:vkdistrib}}
\end{figure}

In the Pleiades, we found a strong correlation between multiple
periodicities and \vmk\ -- most of the earlier stars were multiply
periodic and nearly all the later stars were singly periodic.  The
distribution in Praesepe is different than in the Pleiades; see
Figure~\ref{fig:vkdistrib}.  Through most of the sample, the fraction
that has multiple periods is roughly constant with color. Clearly the
bins for \vmk $\lesssim$ 1 and \vmk$\gtrsim$6.5 are significantly
affected by sample completeness. 

There is a transition between where multiple periods dominate to where
single periods dominate, and it is at \vmk$\sim$1.5  ($\sim$G0).  The
Pleiades sample extends to bluer colors, and the transition between
where multiple periods dominate to where single periods dominate is at
\vmkz$\sim$2.6 (early K).

Inclusion of the likely pulsators in this analysis affects the
results, particularly because of the sample differences for
\vmk$\lesssim$1 (see Fig.~\ref{fig:vkdistrib4}). Dropping the likely
pulsators causes the sample fraction of multi-period sources to
plummet blueward of \vmk$\sim$1 . However, the multi-period sources
still dominate for 1$\lesssim$\vmk$\lesssim$1.5. 

If latitudinal differential rotation dominates blueward of this
\vmk$\sim$1.5 transition point (which was one of our hypotheses in the
Pleiades for the transition point there), then this transition has
moved to more massive types by the age of Praesepe. This transition is
also essentially where the bluest `branch' turns down in the $P$ vs.\
\vmk\ diagram in Fig.~\ref{fig:pvmk} above, again suggesting that
something physically different happens blueward of this color.
Stronger differential rotation is expected for hotter stars (\eg,
Kitchatinov \& Olemskoy 2012), so this may be the dominant effect.
However, identification of a star as multi-period at all may be
limited by the number of complete cycles in the K2 campaign, as
discussed in Sec.~\ref{sec:comparetypestopleiades}.

\subsection{$P$ vs.\ \vmk}
\label{sec:pvmkmultip}

\begin{figure}[ht]
\epsscale{0.8}
\plotone{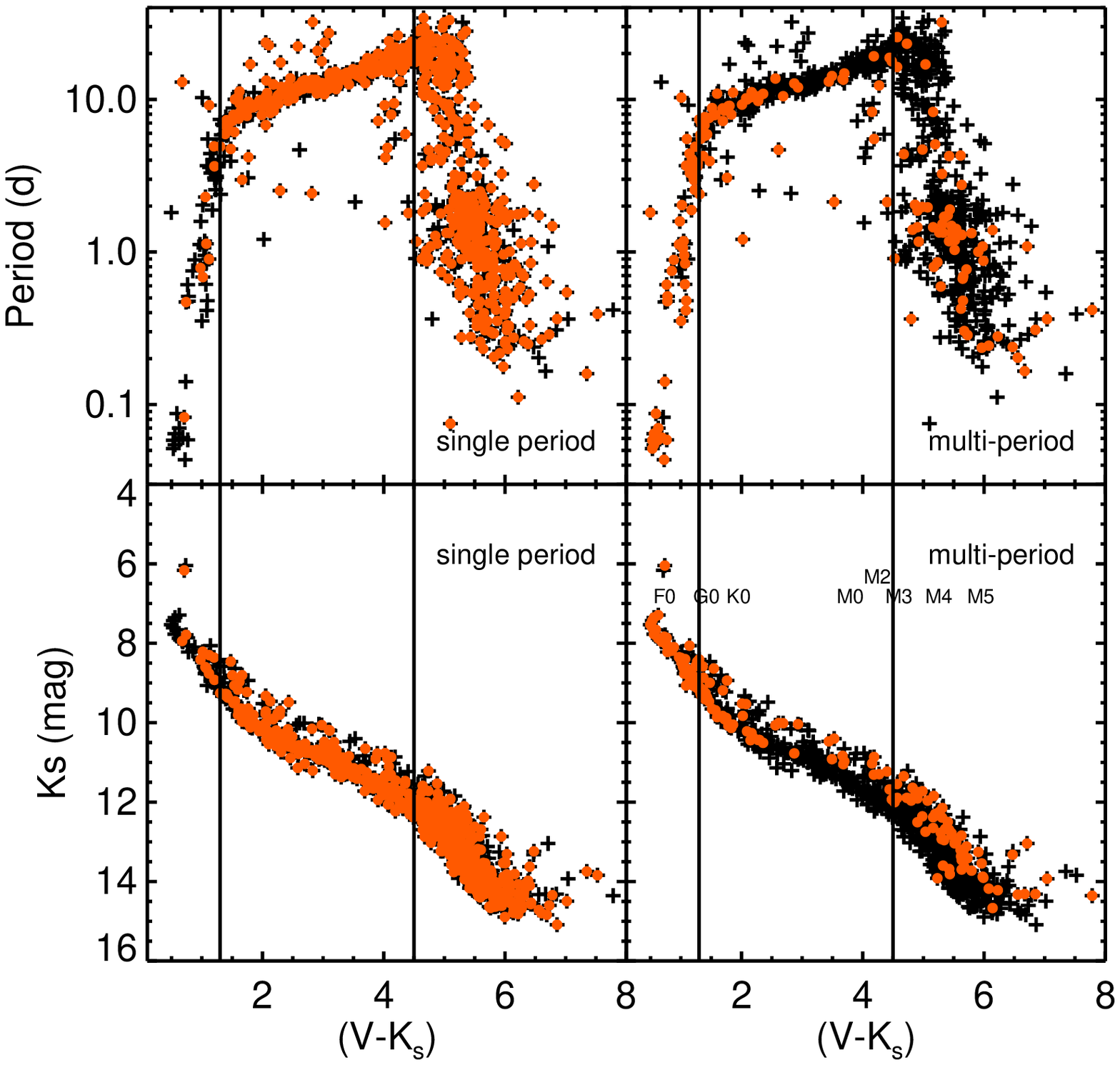}
\caption{Plot of $P$ vs.~\vmk\ (top row) and \ks\ vs.~\vmk\ (bottom
row) highlighting the single-period (left) and multi-period (right)
populations.  The vertical lines are at \vmk=1.3 and 4.5, the
divisions from Fig.~\ref{fig:pvmk2e} above, for reference.   The
multi-period stars dominate for \vmk$\lesssim$1.5-2. The M stars that
are multi-period are more likely to be photometric binaries, as in the
Pleiades.  }
\label{fig:pvmk3freq}
\end{figure}

Figure~\ref{fig:pvmk3freq} shows where the multi-period stars fall in
the  $P$ vs.~\vmk\ and \ks\ vs.~\vmk\ parameter spaces. 
This figure is in direct analogy to Rebull \etal\ (2016b), Figure 12;
in the Pleiades, most of the early-type stars were multi-periodic, and
the later types were nearly all singly periodic, and the
multi-periodic later type stars were nearly all photometric binaries.

In Praesepe, the single-period stars are distributed more uniformly
throughout the diagram, consistent with Fig.~\ref{fig:vkdistrib}. The
earlier type stars are dominantly multiply periodic for
\vmk$\lesssim$1.5-2, which is a smaller range than for the Pleiades,
again consistent with Fig.~\ref{fig:vkdistrib}. It is still true that
most of the multi-period stars with \vmk$\gtrsim$3.5-4 seem to be
dominated by photometric binaries, given the position in the \ks\
vs.~\vmk\ diagram. This is consistent with the M stars still rotating
as solid bodies at Praesepe age. 

\begin{figure}[ht]
\epsscale{1.0}
\plottwo{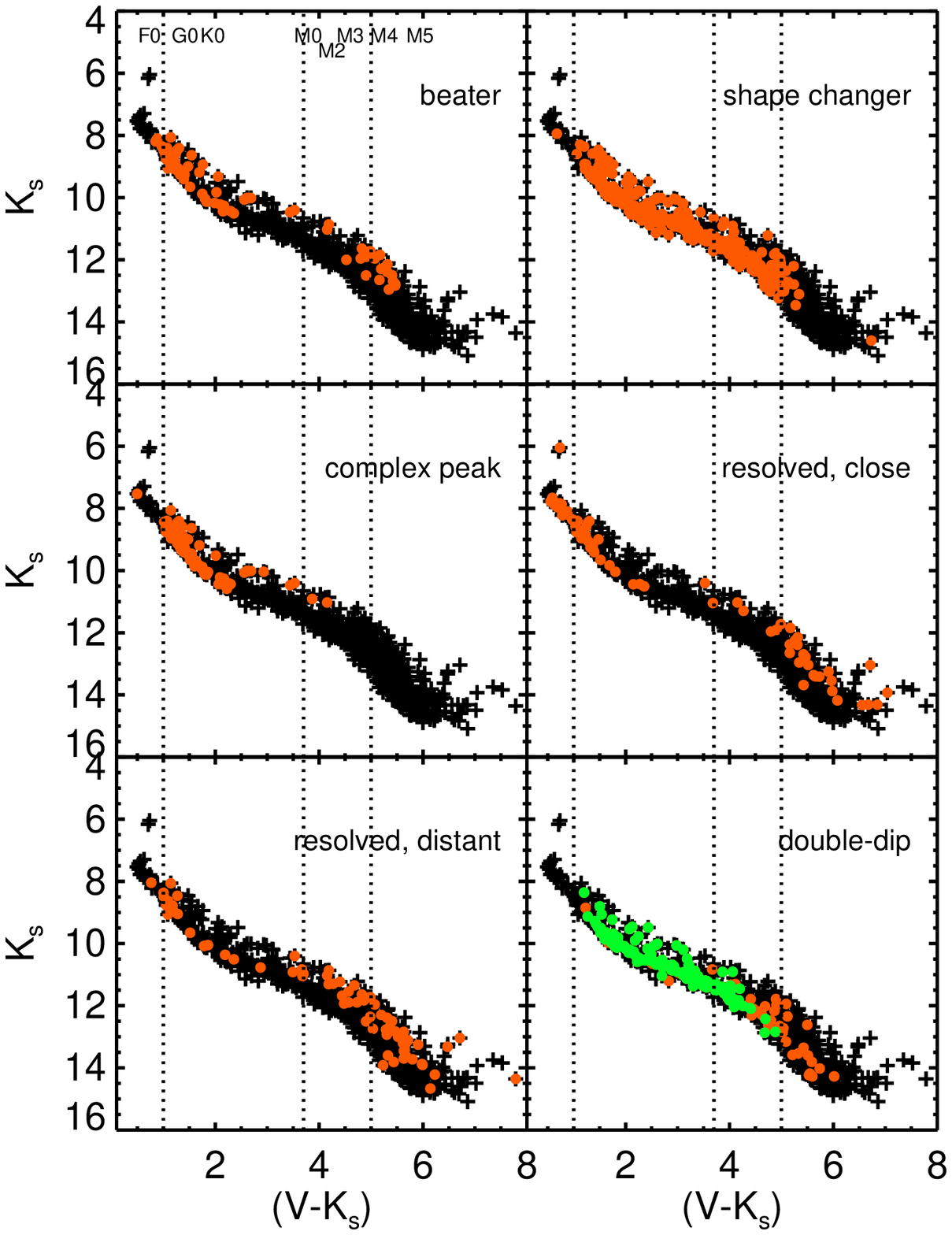}{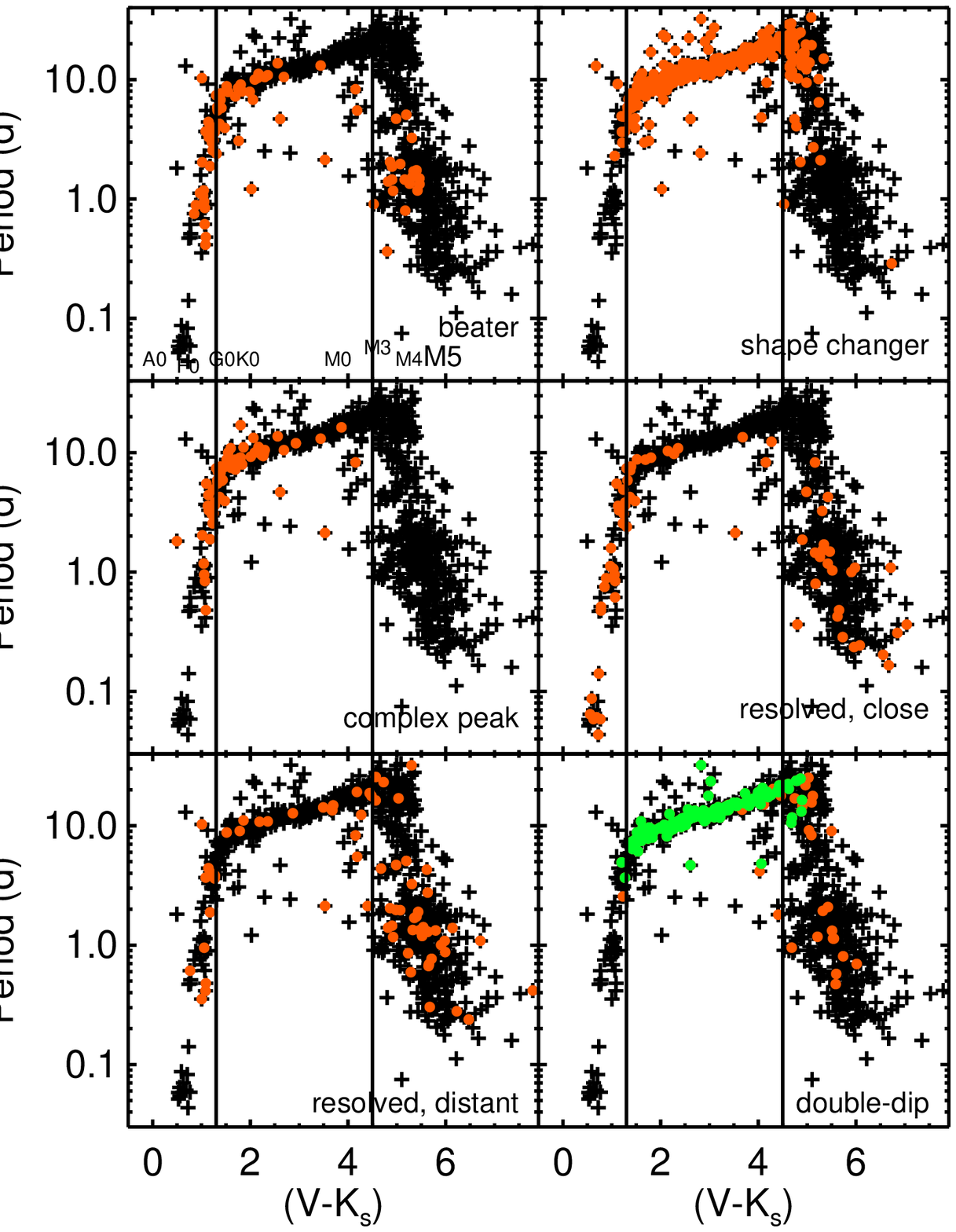}
\caption{Plot of \ks vs.~\vmk\ (left) and $P$ vs.~\vmk\ (right),
highlighting several of the LC categories. Notation (and the location
of the dotted vertical lines) is same as Fig.~\ref{fig:pvmk3freq},
except for the double dip panel, where the moving double-dip stars are
green, and the double-dip stars are orange.  }
\label{fig:optcmdclasses}
\end{figure}

The panels in Figure~\ref{fig:optcmdclasses} break down where the
individual LC classes appear in the \ks\ and $P$ vs.~\vmk\ diagrams;
the analogous Pleiades figures in Rebull \etal\ (2016b) are Figs.~13
and 14. 

Beaters and complex peaks were both, in the Pleiades, found to
dominate the stars with \vmk$<$3.7. In Praesepe, the region where they
dominate has shifted blueward, to \vmk$\lesssim$2.5-3, consistent with
the discussion associated with Fig.~\ref{fig:vkdistrib} above.  In
both cases, beaters that are also M stars are more likely to be
binaries than single stars, based on position in the CMD. Shape
changers dominated in the Pleiades for 1.1$<$\vmk$<$3.7, with a
significant fraction of the stars in the `disorganized region' (with
3.7$<$\vmk$<$5) being in this category. In Praesepe, there are a
higher fraction of shape changers, and they dominate the `middle
branch' (yellow points in Fig.~\ref{fig:pvmk2e}) with
1.3$\leq$\vmk$<$4.5. If the shape changers reflect rapid spot
evolution and/or differential rotation, this happens more frequently
in Praesepe than it does in the Pleiades. Note that the stars that
exhibit shape changing behavior in Praesepe are on average rotating
much more slowly than in the Pleiades. The longer periods on the
slowly rotating branch for 1.3 $<$\vmk$<$ 4.5 for Praesepe relative to
Pleiades combined with the fixed campaign length for K2 makes it
likely that stars will be moved from the resolved close peak category
into the shape-changer or moving double-dip category, explaining at
least some of the differences in light curve class distributions we
see.

The LCs with two distinct periods can be found throughout the $P$
vs.~\vmk\ diagram for both clusters, and, for M stars, tend to be on
the brighter side of the cluster distribution in the CMD. For the M
stars in particular, these are more likely to be binaries. Resolved
distant peaks are distributed throughout the $P$ vs.~\vmk\ diagram,
but resolved close peaks tend to cluster in the bluest portion of the
$P$ vs.~\vmk\ diagram (and to some extent in the M stars'
fast-rotating clump). Resolved close peaks and bluer colors may be
differential rotation and/or spot evolution. (See Sec.~\ref{sec:deltaP} below.)

Finally, moving double-dip stars dominate the `middle branch' (yellow
points in Fig.~\ref{fig:pvmk2e}) of the Praesepe $P$ vs.~\vmk\ diagram
to a much larger extent than in the Pleiades. There are fractionally
many more moving double-dip stars in Praesepe. There are fractionally
far fewer M stars exhibiting any double-dip behavior in Praesepe than
in the Pleiades.

\clearpage

\begin{figure}[ht]
\epsscale{1.0}
\plottwo{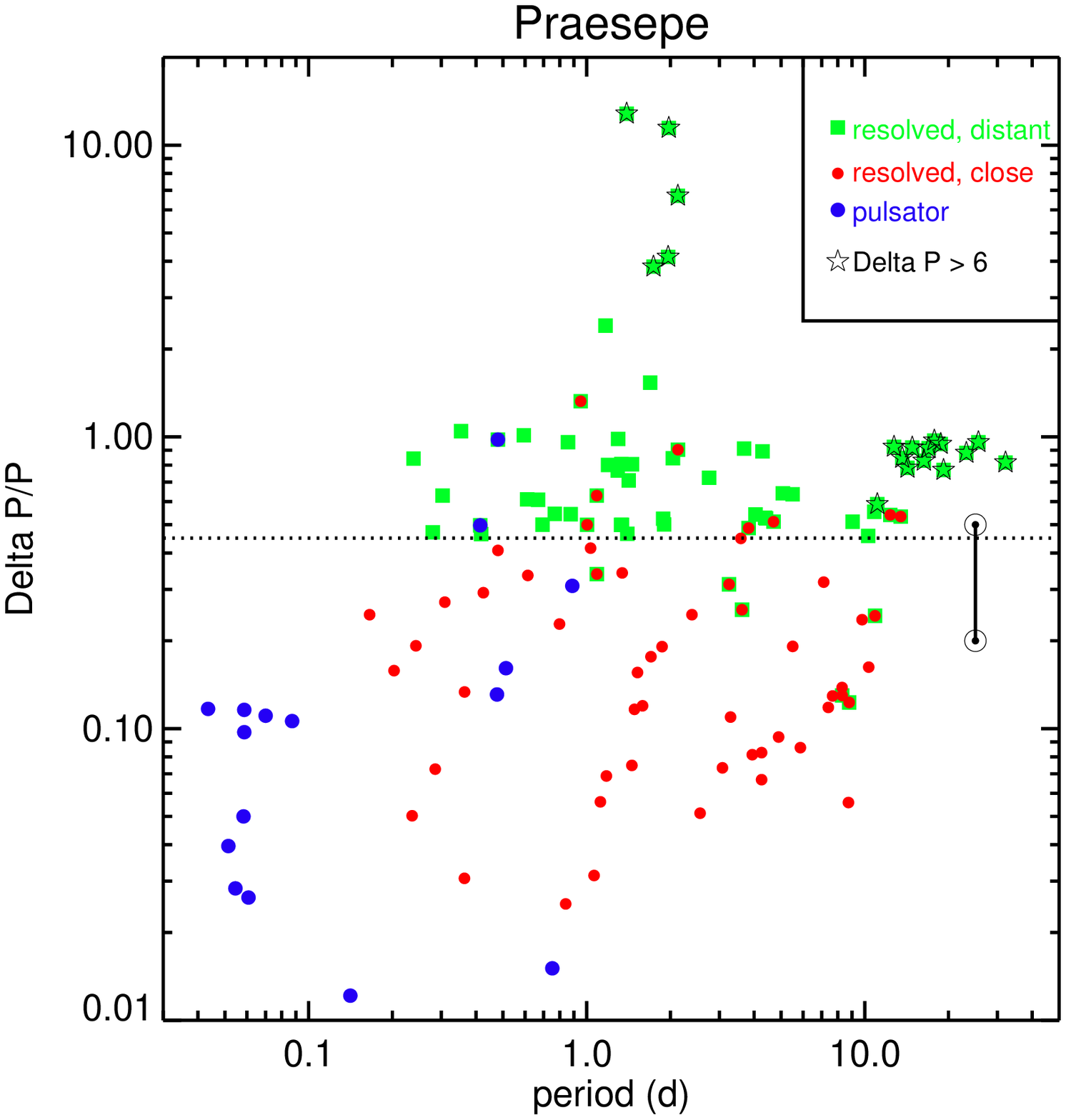}{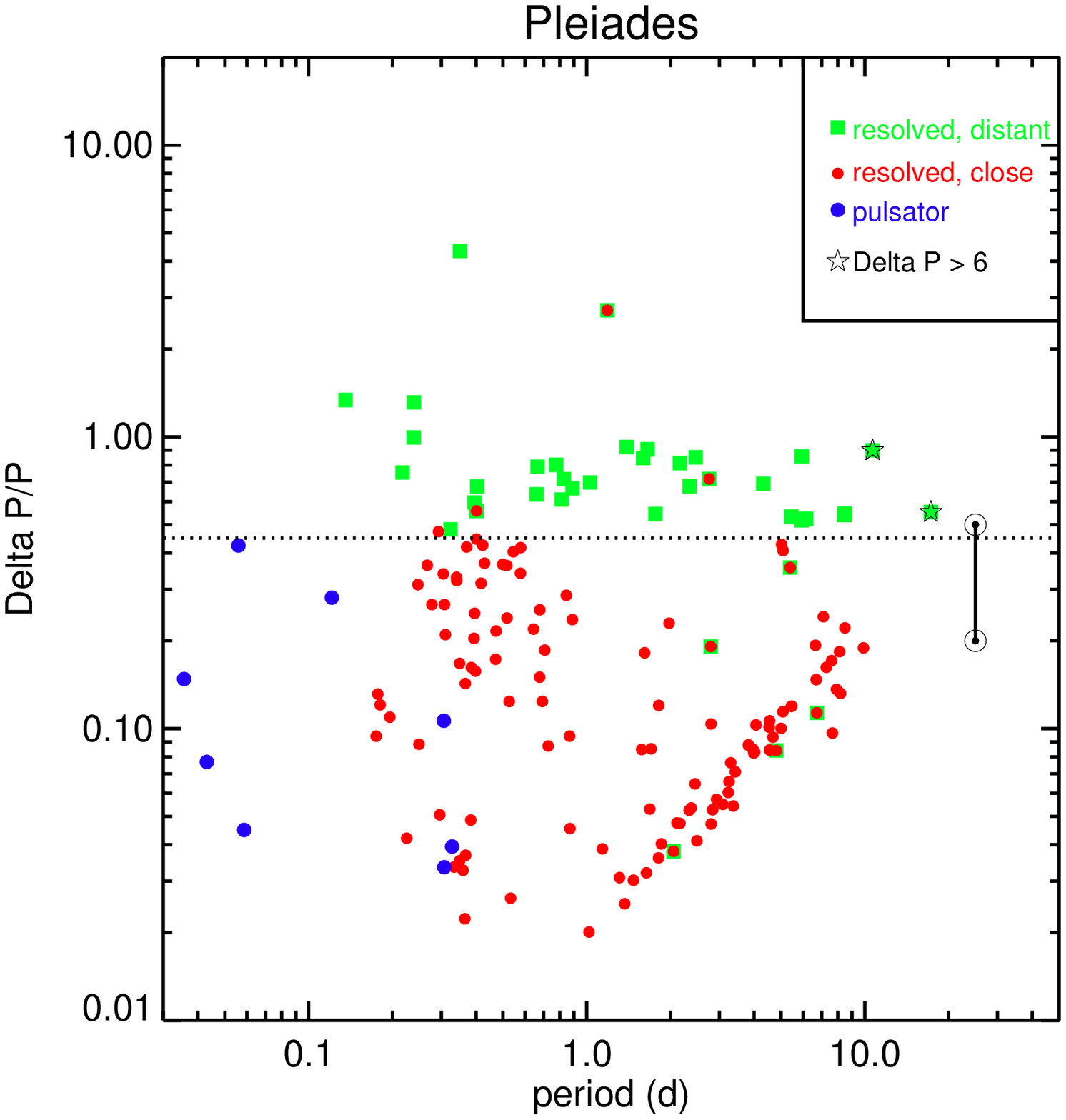}
\caption{For Praesepe (left) and the Pleiades (right): 
Plot of $\Delta P/P_1$ vs.~$P$ for pulsators (blue dots),
resolved distant peaks (green squares), and resolved close peaks (red
dots). An additional black star indicates that $|\Delta P|>$6 d. 
The range of possible values for the 
Sun is included for reference
($\odot$);  if one takes as $\Delta P$ the range of periods measured
where sunspots occur, $\Delta P/P_1 \sim 0.1-0.2$, but if one takes
the full range of $\Delta P$, equator to pole, $\Delta P/P_1 \sim
0.5$. The dotted line is at $\Delta P/P_1$=0.45 and denotes the
boundary between close and distant resolved peaks. The Pleiades data
originally appeared in Fig.~15 from Rebull \etal\ (2016b).}
\label{fig:deltap}
\end{figure}

\section{$\Delta P$ distributions}
\label{sec:deltaP}

 Finally, we calculated the $\Delta P/P_1$ metric for stars
with resolved multi-period peaks in Praesepe.  In the Pleiades work
(Rebull \etal\ 2016b), we took the closest two periods out of those
detected, subtracted the smaller from the larger, and divided by the
period we take to be the rotation period of the primary star in the
system. In this calculation, it could be that the rotation period is
not involved in the numerator.  We have refined our thinking on this
calculation, and now take the closest peak to the $P_{\rm rot}$,
subtract the smaller from the larger, and divide by the $P_{\rm rot}$.
This change ties the $\Delta P/P_1$ metric more closely to the
rotation period of the star.  Our previous calculation is effectively
identical to this calculation for all but 5 stars in the Pleiades
multi-period sample, though it affects 15\% of the Praesepe stars with
at least two periods (which is half of those with at least three
periods.)  

The plot of this corrected $\Delta P/P_1$ vs.\ $P$ for
Praesepe and the Pleiades is shown in Figure~\ref{fig:deltap}. (In
both clusters, most of the likely pulsators are in a different region
of the diagram than other stars.) The morphology of this diagram in
Praesepe is missing two prominent features from the Pleiades version
of this figure. In the Pleiades, there was a very obvious linear
feature in the lower right. Some of this was a selection effect in
that points in the lower right are harder to observe (harder to
distinguish periods and thus harder to obtain two periods that can be
used to place them in this plot). In Praesepe, the feature is not as
obvious.  We suspected that in the Pleiades, this feature was a
signature of differential rotation in stars primarily located on the
slow sequence.  As shown in Fig.~\ref{fig:pvmk3freq}, Praesepe has
very few multiperiod stars on the slow sequence, and for the K and M
dwarfs on that sequence, nearly all the multiperiod stars appear to be
binaries (based on their location in the CMD).  As discussed in
Sec.~\ref{sec:comparetypestopleiades}, the relative paucity of
resolved close peak and moving double dip stars on the slow sequence
in Praesepe (the stars that primarily populate the linear sequence in
the Pleiades $\Delta P/P$ vs.~$P$ plot) is likely a result of their
longer periods -- we simply do not have a long enough sequence of data
to resolve two periods in the periodogram.

The second prominent feature in Fig.~\ref{fig:deltap} that is present
in the Pleiades but absent in Praesepe is a clump of stars near
$P\sim$0.4 and $\Delta P/P_1 \sim$0.3.   Those points are primarily
binary M dwarfs.  We believe the lack of a well-defined peak of binary
M dwarfs in Praesepe is primarily a result of the dependence of
angular momentum loss on mass and period. (See
Figures~\ref{fig:pvmk2e} and~\ref{fig:pvmk3freq}.)   If the Praesepe M
dwarf binaries  were drawn from the same parent distribution at
Pleiades age, they would have started with very few stars with periods
more than 5 days.    Figure~\ref{fig:pvmk2e} shows that many of those
stars must spin down to periods $>$10 days by Praesepe age while
others still maintain quite rapid rotation.   Therefore, it is natural
that the Praesepe M dwarf binaries will have a much large range in
period and a much larger range in $\Delta P/P_1 $ than in the
Pleiades.

Indeed, Praesepe has many more stars for which there are two
very different periods. Those instances where $\Delta P$ is greater
than 6 days are highlighted in the Figure; they clump in two different
locations depending on whether the shorter or the longer period was
taken to be the $P_{\rm rot}$ for the primary. (There are only two
such points in the Pleiades with  $\Delta P >6$.) None of these
Praesepe $\Delta P >6$ points are particularly outliers in the $P$
vs.~\vmk\ diagram; they are distributed between 1.8$<$\vmk$<$6.5
within where the bulk of the distribution is located. The five stars
near $P\sim$1, $\Delta P/P_1 \sim$3-20 are in the M star fast rotating
sequence, and the remainder are within the top, slowly rotating
sequence. They are not particularly clustered in the brighter portion
of the cluster sequence in the CMD, so they are not necessarily
photometric binaries. However, it is hard to imagine a situation in
which latitudinal differential rotation results in a $\Delta P$ of 6
days or more from pole to equator, and we suspect these are all
binaries. The angular momentum loss mechanism operating in these stars
is probably a function of both period and stellar mass.  For the stars
with \vmk$\gtrsim$4, these could be examples of systems where one star
has spun down and the other has not; the difference in periods in
these binaries has significantly increased since the age of the
Pleiades. These pairs could perhaps shed light on the mechanism that
causes some M stars to rapidly spin down (see, \eg, Newton \etal\
2016).

\section{Conclusions}
\label{sec:concl}

We have presented our analysis of nearly a thousand K2 LCs of Praesepe
members, increasing the known Praesepe rotation periods by nearly a
factor of four; 809 of the 941 members (86\%) with K2 LCs have a
measured period from the K2 data.  The distribution of $P$ vs.\ \vmk,
a proxy for mass, has three different regimes:  \vmk$<$1.3
($\lesssim$F8), where the rotation rate rapidly slows as the mass
decreases;  1.3$<$\vmk$<$4.5 ($\sim$F8 to $\sim$M3), where the
rotation rate slows more gradually as the mass decreases; \vmk$>$4.5
($\gtrsim$M3), where the rotation rate rapidly increases as the mass
decreases. Particularly in this last, primarily M star regime, there
is a bimodal distribution of periods, with few between $\sim$2 and
$\sim$10 days. We interpret this to mean that once M stars start to
slow down, they do so rapidly; this is likely a predecessor of the
bimodal distribution of M star rotation rates found in much older
field stars (\eg, Newton \etal\ 2016, Kado-Fong \etal\ 2016).

The distribution of $P$ vs.\ \vmk\ exhibits significant changes
between the Pleiades ($\sim$125 Myr) and Praesepe ($\sim$790 Myr). For
the late F, G, K, and early M stars, the overall distribution slows
considerably compared to the Pleiades, and the higher mass branch
(late F and earliest G) steepens significantly. The transition at
\vmk=1.3 is the Kraft break.  The G and K and early M stars have a
more shallow relationship, with the lower masses rotating more slowly.
For the mid M stars, the period-color relationship changes relatively little
between the Pleiades and Praesepe, though it is steeper in Praesepe
than the Pleiades. Overall these late-type stars rotate significantly
faster as mass decreases, but there is more than an order of magnitude
spread in the rotation rates in Praesepe at any given mass. 

We found the same diversity of LC and periodogram classes in Praesepe
as we did in the Pleiades. About three-quarters of the periodic member
stars in both clusters have only one period; the rest have multiple
periods.  Praesepe has a higher fraction of LC classes we have
interpreted as latitudinal differential rotation and/or spot or spot
group evolution, but this may be influenced by the observing window;
Pleiades stars rotate faster, so there are more complete cycles
observable in the $\sim$70 d K2 campaign for the Pleiades stars than
for Praesepe stars.  Multi-periodic stars dominate the bluest end of
the sample; the transition between where multi-periodic stars dominate
over single-period stars happens at a bluer color in Praesepe
(\vmk$\sim$1.5) than it does in the Pleiades (\vmk$\sim$2.6). 

M stars in both clusters that have multiple periods are likely to be
binaries. In Praesepe, there are many more LCs that have two widely
separated periods, $\Delta P>$6 days. Some of these could be examples
of systems where one M star has spun down but the other has not.

K2 continues to revolutionize the study of rotation in young and
intermediate age open clusters. 

\clearpage

\acknowledgments

Some of the data presented in this paper were obtained from the
Mikulski Archive for Space Telescopes (MAST). Support for MAST for
non-HST data is provided by the NASA Office of Space Science via grant
NNX09AF08G and by other grants and contracts. This paper includes data
collected by the Kepler mission. Funding for the Kepler mission is
provided by the NASA Science Mission directorate. 

This research has made use of the NASA/IPAC Infrared Science Archive
(IRSA), which is operated by the Jet Propulsion Laboratory, California
Institute of Technology, under contract with the National Aeronautics
and Space Administration.    This research has made use of NASA's
Astrophysics Data System (ADS) Abstract Service, and of the SIMBAD
database, operated at CDS, Strasbourg, France.  This research has made
use of data products from the Two Micron All-Sky Survey (2MASS), which
is a joint project of the University of Massachusetts and the Infrared
Processing and Analysis Center, funded by the National Aeronautics and
Space Administration and the National Science Foundation. The 2MASS
data are served by the NASA/IPAC Infrared Science Archive, which is
operated by the Jet Propulsion Laboratory, California Institute of
Technology, under contract with the National Aeronautics and Space
Administration. This publication makes use of data products from the
Wide-field Infrared Survey Explorer, which is a joint project of the
University of California, Los Angeles, and the Jet Propulsion
Laboratory/California Institute of Technology, funded by the National
Aeronautics and Space Administration. 

\facility{Kepler} \facility{K2} 
\facility{2MASS}

\appendix

\section{Two Close Stars}
\label{app:pair}

Out of the 984 LCs, only one pair of stars for which a K2 LC was
requested were within 4$\arcsec$ of each other (within a Kepler
pixel).  EPIC 211934148 is at 08:43:07.40 +19:14:15.4, and as such, is
matched to JS519; EPIC 211934221 is at 08:43:07.44 +19:14:19.2,
3.88$\arcsec$ away, and as such is matched to UGCSJ084307.42+191419.2.
The two stars are of similar brightness in the POSS images, but the
latter is slightly fainter. 

Both of these stars are clearly M stars. There are two obvious,
effectively identical periods in each of the (very similar) light
curves.  The one from the brighter star (211934148) is about double
the counts than the one from the fainter star (211934221).  We suspect
that each star is contributing one periodic signal. However, it is
less clear which period belongs to which star.

The two periods derived from the joint light curves are very different
from each other. A period of 2.954 d is recovered cleanly in both LCs.
The other period, at $\sim$23-24 d, is less reliably recovered,
presumably because there are fewer complete cycles in the K2 campaign.
The longer period derived from the light curves is 23.293 and 24.421,
respectively (with the slightly longer period originating in the LC
with fewer counts). Following the same approach as with the rest of
the light curves, there is more power in the $\sim$23-24 d peaks, so
that period was assigned as $P_1$, the likely rotation period for both
EPIC numbers, with the other period assigned to $P_2$. However, since
these are well-separated peaks, these two LCs were both identified as
coming from likely binaries.

Given our initial data amalgamations above (Sec.~\ref{sec:litphotom}),
we have $K_s$=12.222 and 12.618 mag, and no $V$.  Our estimates of
\vmk\ (see Sec.~\ref{sec:vmk}) are very different for our URAT and
Gaia approaches. Gaia measures $G$=15.601 and 16.422 for the two
stars, and the \vmk\ we thus derive are 4.479 and 5.272. Via the URAT
approach, we obtain $f$ magnitudes of 15.05 and 15.53; this is at
least consistent with the POSS images that suggest the two stars are
very close in brightness. This results in \vmk\ of 4.017 and 4.158. We
looked via Vizier to find additional direct measurements of $V$ for
these stars. NOMAD reports $V$=16.12 and 17.160, respectively, which
would result in \vmk=3.898 and 4.542 mag.

\begin{figure}[ht]
\epsscale{0.7}
\plotone{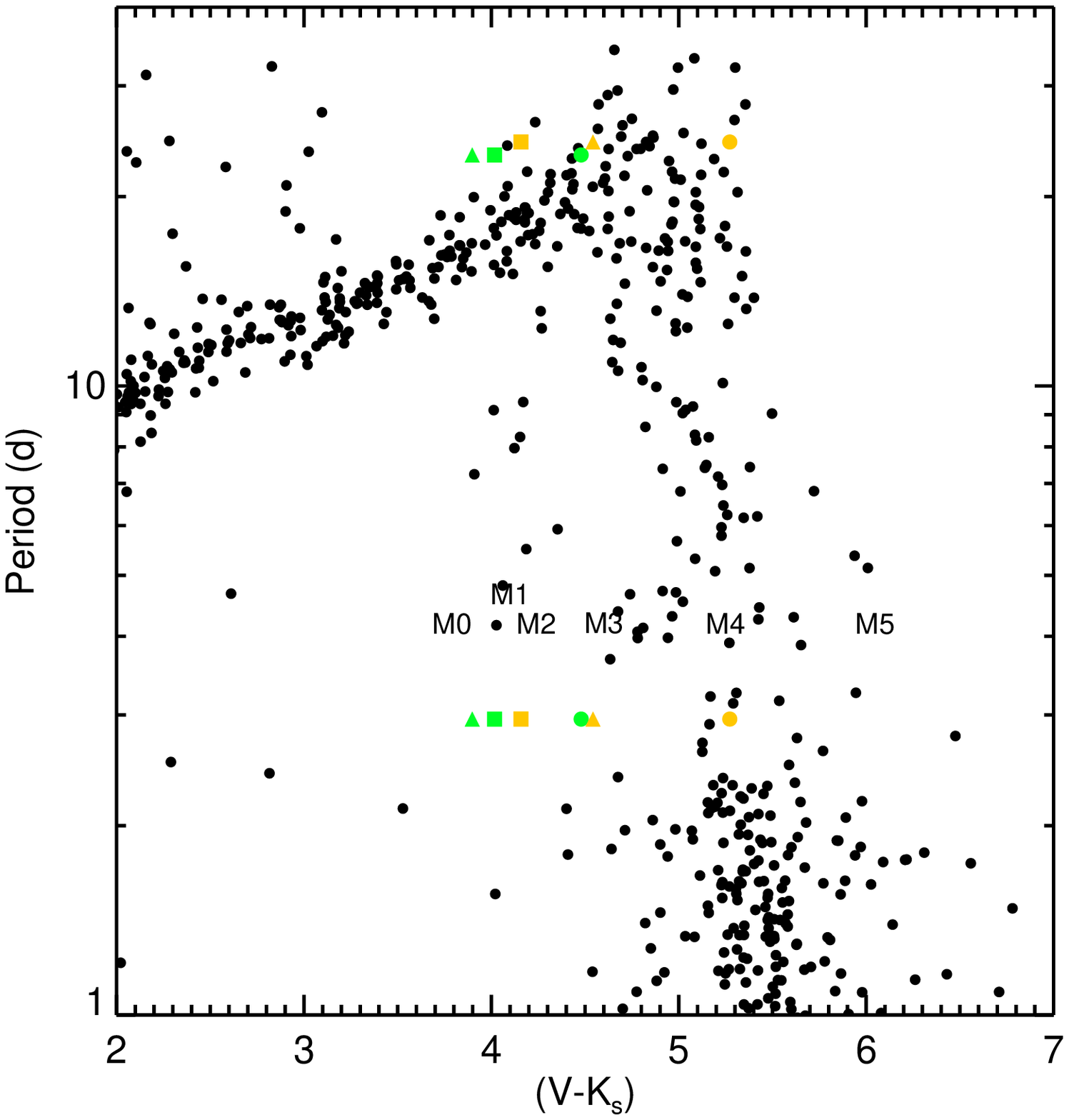}
\caption{Zoom-in on the $P$ vs.\ \vmk\ parameter space for the region
in which EPIC 211934148/JS519 (green symbols) and  EPIC
211934221/UGCSJ084307.42+191419.2 (orange symbols) appear. (Black dots
are the rest of the sample.) Approximate M spectral types are
indicated as a function of \vmk\ near the middle of the plot (in a
relatively sparsely populated region). As described in the text, both periods
appear in both light curves, one near 3 days and one near 24 days.
Colored circles are using the \vmk\ derived via Gaia magnitudes,
triangles are using the \vmk\ derived via URAT magnitudes, and squares
are using the NOMAD-reported $V$ measurements (and 2MASS $K_s$) to
obtain \vmk. We have retained the Gaia-based \vmk\ estimates (circles) and
assigned the $\sim$24 d period to the bluer star (EPIC
211934148/JS519, green circle) and the $\sim$3 d period to the redder star (EPIC
211934221=UGCSJ084307.42+191419.2, orange circle). }
\label{fig:pvmkpair}
\end{figure}

Having no abundantly clear answer with respect to what the \vmk\
should be for each of these stars, we sought input from the $P$ vs.\
\vmk\ plot; see Fig.~\ref{fig:pvmkpair}. If these stars are `typical'
Praesepe members, they should not be outliers. Following the same
approach as with the rest of the sample, these stars would both appear
within the ensemble, even if both use the $\sim$24 d period.  Taking
the shorter period and tying it to the redder (fainter) source, it too
is within the distribution at that location, though it is {\em just}
above the denser population of M stars. The shorter period does not
appear to be appropriate for the brighter (bluer) star; it is an
outlier at that location at any \vmk. Using the URAT \vmk\ estimates
results in the most discrepant points; both stars would be outliers in
those cases. Using the NOMAD $V$, the points move slightly right
(redder) but are still not necessarily `within the pack' of the rest
of the stars. 

We suspect that the Gaia-based \vmk\ estimates are the best we are
going to have at this time. We have retained those, and assigned the
$\sim$24 d period to the bluer star (EPIC 211934148/JS519) and the
$\sim$3 d period to the redder star (EPIC
211934221=UGCSJ084307.42+191419.2).

\section{Binaries}
\label{app:binaries}

\begin{figure}[ht]
\epsscale{1.0}
\plottwo{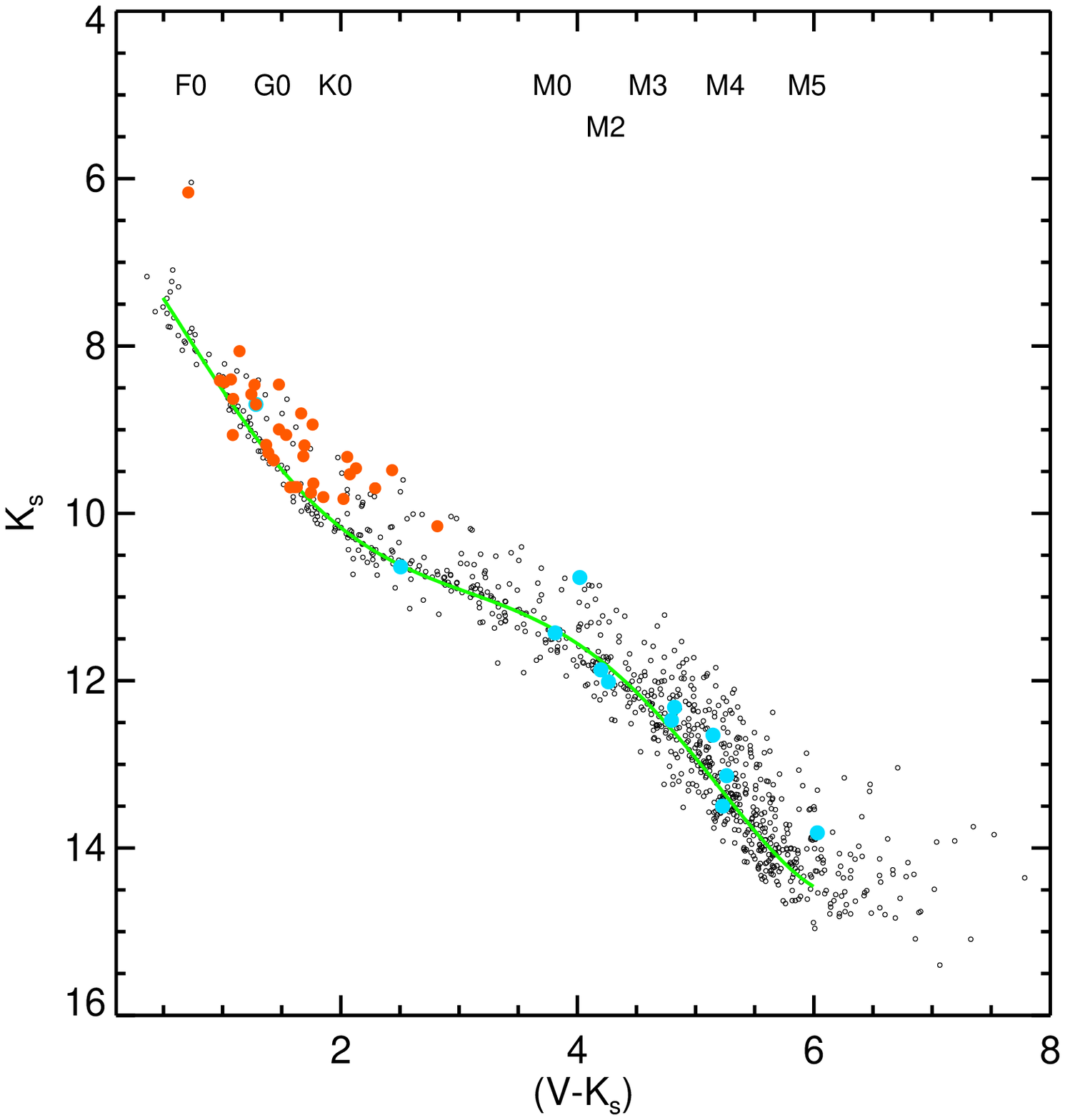}{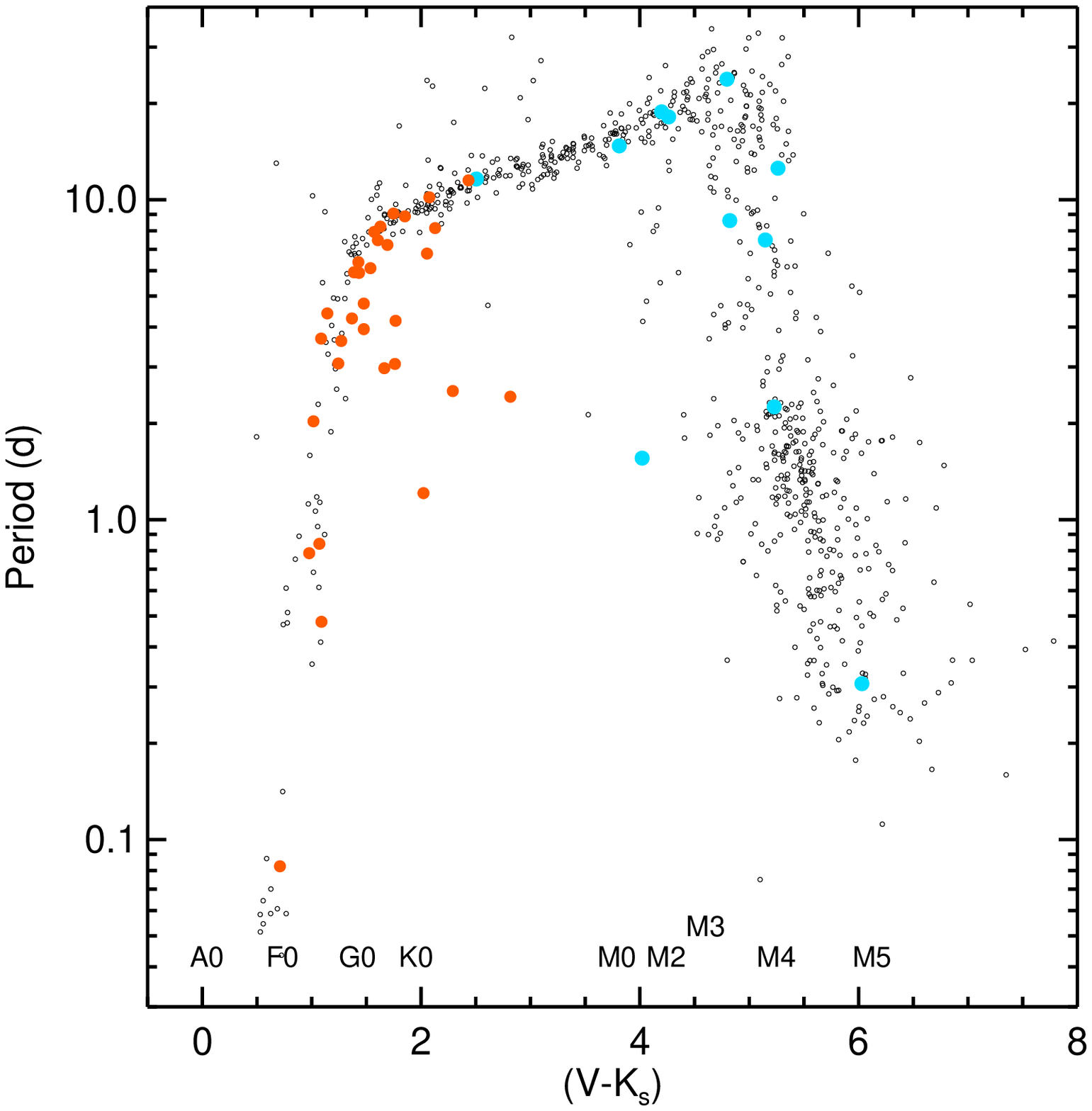}
\caption{Optical CMD (\ks\ vs.~\vmk), left,  and $P$ vs.~\vmk, right,
for the ensemble (small dots), those stars with EBs or planets
(orange dots), and those stars that are literature binaries but for
which we still have a rotation period (blue dots). Non-members have already
been omitted.  Many of the fast-rotating
G and K outliers in the $P$ vs.~\vmk\ plot are known, tidally locked binaries. }
\label{fig:optcmdbin}
\end{figure}

There is limited information in the literature on binaries in
Praesepe. Here we collect what information we have.
Figure~\ref{fig:optcmdbin} has a color-magnitude diagram and a $P$
vs.~\vmk\ diagram of these binaries. We now discuss eclipsing binaries
(EBs) and binaries from the literature.

\subsection{Eclipsing Binaries}

\begin{deluxetable}{clcp{9cm}}
\tabletypesize{\scriptsize}
\tablecaption{Likely Eclipsing Binaries or Planets\label{tab:ebs}}
\tablewidth{0pt}
\tablehead{\colhead{EPIC} & \colhead{Other Name} & \colhead{Coord (J2000)} & \colhead{Notes}}
\startdata
212009427 & AD1508      & 083129.87+202437.5  & sinsuoidal $P$ with dips superimposed; to appear in Gillen \etal\ 2017 in prep\\
211916756 & JS183       & 083727.05+185836.0 & Also Obermeier \etal\ 2016; Mann \etal\ 2016b; Libralato \etal\ 2016; Pepper \etal\ 2017 \\
211969807 & JC126       & 083832.82+194625.7 &     Also Mann \etal\ 2016b; Libralato \etal\ 2016; Pope \etal\ 2016 \\
212002525 & UGCSJ083942.02+201745.0 & 083942.03+201745.0 &        to appear in Gillen \etal\ 2017 in prep\\
211918335 & KW244=TXCnc & 084001.70+185959.4 & W Uma type (Whelan \etal\ 1973)\\
211970147 & JC193       & 084013.45+194643.7 &   Also Mann \etal\ 2016b \\
211901114 & JS441       & 084135.69+184435.0 &  Possible non-member \\
211822797 & HSHJ385     & 084138.48+173824.0 &  Possible non-member, very shallow\\
211946007 & HSHJ430     & 084239.43+192451.9 &   to appear in Gillen \etal\ 2017 in prep\\
211922849 & JS550       & 084344.73+190358.5 &    grazing system \\
211919680 & HSHJ474     & 084403.90+190112.8 &  transit signal may be from an unassociated projected background object  \\
211972086 & AD3814      & 085049.84+194836.4 &   to appear in Gillen \etal\ 2017 in prep\\
\enddata
\end{deluxetable}

There are 12 objects that we identified in the process of light curve
inspection as candidate eclipsing binaries (or planet candidates) in
this dataset. When we report a period here for these objects, we
believe that the $P$ we report and use in our analysis is a rotation
period. All of the EBs are listed in Table~\ref{tab:ebs}. 

We note explicitly here that one object is a known  W Uma type EB
according to Whelan \etal\ (1973): EPIC 211918335/KW244=TX Cnc.
Kov\'acs \etal\ (2014) also report this as an EB. For this LC, we do
not include our 0.191446 d period in our analysis of rotation periods.

\subsection{Literature Binaries}

Mermilliod \etal\ (2009, 1999) includes stars in Praesepe that were
monitored for spectral binarity (radial velocity variations) over
$\sim$20 years. Kov\'acs \etal\ (2014) also reports photometric
periods they attribute to binary components. Barrado \etal\ (1998)
identified spectral binaries from single-epoch observations.
Table~\ref{tab:sb} includes a comparison of the periods derived here
to those derived for the binaries in these papers.  Many of the
periods given in the literature are $>>$35 d, so there is no way that
we could have obtained that period from these data. For cases where
the reported literature periods are both close (and $\lesssim$35 d),
more often than not, we do not recover both periods. It could be that
one of the two stars does not have spots, or organized enough spots,
to create a detectable signal.

\floattable
\begin{deluxetable}{clcccp{9cm}}
\tabletypesize{\scriptsize}
\rotate
\tablecaption{Binaries in the Literature\label{tab:sb}}
\tablewidth{0pt}
\tablehead{\colhead{EPIC} & \colhead{Other Name} & \colhead{Coord} & \colhead{Period(s)} &
\colhead{Period(s)\tablenotemark{a}} & \colhead{Notes} \\
& &\colhead{(J2000)} & \colhead{(here, d)}& \colhead{(literature, d)} }
\startdata
\cutinhead{Kov\'acs \etal\ (2014)}
211935509\tablenotemark{c}  & KW539=S4   & 083648.96+191526.4 &     7.475       & 3.72481,5551 & HAT-269-0001402; our $P_{\rm rot}$ is twice theirs, and we believe ours to be correct\\
211977390\tablenotemark{b}  & KW55=S27   & 083749.95+195328.8 &     6.789 & 7.13827,0.838434 & HAT-269-0001352; our peak is broad and the waveform has changes over the campaign; both 6.8 and 7.1 may be consistent; no peak at 0.8d; also fast outlier in $P$ vs.~\vmk\ diagram \\
211949097\tablenotemark{c}  & BD+19d2061 & 083924.97+192733.6 &   3.940, 4.261& 4.14405,1.41563 & HAT-269-0000761; our peak is broad and both of our periods are probably consistent with theirs; no peak at 1.4d\\
211950081\tablenotemark{c}  & KW184      & 083928.58+192825.1 &        10.175      & 10.4866,4.52284 & HAT-269-0001490; well-matched $P_{\rm rot}$; no peak at 4.5d\\
211918335\tablenotemark{d}  & KW244=TXCnc& 084001.70+185959.4 &   (0.1914) & 0.1915 & HAT-269-000058; WUma EB (Whelan \etal 1973); also listed in EBs table \\
211933215\tablenotemark{c}  & S137       & 084041.90+191325.4 &      6.112       & 3.11867,897 & HAT-269-0000850; our peak is correct for our data \\
211972627\tablenotemark{c}  & KW368=S155 & 084110.30+194907.0 &   9.054       & 9.00657,8.50340 & HAT-269-0001570; our waveform undergoes many changes over the campaign, but we do not have two distinct peaks at 9.0 and 8.5d\\
211935518\tablenotemark{c}  & KW434=S184 & 084154.36+191526.7 &    4.184       & 4.13291,0.951565 & HAT-269-0001333; well-matched to $P_{\rm rot}$; we do not find a 0.95d peak; also fast outlier in $P$ vs.~\vmk\ diagram\\
211947631 & BD+19d2087                   & 084305.93+192615.2 &      4.736       & 4.65095,7.71605 & HAT-269-0000465; well-matched $P_{\rm rot}$; no peak at 7.7d\\
211969494\tablenotemark{c}  & S213       & 084320.19+194608.5 &         6.382       & 6.18544,648 & HAT-269-0000913; well-matched $P_{\rm rot}$\\
211896596 & JS655=FVCnc                  & 084801.74+184037.6  & 2.976       & 2.92535,1.01926 & HAT-318-0000612; well-matched $P_{\rm rot}$; we do not find a 1.02d peak; also appears Mermilliod \etal\ (1990) with $P$=2.98d; also fast outlier in $P$ vs.~\vmk\ diagram \\
 \cutinhead{Mermilliod \etal\ (2009)}
211958646 & HD73081                    & 083702.02+193617.2 &          4.413, 2.093 & 45.97&  longer than we can detect\\
211971690 & S12                        & 083711.48+194813.2 &              8.242&    5.86628 & our waveform undergoes many changes over the campaign, but no peak is apparent at $\sim$6d\\
211929178 & KW34=BD+19d2050            & 083728.19+190944.3 &    0.840, 0.819& 7383 & longer than we can detect\\
211959522 & S16                        & 083727.54+193703.1 &              8.873&    144&  longer than we can detect\\
211955820 & BD+20d2130                 & 083727.94+193345.1 &       3.618, 6.605, 2.572, 2.694 & 47.44&  longer than we can detect\\
211947686 & S25                        & 083746.61+192618.0 &              5.933&    458& longer than we can detect\\
211936163 & HD73210                    & 083746.76+191601.9 &          0.082&      35.90  & we should have seen a $\sim$36d period, but none are apparent\\
211977390\tablenotemark{b}  & KW55=S27 & 083749.95+195328.8 &        6.789& 1268& longer than we can detect\\
211983461 & KW58=S29                   & 083752.09+195913.9 &          7.933& 5567& longer than we can detect\\
211942703 & S38                        & 083814.26+192155.4 &              7.225, 9.554, 6.369 & 117& longer than we can detect\\
211990866 & KW100=S42                  & 083824.29+200621.8 &        4.255, 4.604, 3.971 &   4.92946 & comparable to measured periods, but it is unclear whether these are $P_{\rm rot}$ or $P_{\rm orb}$. \\
\enddata
\tablenotetext{a}{The period given from Mermilliod \etal\ (2009) is
$P_{\rm orb}$. The two periods given from Kov\'acs \etal\ (2014) are in
the order $P_{\rm rot}$, $P_{\rm orb}$.}
\tablenotetext{b}{The same star appears explicitly in Kov\'acs \etal\ (2014) and Mermilliod \etal\ (2009),
as well as Mermilliod \etal\ (1999) and Barrado \etal\ (1998).}
\tablenotetext{c}{Also appears in Mermilliod \etal\ (1999).}
\tablenotetext{d}{Also appears in Barrado \etal\ (1998).}
\end{deluxetable}

\floattable
\begin{deluxetable}{clcccp{9cm}}
\tabletypesize{\scriptsize}
\rotate
\tablecaption{Binaries in the Literature CONTINUED}
\tablewidth{0pt}
\tablehead{\colhead{EPIC} & \colhead{Other Name} & \colhead{Coord} & \colhead{Period(s)} &
\colhead{Period(s)\tablenotemark{a}} & \colhead{Notes} \\
& &\colhead{(J2000)} & \colhead{(here, d)}& \colhead{(literature, d)} }
\startdata
\cutinhead{Mermilliod \etal\ (1999)}
212034371 & JS102           6 & 083556.94+204934.7  & 1.21051 & \ldots & listed as binary; also fast outlier in $P$ vs.~\vmk\ diagram\\
212001830 & S11             & 083711.66+201704.9 &     \ldots &1149.5& $P_{\rm orb}$ very much longer than we can detect. Noisy and not detected as periodic\\
211927313 & KW47=BD+19d2052 & 083742.36+190801.5 &   3.077, 3.303, 3.626 & 34.619 & $P_{\rm orb}$ comparable to our max $P$ but this is not detected\\
211988454 & KW127=S51       & 083850.00+200403.4 &     5.909, 8.514&13.2803 & The waveform is very complicated, with more than one period and dense power spectra; 13.3d isn't recovered, but could be embedded in the signal\\
211898181 & KW547           & 084037.88+184200.4 &   2.524 & \ldots & listed as SB1; also fast outlier in $P$ vs.~\vmk\ diagram \\
211975006\tablenotemark{d} & KW367=BD+19d2077 & 084109.60+195118.6 & 3.0675&1659. & Sinusoidal signal at 3.0675d very strong; also fast outlier in $P$ vs.~\vmk\ diagram\\
211912407 & KW439=HD73994   & 084157.81+185442.2 & 2.0297, 2.102&457.8& longer than we can detect; noisy\\
211989010 & BD+20d2198      & 084407.33+200436.9 &    3.682,7.03816 &40.6649& longer than we can detect but no hint of structure on $\sim$40d\\
 \cutinhead{Barrado \etal\ (1998)}
211958646 & HD73081 & 083702.02+193617.2 &                       4.4134,2.093 & \ldots  & listed as SB\\
211991571 & KW146=HD73429 & 083905.23+200701.8 &                0.7856 & \ldots & listed as SB\\
211956096 & KW236 & 083959.82+193400.2 &                        11.509 & \ldots  & listed as SB\\
212003469 & KW297 & 084028.63+201844.8 &   8.1555 & \ldots  & listed as SB\\
211909748 & KW401 & 084130.70+185218.7 &  2.422 & \ldots & listed as SB; also fast outlier in $P$ vs.~\vmk\ diagram\\
211956984 & BD+20d2193 & 084244.41+193447.8 &                   0.4794,0.9488, 0.9937 & \ldots  & listed as SB\\
211947631 & BD+19d2087 & 084305.93+192615.2 &                   4.7357 &  \ldots  & listed as SB\\
\enddata
\tablenotetext{a}{The period given from Mermilliod \etal\ (2009) is
$P_{\rm orb}$. The two periods given from Kov\'acs \etal\ (2014) are in
the order $P_{\rm rot}$, $P_{\rm orb}$.}
\tablenotetext{b}{The same star appears explicitly in Kov\'acs \etal\ (2014) and Mermilliod \etal\ (2009),
as well as Mermilliod \etal\ (1999).}
\tablenotetext{c}{Also appears in Mermilliod \etal\ (1999).}
\tablenotetext{d}{Also appears in Barrado \etal\ (1998).}
\end{deluxetable}
\clearpage

\section{Timescales}
\label{app:timescales}

\begin{figure}[ht]
\epsscale{1.0}
\plotone{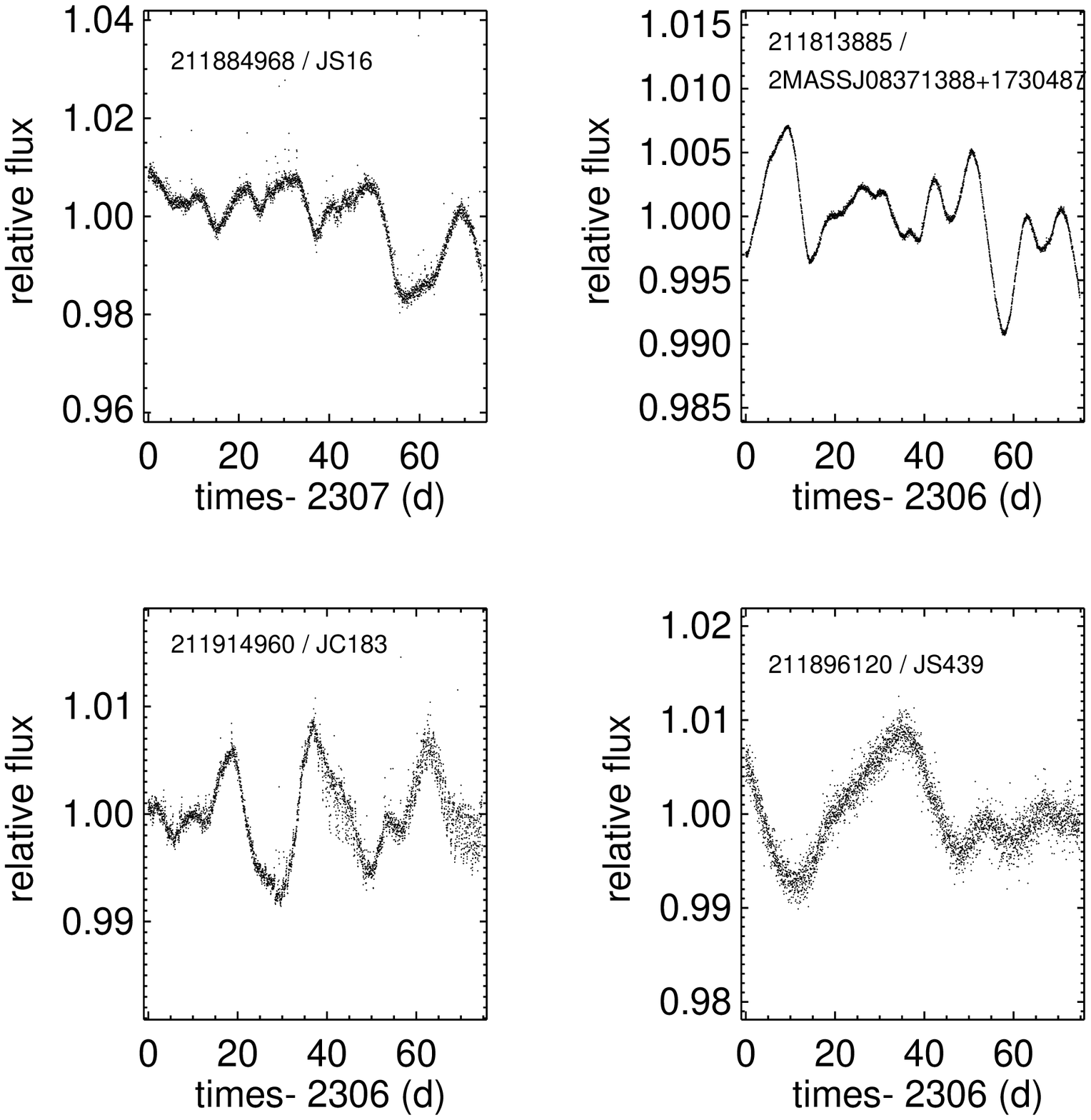}
\caption{Examples of timescales. Upper left: 211884968/JS16,
upper right: 211813885/2MASSJ08371388+1730487,
lower left: 211914960/JC183,
lower right: 211896120/JS439. These stars have repeated patterns that
change considerably with each rotation, so we conclude that they 
may or may not be rotation periods, and have omitted these periods
from the set of rotation periods.  }
\label{fig:timescales}
\end{figure}

\begin{figure}[ht]
\epsscale{1.0}
\plottwo{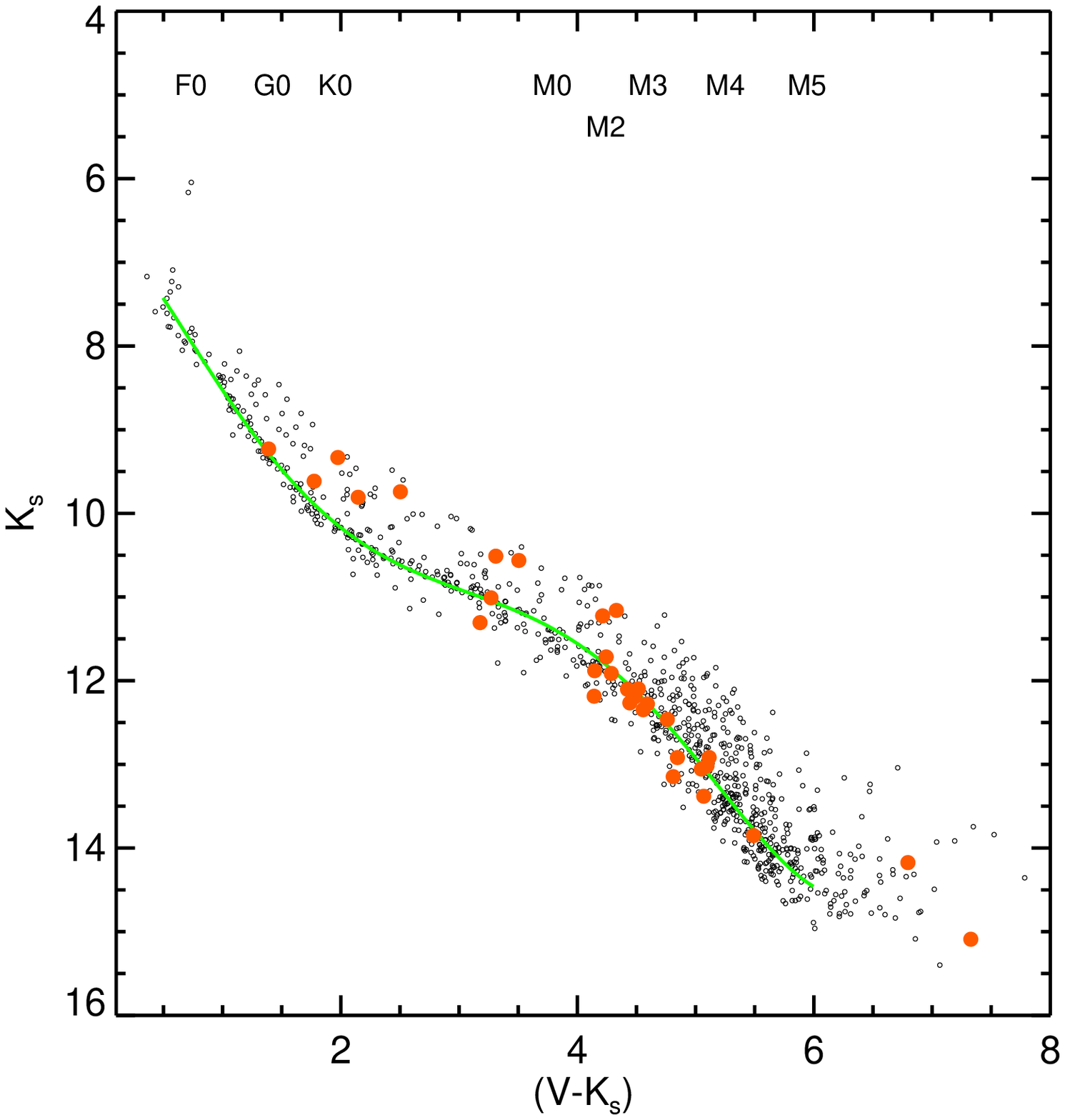}{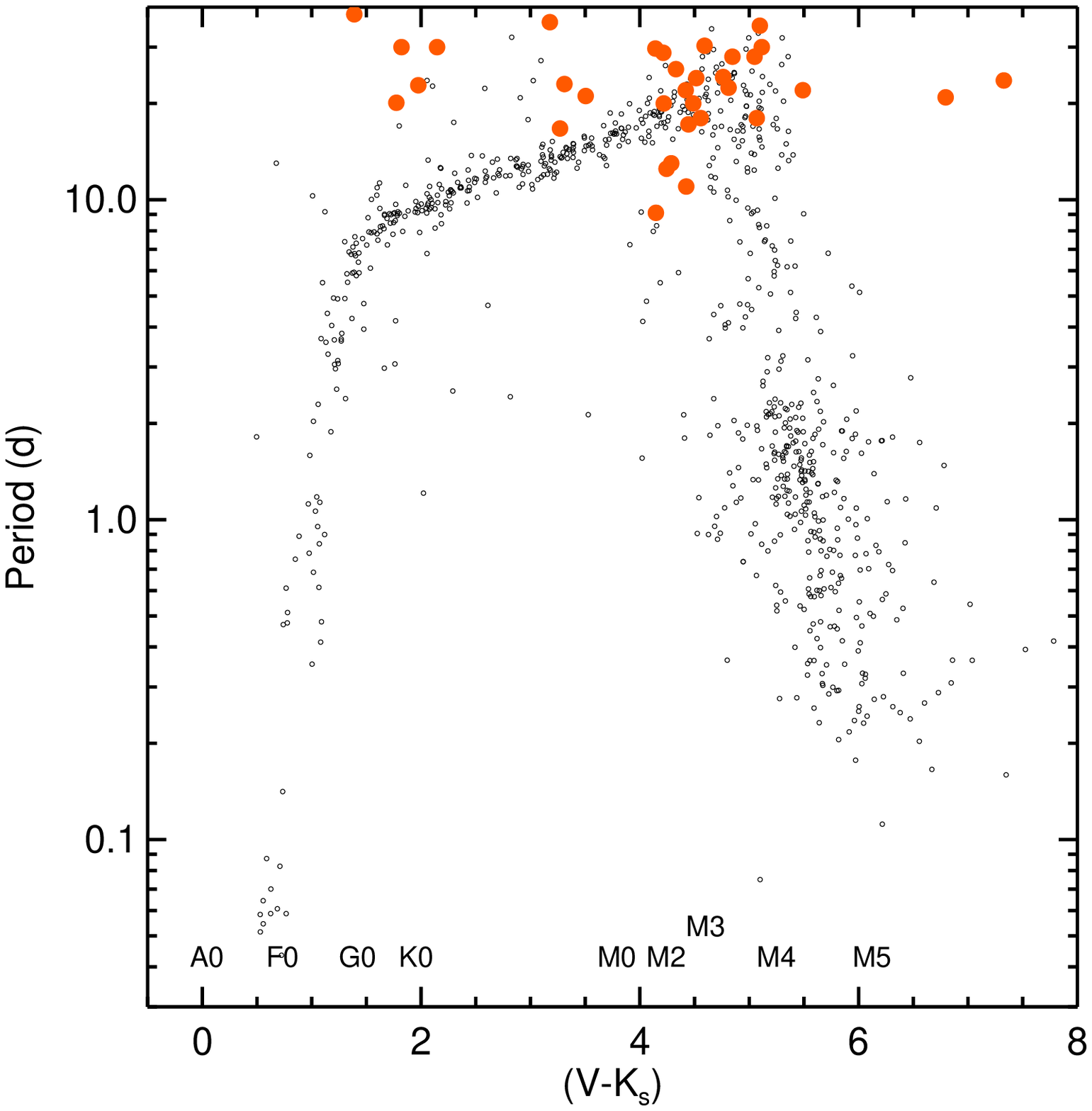}
\caption{Optical CMD (\ks\ vs.~\vmk), left,  and $P$ vs.~\vmk, right,
for the ensemble (small dots), and for stars we identify as having
timescales.  Non-members are included in these plots because many of the
timescale sources are NM. For these timescales, it is not clear if these signatures
come from rotationally-modulated spots or not. These objects are
plotted on the right as if their timescales were periods. Most are
long period outliers. }
\label{fig:optcmdts}
\end{figure}

As for the Pleiades, we identified some objects in Praesepe that have
a repeated pattern, but that we do not think are necessarily due to
spot-modulated rotation periods; see Figure~\ref{fig:timescales} for
four examples.  We describe these as `timescales'; these objects are
listed in Table~\ref{tab:nm}. They do not have a preferential color
range; see Fig.~\ref{fig:optcmdts} (left). They do, however, tend to
have longer periods; see Fig.~\ref{fig:optcmdts} (right). 

\floattable
\begin{deluxetable}{cccc}
\tabletypesize{\scriptsize}
\tablecaption{Timescales\label{tab:timescales}}
\tablewidth{0pt}
\tablehead{\colhead{EPIC} & \colhead{RA, Dec (J2000)} &\colhead{Other name}
&\colhead{Timescale (d)} }
\startdata
 211985684       & 082409.33+200124.9      &            2MASSJ08240933+2001249  &       $\sim$30        \\
 211995609       & 082635.20+201056.6      &            2MASSJ08263520+2010567  &       $\sim$22        \\
 211935588       & 082820.20+191530.7      &            2MASSJ08282020+1915307  &       25.6    \\
 212025207       & 082944.41+204023.1      &            AD1268=TYC1391-40-1     &       $\sim$13        \\
 212173061       & 083008.81+233336.4      &            2MASSJ08300882+2333365  &       $\sim$24        \\
 211787610       & 083055.77+170822.6      &            2MASSJ08305580+1708223  &       $\sim$18        \\
 211852745       & 083147.19+180322.4      &            2MASSJ08314719+1803225  &       $\sim$23        \\
 211981509       & 083239.47+195722.2      &            JC10    &       28.8    \\
 211884968       & 083253.10+183029.3      &            JS16    &       $\sim$22        \\
 211993704       & 083550.36+200904.5      &            JS98    &       29.7    \\
 211975682       & 083627.12+195154.6      &            JS132=JC63      &       $\sim$20.1      \\
 211813885       & 083713.83+173048.7      &            2MASSJ08371388+1730487  &       $\sim$20        \\
 211897713       & 083729.40+184135.5      &            2MASSJ08372941+1841355  &       28.00   \\
 211795569       & 083813.64+171515.7      &            2MASSJ08381365+1715158  &       $\sim$24.2      \\
 211776073       & 083824.86+165836.3      &            AD2396=2MASSJ08382489+1658360   &       22.8    \\
 211744621       & 083848.16+163155.9      &            \ldots  &       9.7     \\
 212030517       & 083903.86+204547.6      &            JS264   &       $\sim$30        \\
 212005623       & 083906.87+202054.2      &            JS270=2MASSJ08390688+2020542    &       $\sim$20        \\
 212103507       & 083921.31+220520.7      &            HSHJ259 &       $\sim$38        \\
 211935711       & 083936.43+191537.8      &            JS302=JC166     &       $\sim$18        \\
 211950716       & 083941.66+192900.4      &            HSHJ283 &       $\sim$11        \\
 211914960       & 084002.22+185656.9      &            JC183   &       22.4    \\
 211968228       & 084013.78+194455.9      &            JC194   &       20.9    \\
 212112321       & 084111.62+221551.7      &            HSHJ350 &       $\sim$35        \\
 211896120       & 084132.35+184010.7      &            JS439   &       30.3    \\
 211812292       & 084135.58+172927.1      &            \ldots  &       35.9    \\
 212005503       & 084151.91+202047.8      &            JS459=2MASSJ08415192+2020479    &       21.1    \\
 212108286       & 084207.85+221105.1      &            2MASSJ08420785+2211051  &       $\sim$28        \\
 211886612       & 084312.92+183150.9      &            JS525   &       17.2    \\
 211983811       & 084332.62+195933.0      &            HSHJ458=2MASSJ08433262+1959330  &       $\sim$30        \\
 211939484       & 084552.80+191900.6      &            2MASSJ08455280+1919006  &       16.7    \\
\enddata
\end{deluxetable}

\section{Likely Pulsators}
\label{app:pulsator}

There are 18 stars that we suspect are pulsators; that is, their power
spectra have a `forest' of short-period peaks (see
Sec.~\ref{sec:LCPcats} and Rebull \etal\ 2016b). One of these 18 is one of the
stars that is too bright for our final best sample. All of these 
pulsators are listed in
Table~\ref{tab:pulsator}. Most of them are bright (\ks$\lesssim$8.5),
and blue (\vmk$\lesssim$1.1). There are some stars that are comparably
bright and blue, but that do not have a comparable power spectrum; if
we derived single periods for them, we retained that period as the
rotation period.

\floattable
\begin{deluxetable}{clcclp{4cm}}
\tabletypesize{\scriptsize}
\tablecaption{Likely Pulsators\label{tab:pulsator}}
\tablewidth{0pt}
\tablehead{\colhead{EPIC} & \colhead{Other Name}& \colhead{Coord (J2000)}& \colhead{\vmk\
(mag)} &\colhead{Periods (d)}  & \colhead{Notes}}
\startdata
211953002&BR Cnc=HD 73175 & 083740.70+193106.3&   0.588&    0.0872, 0.1219, 0.0965 &  F0; $\delta$ Scuti (\eg, Breger 1970)\\
211983602&CY Cnc=HD 73345 & 083837.86+195923.1& 0.530&  0.051,  0.058,  0.053,  0.067&F0V; $\delta$ Scuti (\eg, Hauck 1971)\\
211951863&HD 73397        & 083846.95+193003.3&     0.779&     0.5125, 0.5950, 0.6150, 0.7568 & F4 \\
211957791&BS Cnc= HD 73450& 083909.09+193532.7&   0.625&    0.0587, 0.0644 & A9; $\delta$ Scuti (\eg, Breger 1970, Breger \etal\ 2012)\\
211994121&HD 73616        & 083958.37+200929.6& 0.884 &   0.8876, 0.2713, 0.5718, 0.6137 & F2 \\
211984704&39 Cnc=HD 73665 & 084006.41+200028.0& 2.219& 1.2411 & G8III (dropped as too bright)\\
211941583&HD 73712        & 084020.13+192056.4 &0.735&     0.1413, 0.4815, 0.1179, 0.1395 & Spectroscopic binary; A9V \\
211931309&BV Cnc=HD 73746 & 084032.96+191139.5 &  0.686&    0.0608, 0.0637, 0.0494, 0.0624 & F0; $\delta$ Scuti (\eg, Hauck 1971)\\
211973314&HD 73854        & 084110.67+194946.5& 0.850&    0.7522, 0.6970, 0.7937, 0.7409 & F5 \\
211979345&HD 73872        & 084113.76+195519.1& 0.556&    0.0644, 0.0437, 0.0771, 0.0649 & A5; $\delta$ Scuti (\eg, Breger \etal\ 2012) \\
211935741&HI Cnc=HD 73890 & 084118.40+191539.4 & 0.627&  0.070,  0.078,  0.089,  0.059&A7V; $\delta$ Scuti (\eg, Hauck 1971)\\
211945791&BX Cnc=HD 74028 & 084206.49+192440.4& 0.529&  0.058,  0.126,  0.136,  0.055&A7V; $\delta$ Scuti (\eg, Breger 1970)\\
211914004&BY Cnc=HD 74050 & 084210.80+185603.7 & 0.557&  0.055,  0.053,  0.047,  0.052&A7V; $\delta$ Scuti (\eg, Breger 1970)\\
211954593&BD+20d2192      & 084240.71+193235.4 &1.081&   0.4142, 0.6202, 1.2397, 1.2719 & F2III \\
211956984&BD+20d2193      & 084244.41+193447.8 &1.089&   0.4794, 0.9488, 0.9937 & F6 \\
212033939&HD 74135        & 084253.07+204909.1& 0.777&    0.4757, 0.3128, 0.4135, 0.2734 & A9III \\
211909987&HD 74589        & 084520.53+185231.3& 0.724&    0.0434, 0.0486, 0.0552, 0.0535 & F0\\
212008515&HD 74587        & 084528.25+202343.5& 0.767&    0.0587, 0.0513, 0.0519, 0.0801 & A5; $\delta$ Scuti in SIMBAD and Kov\'acs \etal\ (2014) but unclear if identified as $\delta$ Scuti prior to Kov\'acs \etal\ (2014) \\
\enddata
\end{deluxetable}

\section{Non-Members}
\label{app:nm}

\begin{figure}[ht]
\epsscale{1.0}
\plottwo{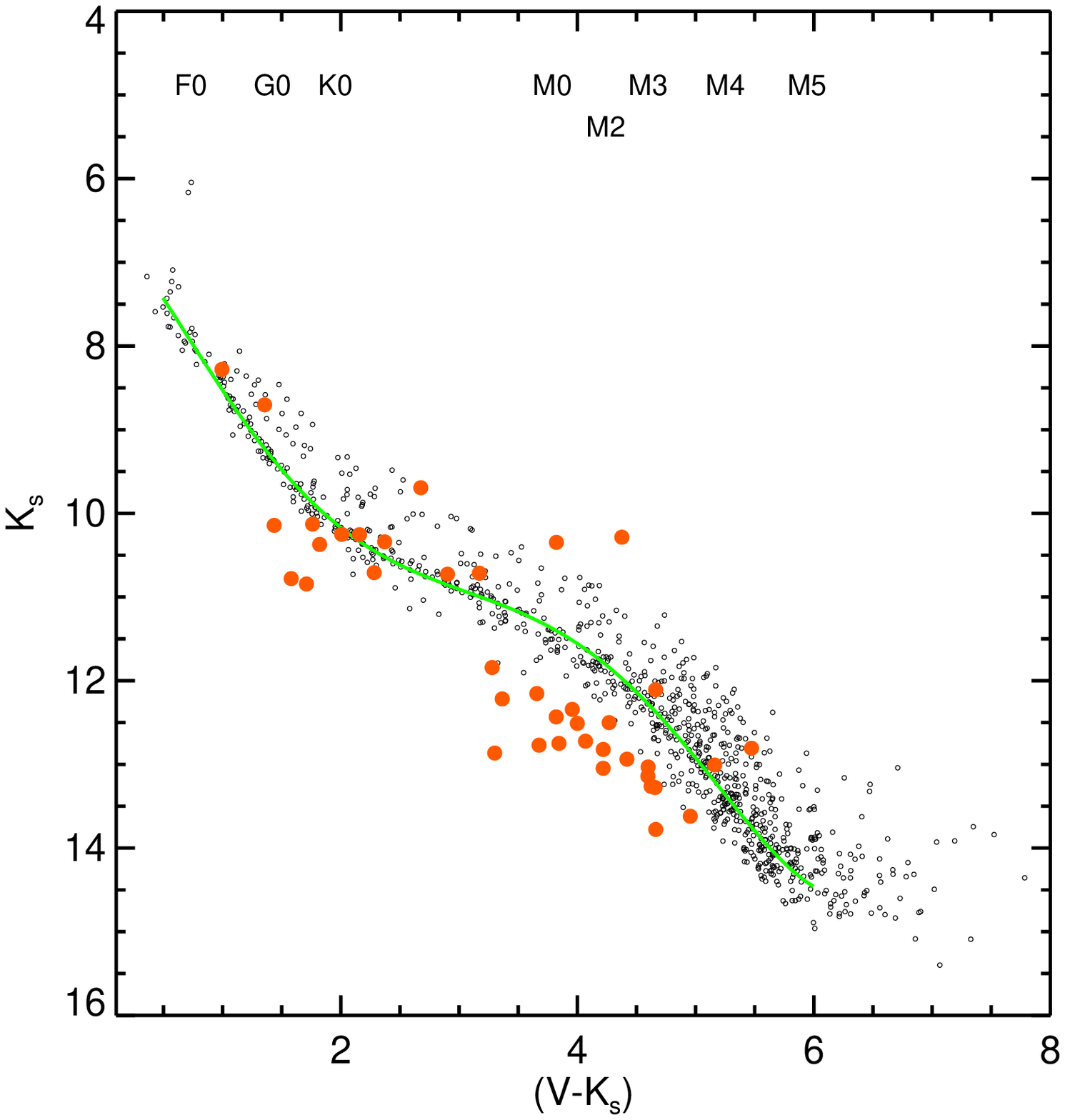}{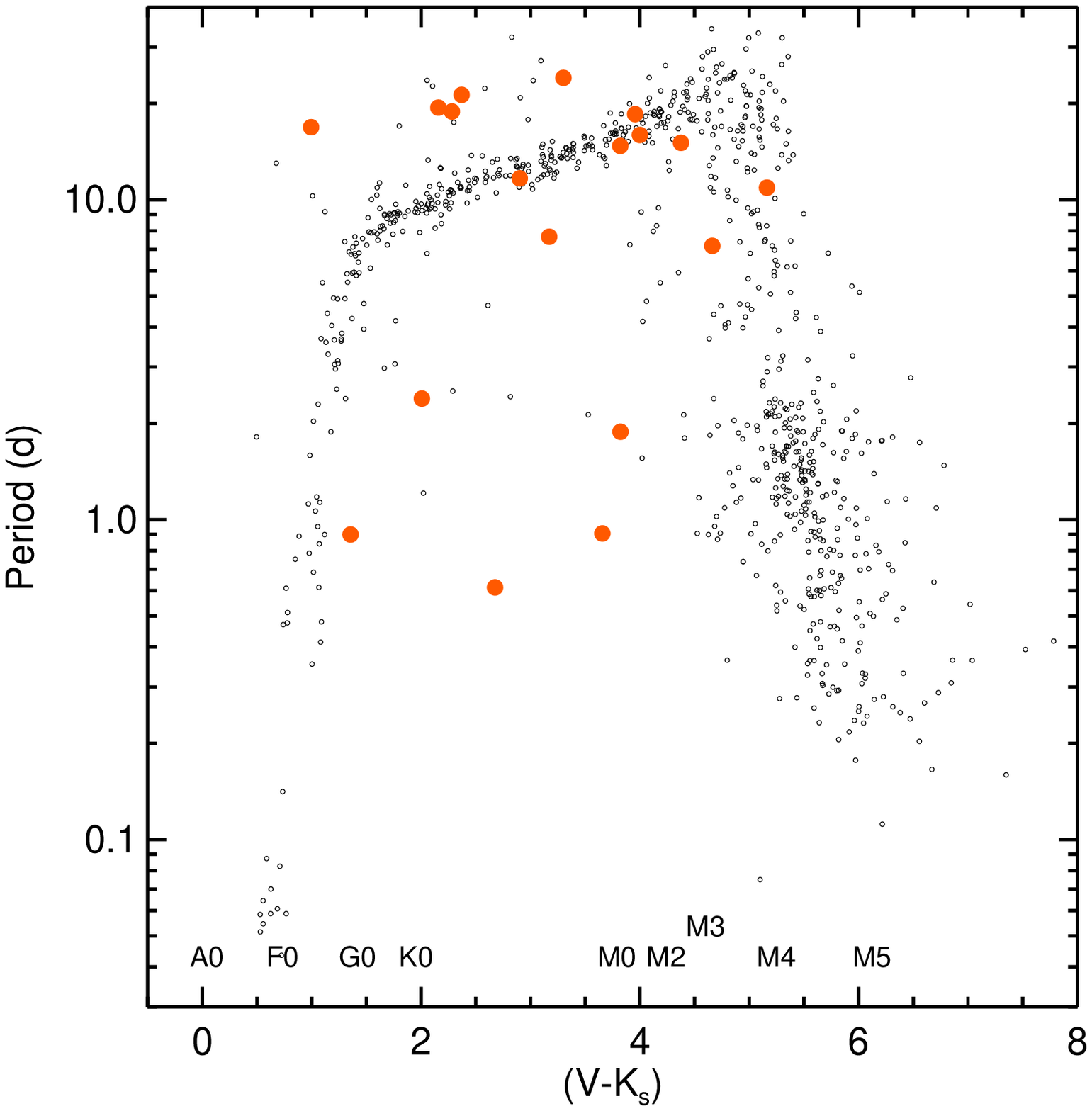}
\caption{Optical CMD (\ks\ vs.~\vmk), left,  and $P$ vs.~\vmk, right,
for the ensemble (small dots), and for stars we identify as
non-members.  Some do not have identifiable periods, and thus cannot
appear on the right. Most are outliers in both diagrams. }
\label{fig:optcmdnm}
\end{figure}

We identified stars that we took to not be members of Praesepe (see
Sec.~\ref{sec:membership}, with  modifications as per the outliers
discussion in Sec.~\ref{sec:outliers} and App.~\ref{app:pvmkoutliers}).
Table~\ref{tab:nm} lists these stars. They appear in a color-magnitude
diagram and a $P$ vs.\ \vmk\ diagram in Fig.~\ref{fig:optcmdnm}. In
general, they are not as well clustered along the main sequence in
Praesepe as the members. If they are periodic, these stars generally
are outliers in the $P$ vs.\ \vmk\ diagram, consistent with being non-members.

\floattable
\begin{deluxetable}{cclccp{4cm}}
\tabletypesize{\scriptsize}
\tablecaption{Non-Members\label{tab:nm}}
\tablewidth{0pt}
\tablehead{\colhead{EPIC} & \colhead{RA, Dec (J2000)} &\colhead{Other name}
&\colhead{\vmk\ (mag)}&\colhead{Period(s) (d)}&\colhead{Notes} }
\startdata
212158768&083012.13+231336.9&2MASSJ08301213+2313370& 2.158&31.220& New spectrum has RV inconsistent with membership; also too slow for this color\\
211897926&083150.92+184147.1&2MASSJ08315092+1841470& 2.282&24.521& New spectrum has RV inconsistent with membership; also too slow for this color\\
211981509&083239.47+195722.2&JC10& 1.821&\ldots&\ldots\\
211803269&083300.85+172146.6&\ldots& 1.709&\ldots&\ldots\\
211933061&083316.38+191317.3&JS666& 4.269&\ldots&\ldots\\
211885783&083334.80+183108.9&2MASSJ08333480+1831089& 3.279&\ldots&\ldots\\
211993704&083550.36+200904.5&JS98& 4.419&\ldots&\ldots\\
212033649&083608.34+204852.8&\ldots& 4.663&\ldots&\ldots\\
212099156&083626.89+220016.8&2MASSJ08362690+2200168& 3.845&\ldots&\ldots\\
211955417&083649.56+193322.8&JC75& 4.376&12.019, 5.637, 1.235&\ldots\\
211801256&083701.08+172005.1&2MASSJ08370108+1720051& 4.597&\ldots&\ldots\\
211930233&083703.45+191041.1&JC85& 2.902&18.941& New spectrum has RV inconsistent with membership; also too slow for this color\\
211833885&083707.22+174748.8&2MASSJ08370722+1747487& 3.676&\ldots&\ldots\\
212021664&083713.33+203653.6&JS170& 3.957&22.048& also too slow for this color \\
212072324&083816.83+212953.7&JS223& 4.069&\ldots&\ldots\\
211834700&083844.47+174829.4&ANM690& 1.436&\ldots&\ldots\\
211990313&084028.47+200551.2&JS363& 4.598&\ldots&\ldots\\
212066519&084030.72+212333.1&2MASSJ08403072+2123331& 4.624&\ldots&\ldots\\
212072257&084048.52+212949.5&2MASSJ08404852+2129495& 3.365&\ldots&\ldots\\
211984180&084053.83+195956.0&JS392& 4.662&23.994, 1.863&also too slow for this color\\
211988382&084058.92+200359.2&JS399& 4.000&25.393& also too slow for this color\\
211954582&084113.18+193234.9&JC243& 3.823& 3.191& also too fast for this color\\
212005583&084113.73+202051.7&JS417& 3.301&21.610& also too slow for this color\\
212094510&084120.89+215454.0&2MASSJ08412090+2154540& 3.821&30.708& also too slow for this color\\
212112522&084128.93+221605.3&\ldots& 1.355&16.018, 5.011& also too slow for this color\\
211812292&084135.58+172927.1&\ldots& 4.218&\ldots&\ldots\\
212084898&084136.38+214353.8&JS732& 1.580&\ldots&\ldots\\
211784450&084155.14+170540.8&ANM1065& 2.676&10.992,13.714&\ldots\\
211995547&084221.63+201053.1&\ldots& 0.994& 0.591&\ldots\\
212137243&084221.95+224552.2&2MASSJ08422195+2245521& 3.657& 8.950& also too fast for this color\\
211931928&084303.65+191215.9&JS739& 5.473&\ldots&\ldots\\
212112578&084338.80+221609.1&KW529& 2.007& 8.878&\ldots\\
212029850&084430.66+204505.0&SDSSJ084430.65+204504.7& 5.162& 0.213, 0.227& also too fast for this color\\
212011328&084501.06+202631.9&JS585& 1.761&\ldots&\ldots\\
212069325&084811.53+212635.0&\ldots&\ldots&\ldots&\ldots\\
212120476&084850.34+222531.9&HSHJ506& 3.171&17.097& New spectrum has RV inconsistent with membership; also too slow for this color\\
211921444&084852.61+190247.0&2MASSJ08485261+1902470& 4.219&\ldots&\ldots\\
211905939&085001.55+184852.6&2MASSJ08500156+1848526& 4.955&\ldots&\ldots\\
211939409&085125.81+191856.3&\ldots&\ldots&\ldots&\ldots\\
211935447&085528.96+191523.3&2MASSJ08552896+1915234& 4.657&\ldots&\ldots\\
211875602&090222.36+182223.8&2MASSJ09022236+1822238& 2.371&15.490& New spectrum has RV inconsistent with membership; also too slow for this color\\
\enddata
\end{deluxetable}

\section{Bright Giants}
\label{app:bright}

The two stars with \ks$>$6 are EPIC 211984704/39 Cnc=HD 73665  and EPIC
211976270/HD 73974. These were discarded from our sample as too
bright, but we note some characteristics of their LCs here.

The star for which we do not determine a periodic signal using these
reduced LCs is EPIC 211976270/HD 73974; it is a giant (K0III; Yang
\etal\ 2015). It is very bright, and the artifacts in the light curve
reductions we have reflect that. The dominant structure in all the LC
versions seems to be the 0.245 day thruster firings.

We find a period in the other star, which is is EPIC 211984704/39
Cnc=HD 73665. It is also a giant (G8III; Yang \etal\ 2015).  We
determine only one period, 1.2411 d. While this object is also very
bright, there seems to be a readily apparent $\sim$1 d oscillation
even in the raw LC versions, which is why we retained this period. The
phased LC has a lot of scatter in any of the LC versions, admittedly,
but there is a peak near 1.24 d that persists across LC versions. But,
what does that periodicity represent?

There are 4 giants in Praesepe, but there are only 2 K2 LCs. The other
two giants are HD73598 (K0III) and HD73710 (G9III).  Pasquini \etal\
(2000)  provide $v \sin i$ values for two of them, and give an average
$v \sin i$ for all four, and they are between 1 and 3 km s$^{-1}$,
comparable to field red giants.  Choi \etal\ (1995)  used long
duration monitoring of Calcium HK emission to find rotation
periods for three of them.  (HD 73598 has $P\sim$159 d; HD73974 has
$P\sim$112 d; HD73710 has $P\sim$155 d.) All of those are more or less
consistent with the $v \sin i$ values. So, the 1.2 d period we see in
the K2 LC cannot be the rotation period.

An estimate of the max power in the oscillations for stars is (Brown
1991) $\nu_{\rm max}/\nu_{\rm max,\odot}  \sim (M/M_{\odot} \times
(T_{\rm eff}/T_{\rm eff,\odot})^{3.5} \times  (L/L_{\odot})^{-1}$.
For the Sun, $\nu_{\rm max,\odot}$ is about  3100 $\mu$Hz  (which
converts to about 5 minutes for a period).  Assuming that the \teff\
for the giants in Praesepe is about 4800 K, and taking their
luminosity to be $\sim$60 $L_{\odot}$, and their masses as
$\sim$2.5 $M_{\odot}$, the max oscillation power for a Praesepe giant
should be at about 4 hours.  So, the 1.2 d period from the K2 light
curve is also not likely to be pulsation.

The star is saturated in the K2 thumbnail, but in general that should
result in no derived period, or a period at 0.245 d (or a multiple of
that), reflecting the thruster firings (as for the other giant with a
K2 LC), not a 1.2 d period. We are not sure if this 1.2 d period is
real, nor how to interpret it.

\section{Outliers in the $P$ vs.~\vmk\ Plot: Detailed Notes}
\label{app:pvmkoutliers}

\begin{figure}[ht]
\epsscale{1.0}
\plottwo{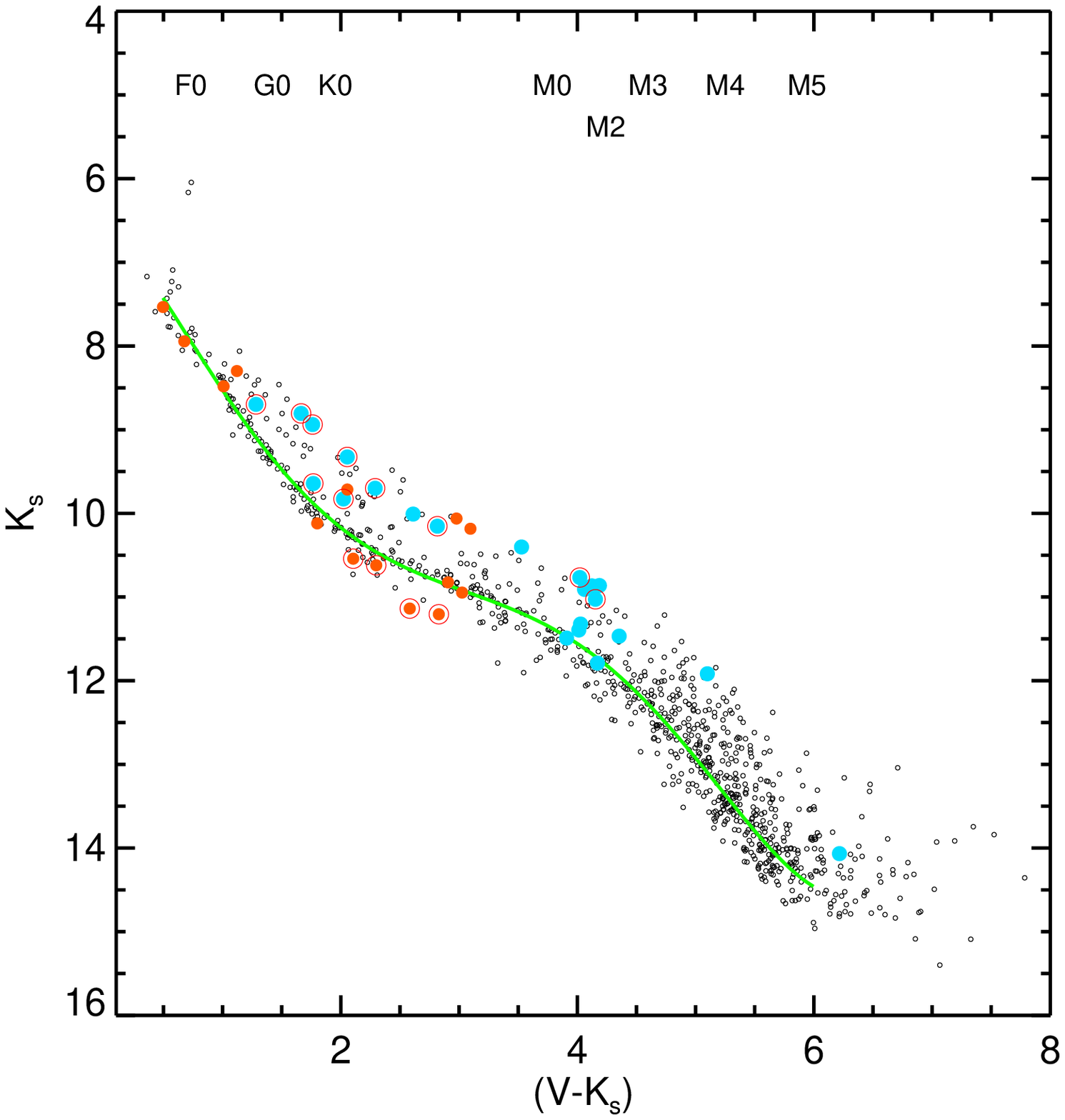}{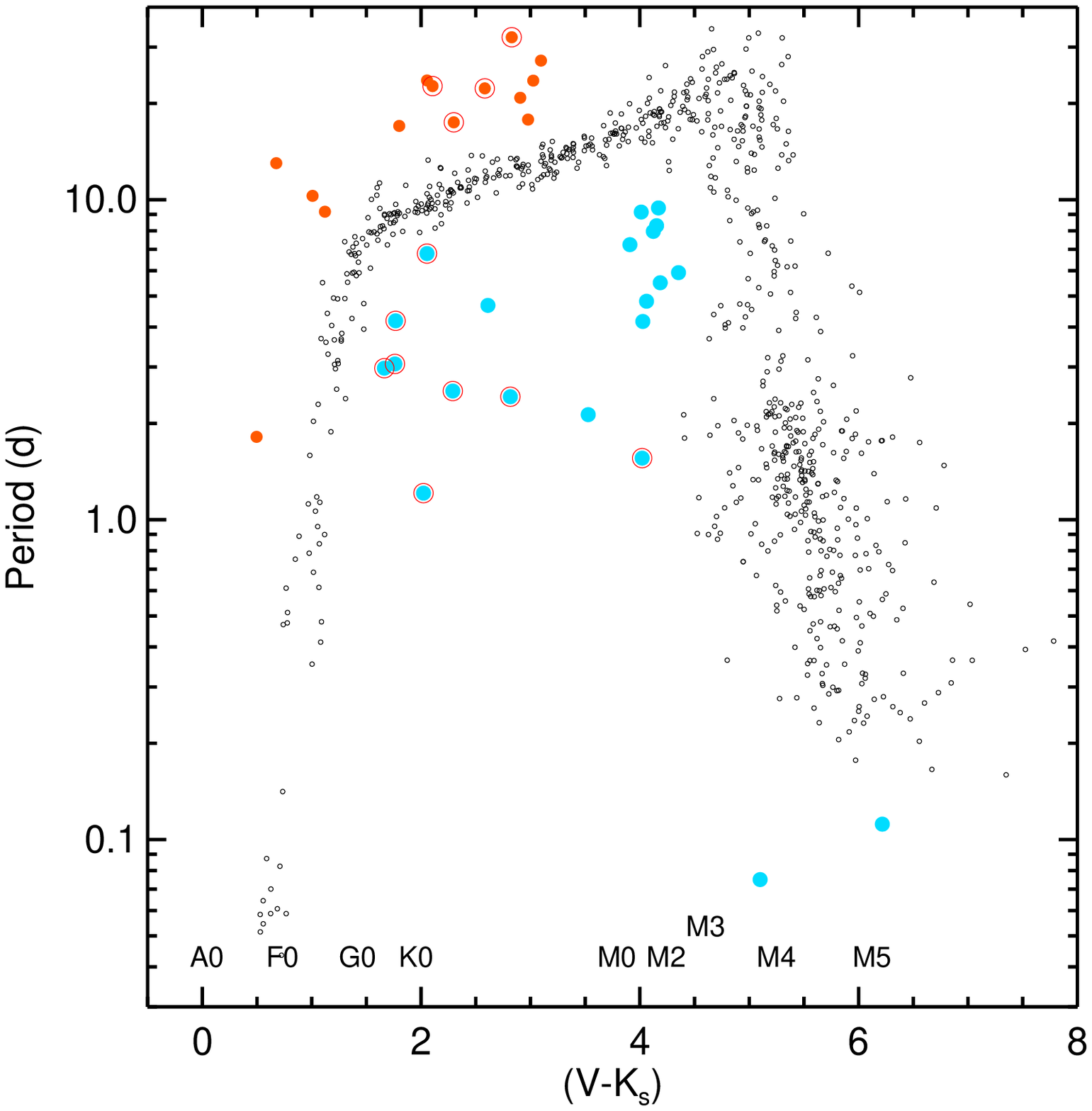}
\caption{Optical CMD (\ks\ vs.~\vmk), left,  and $P$ vs.~\vmk, right,
for the ensemble (small dots), and for stars we identify as rapidly
rotating outliers (blue dots) and slowly rotating outliers (orange
dots). Non-members have already been removed from these diagrams. An
additional red circle around an orange point means that there is some
doubt that the star is a member; an additional red circle around a
blue point means that the star is a known or suspected short-period
spectroscopic binary, where tidal synchronization may have affected the 
star's rotation period. }
\label{fig:optcmdoutliers}
\end{figure}

Section~\ref{sec:outliers} above highlights some of the outliers in
the $P$ vs.~\vmk\ plot, focusing on both the objects rotating too
quickly for their color and those that are going too slowly. Here we
focus on individual objects that were covered more broadly above.
Fig.~\ref{fig:optcmdoutliers} includes an optical CMD and the $P$
vs.~\vmk\ plot with these objects highlighted; they are listed in
Tables~\ref{tab:fastoutliers} and \ref{tab:slowoutliers}.

For the fast rotating stars, we checked the veracity of their colors
and periods, and all of them seemed consistent with the observations. 
Among the fast rotating K stars, many of them are known short-period
binaries, and we infer that the rest of them are probably binaries as
well. Three of the ones with published orbital periods of less than 4
days are 211896596/JS655=FV Cnc, 211975006/KW 367, and 211935518/KW
434=S184 (Mermilliod \& Mayor 1999; Mermilliod \etal\ 1990).  One more
(211909748/KW401) is a known SB with a large radial velocity 
amplitude (Barrado \etal\ 1998), and another is simply listed as an SB
in Mermilliod \& Mayor (1999).  Note that there is little literature
binary monitoring work done for \vmk$>$3.

For the slowly rotating stars, we also checked colors and periods. The
assessment of their periods is more difficult than other
shorter-period stars, as described in Sec.~\ref{sec:outliers} above.
Stars particularly subject to this are noted as such in
Table~\ref{tab:slowoutliers}.

Most of the slow outliers with 1.7$<$\vmk$<$3.4 and $P >$ 15 are
relatively anonymous stars, with only a few references in SIMBAD.  
Particularly for the few slowly rotating stars that are photometric
binaries in Fig.~\ref{fig:optcmdoutliers}, tidal synchronization in a
short-period binary system may be a good explanation for the slow
rotation. In most cases, they are listed as candidate members in only
one or two Praesepe proper motion membership studies.  Only one of
them has a published radial velocity; that star (212034762/JS409)
appears to be a member.  We have obtained Keck HIRES spectra for 11 of
these stars, which are noted in Table~\ref{tab:slowoutliers}  (see
App.~\ref{app:keck} for details). About half of the ones for which we
have spectroscopy appear to have radial velocities inconsistent with 
membership. Many of these slow outliers   have CMD locations and
proper motions generally consistent with being Praesepe members.  Some
have proper motions consistent with  membership, but are slightly low
in the CMD; these are noted in  Table~\ref{tab:slowoutliers}.  The
lack of detailed information for these stars makes it more likely that
they might be non-members, but more information is needed.

\floattable
\begin{deluxetable}{cclccp{7cm}}
\tabletypesize{\scriptsize}
\rotate
\tablecaption{Fast Outliers in the $P$ vs.~\vmk\ Plot\label{tab:fastoutliers}}
\tablewidth{0pt}
\tablehead{\colhead{EPIC} & \colhead{RA, Dec (J2000)} &\colhead{Other name}
&\colhead{\vmk\ (mag)}&\colhead{Period(s) (d)}&\colhead{Notes} }
\startdata
211898878&081903.68+184236.1&2MASSJ08190368+1842361& 5.100& 0.075& fastest, bluest M star; it is one of the furthest west sources in Fig.~\ref{fig:where}; it may be a CV\\
212009427&083129.87+202437.5&AD1508=2MASSJ08312987+2024374& 4.021& 1.557& likely EB (also listed in EBs table)\\
211971354&083140.45+194754.2&HSHJ15& 4.153& 8.297, 9.374, 0.798& $\Delta P>$6 d, likely binary; bluest end of sparsely populated $P$ range of M stars \\
212034371&083556.94+204934.7&JS102=2MASSJ08355696+2049346& 2.022& 1.211& binary (also listed in binaries table)\\
211915860&083656.25+185747.9&JS159=JC80& 4.061& 4.816&bluest end of sparsely populated $P$ range of M stars \\
212021253&083728.46+203628.5&JS181& 3.909& 7.238& no RVs; bluest end of sparsely populated $P$ range of M stars \\
211977390&083749.95+195328.8&KW55=S27& 2.054& 6.789&  binary (also listed in binaries table)\\
211988628&083915.79+200414.0&KW566=JS280& 4.123& 7.964&bluest end of sparsely populated $P$ range of M stars \\
211983725&083941.03+195928.8&KW570=JS309& 4.027& 4.165&bluest end of sparsely populated $P$ range of M stars \\
211918335&084001.70+185959.4&KW244=TXCnc& 1.282&(0.1914)& already removed $P$ as WUma binary (also listed in binaries table)\\
211920022&084005.71+190130.6&KW256=2MASSJ08400571+1901307& 2.611& 4.675& \ldots \\
211984704&084006.41+200028.0&39Cnc& 2.219& 1.241& already removed from sample as too bright (G8III giant)\\
211915202&084029.21+185709.4&\ldots& 4.186& 5.503, 2.004&bluest end of sparsely populated $P$ range of M stars \\
211898181&084037.88+184200.4&KW547& 2.290& 2.524&  binary (also listed in binaries table)\\
212013132&084044.25+202818.6&JS379=2MASSJ08404426+2028187& 3.528& 2.129, 4.367, 4.053& no RVs known\\
211975006&084109.60+195118.6&KW367=BD+19d2077& 1.761& 3.068&  binary (also listed in binaries table)\\
211954582&084113.18+193234.9&JC243& 3.823& 3.191& already NM (also listed in NM table)\\
211909748&084130.70+185218.7&KW401=2MASSJ08413070+1852188& 2.816& 2.422&  binary (also listed in binaries table)\\
211935518&084154.36+191526.7&KW434=S184& 1.767& 4.184&  binary (also listed in binaries table)\\
212137243&084221.95+224552.2&2MASSJ08422195+2245521& 3.657& 8.950&already NM (also listed in NM table)\\
211890774&084232.06+183528.0&JS488& 4.170& 9.428& marginal outlier; bluest end of sparsely populated $P$ range of M stars \\
212066424&084300.54+212328.1&2MASSJ08430054+2123281& 6.217& 0.112& fastest M star in clump; everything about it looks fine\\
212029850&084430.66+204505.0&SDSSJ084430.65+204504.7& 5.162& 0.213, 0.227& already NM (also listed in NM table) \\
211896596&084801.74+184037.6&JS655=FVCnc& 1.664& 2.975&  binary (also listed in binaries table)\\
211773459&084832.70+165623.6&2MASSJ08483271+1656236& 4.353& 5.919&bluest end of sparsely populated $P$ range of M stars \\
211885995&084926.76+183119.5&2MASSJ08492676+1831195& 4.013& 9.153& marginal outlier; bluest end of sparsely populated $P$ range of M stars  \\
\enddata
\end{deluxetable}

\floattable
\begin{deluxetable}{cclccp{7cm}}
\tabletypesize{\scriptsize}
\rotate
\tablecaption{Slow Outliers in the $P$ vs.~\vmk\ Plot\label{tab:slowoutliers}}
\tablewidth{0pt}
\tablehead{\colhead{EPIC} & \colhead{RA, Dec (J2000)} &\colhead{Other name}
&\colhead{\vmk\ (mag)}&\colhead{Period(s) (d)}&\colhead{Notes} }
\startdata
212032123&082507.04+204725.1&2MASSJ08250705+2047252& 2.105&22.662& just below MS, and $>$3.6$\arcdeg$ from cluster center; judgement call as to whether $P$ is $P_{\rm rot}$ or timescale \\
212158768&083012.13+231336.9&2MASSJ08301213+2313370& 2.158&31.220& new spectrum has RV inconsistent with membership (also listed in NM table)\\
212077235&083110.44+213522.4&2MASSJ08311044+2135224& 2.829&32.210& slightly below MS; judgement call as to whether $P$ is $P_{\rm rot}$ or timescale\\
211897926&083150.92+184147.1&2MASSJ08315092+1841470& 2.282&24.521& slightly below MS; new spectrum has RV inconsistent with membership (also listed in NM table); judgement call as to whether $P$ is $P_{\rm rot}$ or timescale\\
211898294&083249.71+184206.2&JS14=JC12& 2.978&17.808&  new spectrum is double lined and has RV consistent with membership\\
211916015&083629.84+185757.0&KW536=BD+19d2045& 1.121& 9.175& upper left of diagram (near blue bend in distribution); judgement call as to whether $P$ is $P_{\rm rot}$ or timescale \\
211910450&083647.99+185258.0&\ldots& 0.676&13.011&judgement call as to whether $P$ is $P_{\rm rot}$ or timescale; far upper left of diagram \\
211930233&083703.45+191041.1&JC85& 2.902&18.941& new spectrum has RV inconsistent with membership (also listed in NM table)\\
212021664&083713.33+203653.6&JS170& 3.957&22.048& already identified as NM (also listed in NM table)\\
211946055&083751.85+192453.4&\ldots& 2.299&17.459& just below MS\\
211892153&083821.66+183639.9&JC123& 3.096&27.225& judgement call as to whether $P$ is $P_{\rm rot}$ or timescale\\
211931736&083958.06+191205.9&KW227=HD73641& 1.008&10.285,14.992, 1.536& upper left of diagram (near blue bend in distribution); noisy LC\\
211984180&084053.83+195956.0&JS392& 4.662&23.994, 1.863& already identified as NM (also listed in NM table); $\Delta P >$6d\\
211988382&084058.92+200359.2&JS399& 4.000&25.393& already identified as NM (also listed in NM table)\\
212034762&084111.05+204957.8&JS409& 1.800&17.016&new spectrum has RV consistent with membership; judgement call as to whether $P$ is $P_{\rm rot}$ or timescale \\
212005583&084113.73+202051.7&JS417& 3.301&21.610& already identified as NM (also listed in NM table)\\
211990785&084119.43+200618.2&\ldots& 2.582&22.292& slightly below MS; judgement call as to whether $P$ is $P_{\rm rot}$ or timescale\\
212094510&084120.89+215454.0&2MASSJ08412090+2154540& 3.821&30.708& already identified as NM (also listed in NM table); judgement call as to whether $P$ is $P_{\rm rot}$ or timescale \\
212112522&084128.93+221605.3&\ldots& 1.355&16.018, 5.011& already identified as NM (also listed in NM table); $\Delta P >$6d\\
212098754&084212.66+215948.8&\ldots& 2.055&23.601& new spectrum has RV consistent with membership; judgement call as to whether $P$ is $P_{\rm rot}$ or timescale\\
211921647&084546.53+190258.1&\ldots& 0.496& 1.815, 1.700& above the bluest branch\\
212120476&084850.34+222531.9&HSHJ506& 3.171&17.097& new spectrum has RV inconsistent with membership (also listed in NM table); judgement call as to whether $P$ is $P_{\rm rot}$ or timescale \\
212027750&084914.76+204300.9&2MASSJ08491476+2043009& 2.907&20.837& new spectrum has RV consistent with membership\\
212008710&085502.23+202354.0&2MASSJ08550224+2023540& 3.025&23.573& new spectrum has RV consistent with membership; judgement call as to whether $P$ is $P_{\rm rot}$ or timescale \\
211875602&090222.36+182223.8&2MASSJ09022236+1822238& 2.371&15.490& new spectrum has RV inconsistent with membership (also listed in NM table)\\
\enddata
\end{deluxetable}

\clearpage

\section{Halo outliers}
\label{app:halo}

\begin{figure}[ht]
\epsscale{1.0}
\plottwo{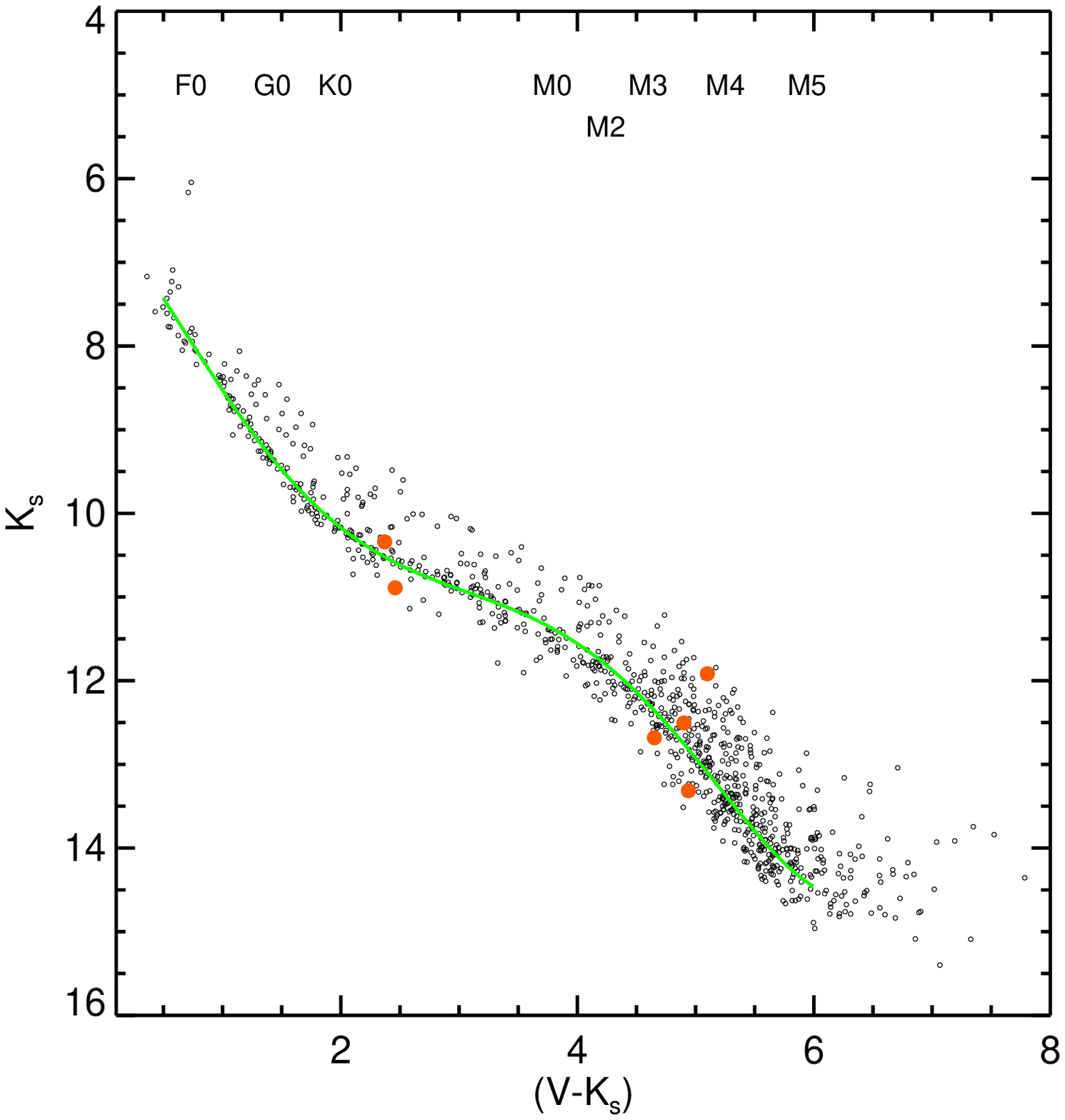}{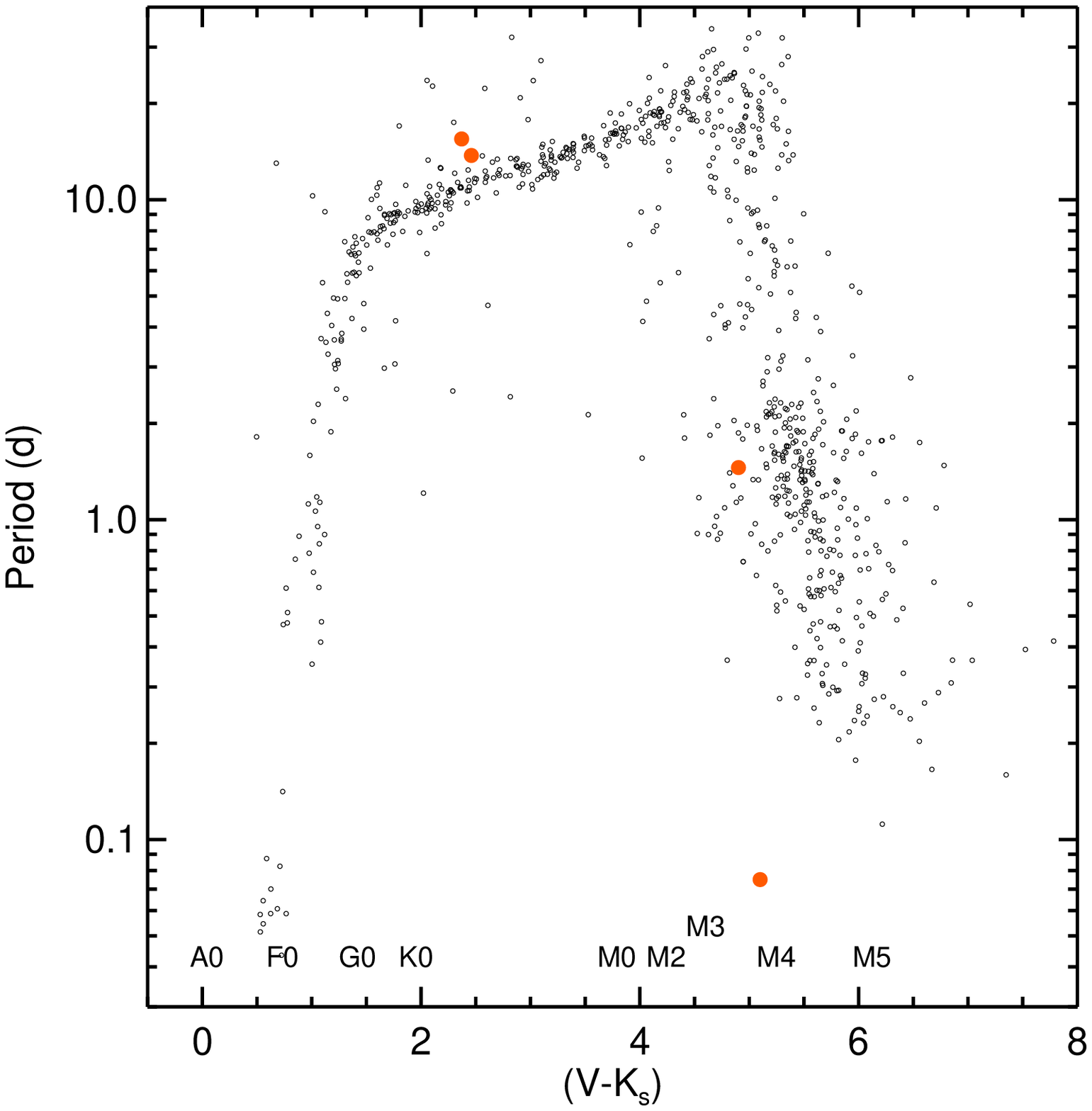}
\caption{Optical CMD (\ks\ vs.~\vmk), left,  and $P$ vs.~\vmk, right,
for the ensemble (small dots), and for stars more than 5$\arcdeg$ from the cluster center.   }
\label{fig:halo}
\end{figure}

The tidal radius of Praesepe is 12.1 pc (Holland \etal\ 2000), which
at a distance of 184 pc is $\sim$3.8$\arcdeg$ across. Six of the stars
in the original sample of targets with K2 light curves are more than
5$\arcdeg$ away from the cluster center; see Table~\ref{tab:halo}.
They appear in a CMD and the $P$ vs.~color diagram in
Fig.~\ref{fig:halo}. One is dropped as a non-member
(Sec.~\ref{sec:membership} and App.~\ref{app:nm}). Two more are listed
as outliers in the $P$ vs.~\vmk\ diagram
(App.~\ref{app:pvmkoutliers}). In general, these stars are poorly
studied, but we have retained those stars that met the rest of our
criteria.

\floattable
\begin{deluxetable}{cccccp{5cm}}
\tabletypesize{\scriptsize}
\tablecaption{Outer halo of cluster\label{tab:halo}}
\tablewidth{0pt}
\tablehead{\colhead{EPIC} & \colhead{RA, Dec (J2000)} &\colhead{Other name}
&\colhead{Period(s) (d)} &\colhead{Distance (deg)} &\colhead{Notes}  }
\startdata
211999629&081259.31+201453.6&2MASSJ08125931+2014537&\ldots&6.4 & little additional relevant information found; perhaps below single-star MS  \\
211756862&081551.30+164215.9&\ldots& 1.455, 2.627&6.4 & little additional relevant information found\\
211984209&081615.53+195956.9&2MASSJ08161554+1959570&13.754&5.6 & Adams \etal\ (2002) has membership probability of 0.41; perhaps below single-star MS \\
211898878&081903.68+184236.1&2MASSJ08190368+1842361& 0.075&5.0 & One of the fastest rotating M stars (listed as fast outlier); Adams \etal\ (2002) has membership probability of 0.39; a CV rather than member?\\
212148559&082333.40+230007.9&2MASSJ08233340+2300079&\ldots&5.1 & Adams \etal\ (2002) has membership probability of 0.31; perhaps below single-star MS \\
211875602&090222.36+182223.8&2MASSJ09022236+1822238&15.490&5.4 & already dropped as a NM and long-P outlier (listed in corresponding tables); above single-star MS \\
\enddata
\end{deluxetable}

\clearpage

\section{New Keck Spectra}
\label{app:keck}

We obtained high resolution spectra for several of the anomalously
slowly rotating stars (App.~\ref{app:pvmkoutliers}) and all of the
objects with odd LCs (Sec.~\ref{sec:batwingy}) using the Keck HIRES
spectrograph Vogt \etal\ 1994). The observations were taken on one of:
2016 October 14, 2016 December 22, 2016 December 26, or 2017 January
13, UT, and  cover the wavelength range roughly 4800 to 9200 \AA at a
spectral resolution of  $R \approx 45,000$; the spectra were were
reduced using the {\em makee} software written by Tom Barlow. We
measured radial velocities using the cross correlation techniques in
the {\em rv} package in {\em IRAF}, with absolute reference to between
3 and 5 (depending on the night) late type radial velocity standards.
Because these are slowly rotating stars, the errors are all under 0.5
km s$^{-1}$, and are determined from the empirical scatter among
orders and reference stars for each observation.

\floattable
\begin{deluxetable}{cclccp{7cm}}
\tabletypesize{\scriptsize}
\rotate
\tablecaption{Objects with New  Keck Spectra\label{tab:keck}}
\tablewidth{0pt}
\tablehead{\colhead{EPIC} & \colhead{RA, Dec (J2000)} &\colhead{Other name}
&\colhead{RV (km s$^{-1}$)} &\colhead{Notes} }
\startdata
212158768&083012.13+231336.9&2MASSJ08301213+2313370& 26.36$\pm$0.32 & non-member long-P outlier\\
211897926&083150.92+184147.1&2MASSJ08315092+1841470& 9.08$\pm$0.32 & non-member long-P outlier \\
211898294&083249.71+184206.2&JS14=JC12& 27.12$\pm$0.26 and 41.33$\pm$0.39 & double-lined, member long-P outlier\\
212011416&083308.44+202637.3& 2MASSJ08330845+2026372 & 35.08$\pm$0.31 & unusual LC shape, member\\
211930233&083703.45+191041.1&JC85&23.88$\pm$0.26 & non-member long-P outlier \\
211915940&083800.61+185752.9& JS208 & 35.99$\pm$0.47 & unusual LC shape, member\\
212034762&084111.05+204957.8&JS409& 34.32 $\pm$0.30 & non-member long-P outlier \\
212098754&084212.66+215948.8& HSHJ407& 35.22$\pm$0.35 & member long-P outlier\\
211931651&084322.40+191200.7& AD3196=CP Cnc & 34.48$\pm$0.26 & unusual LC shape, member\\
212120476&084850.34+222531.9&HSHJ506& 16.61$\pm$0.31 & non-member long-P outlier \\
212027750&084914.76+204300.9&2MASSJ08491476+2043009& 35.95$\pm$0.48 & member long-P outlier\\
212008710&085502.23+202354.0&2MASSJ08550224+2023540& 37.07$\pm$0.40 & member long-P outlier\\
211875602&090222.36+182223.8&2MASSJ09022236+1822238& 28.57$\pm$0.42 & non-member long-P outlier\\
\enddata
\end{deluxetable}

\end{document}